\documentclass[traditabstract]{aa}

\usepackage{graphicx}
\usepackage{latexsym}
\usepackage{amssymb}
\usepackage{amsmath}
\usepackage{morefloats}
\usepackage{color}
\usepackage{footnote}
\usepackage{tablefootnote}
\usepackage{subfig}
\usepackage{appendix}
\usepackage{natbib,twoopt}

\newcommand{\mpc}{\mathrm{Mpc}}
\newcommand{\kpc}{\mathrm{kpc}}
\newcommand{\ghz}{\mathrm{GHz}}
\newcommand{\kelvin}{\mathrm{K}}
\newcommand{\mum}{\mathrm{\mu m}}

\newcommand{\niitof}{\mathrm{[N\textsc{II}]\,205}\,\mum}
\newcommand{\cits}{\mathrm{[C\textsc{I}]\,370}\,\mum}
\newcommand{\cison}{\mathrm{[C\textsc{I}]\,609}\,\mum}
\newcommand{\sfrd}{\Sigma_{\mathrm{SFR}}}
\newcommand{\sfrdunit}{\mathrm{M_{\odot}}\,\mathrm{yr}^{-1}\,\mathrm{kpc}^{-2}}
\newcommand{\htwod}{\mathrm{\Sigma_{H_{2}}}}
\newcommand{\scoone}{I_{\mathrm{CO(1-0)}}}
\newcommand{\scowarm}{I_{\mathrm{CO(J_{up}=4\ to\ 8)}}}

\newcommand{\uav}{$\langle U\rangle$}
\newcommand{\cmcube}{\mathrm{cm^{-3}}}
\newcommand{\emico}{\epsilon_{\mathrm{CO(J=1-0)}}}
\newcommand{\mhtwoico}{M(\mathrm{H_{2}})_{\mathrm{J}=1-0}}
\newcommand{\mhtwonco}{M(\mathrm{H_{2}})_{N(\mathrm{CO})}}
\newcommand{\ghtwodep}{\tau^{\mathrm{H_{2}}}_{dep}}
\bibpunct{(}{)}{;}{a}{}{,} 
\makeatletter
\newcommandtwoopt{\citeads}[3][][]{\href{http://adsabs.harvard.edu/abs/#3}%
{\def\hyper@linkstart##1##2{}%
\let\hyper@linkend\@empty\citealp[#1][#2]{#3}}}
\newcommandtwoopt{\citepads}[3][][]{\href{http://adsabs.harvard.edu/abs/#3}%
{\def\hyper@linkstart##1##2{}%
\let\hyper@linkend\@empty\citep[#1][#2]{#3}}}
\newcommandtwoopt{\citetads}[3][][]{\href{http://adsabs.harvard.edu/abs/#3}%
{\def\hyper@linkstart##1##2{}%
\let\hyper@linkend\@empty\citet[#1][#2]{#3}}}
\newcommandtwoopt{\citeyearads}[3][][]%
{\href{http://adsabs.harvard.edu/abs/#3}
{\def\hyper@linkstart##1##2{}%
\let\hyper@linkend\@empty\citeyear[#1][#2]{#3}}}
\makeatother

\begin{document}
\sloppy

\title{Spatially resolved physical conditions of molecular gas and potential star formation tracers in M83, revealed by the \textsl{Herschel} SPIRE FTS}
\titlerunning{Physical conditions in M83 revealed by the \textsl{Herschel} SPIRE FTS}

\author{
R. Wu\inst{1,2}
\and
S. C. Madden\inst{1}
\and
F. Galliano\inst{1}
\and
C. D. Wilson\inst{3}
\and
J. Kamenetzky\inst{4}
\and
M.-Y. Lee\inst{1}
\and
M. Schirm\inst{3}
\and
S. Hony
\and
V. Lebouteiller\inst{1}
\and
L. Spinoglio\inst{5}
\and
D. Cormier\inst{6}
\and
J. Glenn\inst{4}
\and
P. R. Maloney\inst{4}
\and
M. Pereira-Santaella\inst{5,6}
\and
A. R\'emy-Ruyer\inst{1}
\and
M. Baes\inst{8}
\and
A. Boselli\inst{9}
\and
F. Bournaud\inst{1}
\and
I. De Looze\inst{8}
\and
T. M. Hughes\inst{8}
\and
P. Panuzzo\inst{10}
\and
N. Rangwala\inst{4}
}

\institute{
Laboratoire AIM, CEA/Saclay, L'Orme des Merisiers, 91191 Gif-sur-Yvette, France
\and
International Research Fellow of the Japan Society for the Promotion of Science (JSPS)\\
Department of Astronomy, the University of Tokyo, Bunkyo-ku, Tokyo 113-0033, Japan\\
\email{ronin.wu@astron.s.u-tokyo.ac.jp}
\and
McMaster University, Department of Physics and Astronomy, Hamilton, Ontario, L8S 4M1, Canada
\and
Center for Astrophysics and Space Astronomy, 389-UCB, University of Colorado, Boulder, CO 80303, USA
\and
Istituto di Astrofisica e Planetologia Spaziali, INAF, Via Fosso del Cavaliere 100, I-00133 Roma, Italy
\and
Centro de Astrobiología (CSIC), Ctra de Torrejón a Ajalvir, km 4, 28850 Torrej\'on de Ardoz, Madrid, Spain
\and
Zentrum f\"ur Astronomie der Universit\"at Heidelberg, Institut f\"ur Theoretische Astrophysik, Albert-Ueberle-Str. 2, 69120 Heidelberg, Germany
\and
Sterrenkundig Observatorium, Universiteit Gent, Krijgslaan 281 S9, 9000 Gent, Belgium
\and
Laboratoire dAstrophysique de Marseille, Universit\'e d’Aix-Marseille \& CNRS, UMR7326, 38 rue F. Joliot-Curie, 13388 Marseille Cedex 13, France
\and
Observatoire de Paris - Lab. GEPI, Bat. 11, 5, place Jules Janssen, 92195 Meudon Cedex, France
}
\authorrunning{Ronin Wu et al.}

\abstract{We investigate the physical properties of the molecular and ionized gas, and their relationship to the star formation and dust properties in M83, based on submillimeter imaging spectroscopy from within the central $3.5'$~($\sim\,4\ \mathrm{kpc}$ in diameter) around the starburst nucleus. The observations use the Fourier Transform Spectrometer~(FTS) of the Spectral and Photometric Imaging REceiver (SPIRE) onboard the \textsl{Herschel} Space Observatory. The newly observed spectral lines include $\cits$, $\cison$, $\niitof$, and CO transitions from $\mathrm{J}=4-3$ to $\mathrm{J}=13-12$. Combined with previously observed $\mathrm{J}=1-0$ to $\mathrm{J}=3-2$ transitions, the CO spectral line energy distributions are translated to spatially resolved physical parameters, column density of CO, $N(\mathrm{CO})$, and molecular gas thermal pressure, $P_{\mathrm{th}}$, with a non-local thermal equilibrium~(non-LTE) radiative transfer model, \texttt{RADEX}. Our results show that there is a relationship between the spatially resolved intensities of $\niitof$ and the surface density of the star formation rate~(SFR), $\sfrd$. This relation, when compared to integrated properties of ultra-luminous infrared galaxies~(ULIRGs), exhibits a different slope, because the $\niitof$ distribution is more extended than the SFR. The spatially resolved $\cits$, on the other hand, shows a generally linear relationship with $\sfrd$ and can potentially be a good SFR tracer. Compared with the dust properties derived from broad-band images, we find a positive trend between the emissivity of CO in the $\mathrm{J}=1-0$ transition with the average intensity of interstellar radiation field~(ISRF), \uav. This trend implies a decrease in the CO-to-H$_{2}$ conversion factor, $X_{\mathrm{CO}}$, when \uav~increases. We estimate the gas-to-dust mass ratios to be $77\pm33$ within the central $2\ \mathrm{kpc}$ and $93\pm19$ within the central $4\ \kpc$ of M83, which implies a Galactic dust-to-metal mass ratio within the observed region of M83. The estimated gas--depletion time for the M83 nucleus is $1.13\pm0.6\ \mathrm{Gyr}$, which is shorter than the values for nearby spiral galaxies found in the literature~($\sim2.35\ \mathrm{Gyr}$), most likely due to the young nuclear starbursts. A linear relationship between $P_{\mathrm{th}}$ and the radiation pressure generated by \uav, $P_{\mathrm{rad}}$, is found to be $P_{\mathrm{th}}\,\approx\,30
\,P_{\mathrm{rad}}$, which signals that the ISRF alone is insufficient to sustain the observed CO transitions. The spatial distribution of $P_{\mathrm{th}}$ reveals a pressure gradient, which coincides with the observed propagation of starburst activities and the alignment of (possibly background) radio sources. We discover that the off-centered~(from the optical nucleus) peak of the molecular gas volume density coincides well with a minimum in the relative aromatic feature strength, indicating a possible destruction of their carriers.  We conclude that the observed CO transitions are most likely associated with mechanical heating processes that are directly or indirectly related to very recent nuclear starbursts.}

\date{Received; accepted\vspace{-2mm}
}
\maketitle


\section{Introduction}

	\indent\par{
	Stars are nurtured in the interstellar medium~(ISM) of galaxies. Observational results from the past few decades have established that interstellar molecular clouds and the process of star formation are very closely associated. It has been reported from Galactic observations that young stars are often associated with dense molecular clouds~\citep{Loren1978, Blitz1982}. In extragalactic systems, observations have shown that the surface density of molecular gas, as traced by CO, strongly correlates with the star formation rate~(SFR) at the $\lesssim1\ \mathrm{kpc}$ scale~\citep{Kennicutt2007, Bigiel2011, Schruba2011, Leroy2013a}. These results have led to a widely accepted picture that stars are formed in molecular clouds and that molecular gas is the fuel for star formation~\citep{Elmegreen1998, Elmegreen2007}. However, results from numerical simulations have provided an alternative picture that the production of a dense layer of molecular hydrogen is a consequence of the thermal instability created by shocks associated with supernova remnants~(SNR,~\citealt{Koyama1999, MacLow2004}), and that star formation can occur as easily in atomic gas as in molecular gas~\citep{Glover2012}. Nonetheless, if one would understand the interrelation between molecular gas and star formation, it is crucial to identify the main mechanisms that stimulate the observed transitions of molecular gas and to understand the relationships between molecular gas and the characteristics of its surrounding ISM, such as gas-to-dust mass ratio, interstellar radiation field~(ISRF), and metallicity. }
	
	\par{
	Because of its perfectly symmetric structure, direct detection of H$_{2}$ is difficult. For decades, studies of molecular clouds have relied mainly on the observations of CO, the second most abundant species in molecular clouds. The most frequently used CO emission line stems from the relaxation of the $J=1$ rotational state of the predominant isotopologue $^{12}\mathrm{C}^{16}\mathrm{O}$. The H$_{2}$ column density is derived from the CO intensity via $X_{\mathrm{CO}}$, where $X_{\mathrm{CO}}\,\equiv\,(N_{\mathrm{H_{2}}}/I_{\mathrm{CO}})\ (\mathrm{cm^{-2}}\ (\mathrm{K~km~s^{-1}})^{-1})$. In the solar neighborhood, measurements of the mass of the molecular clouds and the CO intensity suggest that $X_{\mathrm{CO}}$ ranges from $(\,2\,-\,3\,)\,\times\,10^{20}$~\citep{Young1991, Strong1996, Bolatto2013}. For extragalactic studies of the total H$_{2}$ mass, a constant value of $X_{\mathrm{CO}}$ around the value estimated in the solar neighborhood is commonly adopted. However, observations of CO in the Small Magellanic Cloud~(SMC) suggest a dependence of $X_{\mathrm{CO}}$ on the UV radiation fields~\citep{Lequeux1994}. The same dependence has also been confirmed through modeling photo-dissociation regions~(PDRs)~\citep{Kaufman1999}. In nearby galaxies, where $X_{\mathrm{CO}}$ cannot be estimated by $\gamma$-ray observations as in the Milky Way~\citep{Grenier2005, Abdo2010}, resolved CO observations with the assumption of virial equilibrium, or a metallicity-dependent gas-to-dust mass ratio, show a strong anti-correlation between $X_{\mathrm{CO}}$ and metallicity~\citep{Boselli2002, Leroy2011}. Furthermore, although both H$_{2}$ and CO are capable of self-shielding, or can be shielded by dust, from UV photo-dissociation, the H$_{2}$ molecule, thanks to its higher abundance, can remain undissociated closer to the surface of the cloud where the CO molecules are dissociated~\citep{Dishoeck1988, Liszt1998}. It is expected that part of the molecular gas is not cospatial with CO molecules~\citep{Wolfire2010, Levrier2012}. These results suggest that understanding the physics driving the emission of CO is fundamental if one wants to understand how reliable the CO molecule is as a tracer of molecular gas. Such an analysis is only possible when one can derive physical conditions with multiple CO transitions.
	}

	\par{
	The launch of the \textsl{Herschel Space Observatory}~\citep{Pilbratt2010} has turned over a new leaf for the observation of CO. With its designed bandwidth, $194\,<\,\lambda\,<\,671\ \mum$, the Fourier Transform Spectrometer~(FTS) of the Spectral and Photometric Imaging REceiver~(SPIRE, \citealt{Griffin2010}) onboard \textsl{Herschel} gives us an unprecedented way to observe CO in its transitions from $\mathrm{J}=4-3$ to $\mathrm{J}=13-12$. Based on analyses of the CO spectral line energy distribution~(SLED), observations using the SPIRE FTS of a nearby starburst galaxy, M82, have shown that the mechanical energy provided by supernovae and stellar winds is a dominant heating source, rather than the ISRF, for the observed CO transitions~\citep{Panuzzo2010, Kamenetzky2012}. Such results are supported by the study of other nearby galaxies, Arp220~\citep{Rangwala2011}, NGC1266~\citep{Pellegrini2013} and the Antennae~\citep{Schirm2013}. However, observations of a nearby spiral galaxy, IC342, suggest that most observed CO emission should originate in PDRs~\citep{Rigopoulou2013}. Studies of Seyfert galaxies and luminous infrared galaxies~(LIRGs) also show a degeneracy between models of PDR, X-ray dominated regions~(XDR), and shocks for explaining the observed CO SLED~\citep{Spinoglio2012, Pereira-Santaella2013, Lu2014}.
	}
	
	\par{
	Motivated by the previous results, we study the properties of molecular gas and dust, and their relationships with star formation, in M83~(NGC5236) with the SPIRE FTS. M83, the so--called Southern Pinwheel galaxy, is the grand-design galaxy closest to the Milky-Way. Its face-on orientation ($24^{\circ}$, \citealt{Comte:1981tw}), starburst nucleus, active star formation along the arms, and prominent dust lanes~\citep{Elmegreen1998} make it one of the most intensely studied galaxies in the nearby Universe. spatially resolved studies of this galaxy in a wide range of wavelengths, e.g. radio~\citep{Maddox:2006ey}, CO~\citep{Crosthwaite:2002cr, Lundgren2004a}, infrared~\citep{Vogler2005, Rubin:2007ig}, optical~\citep{Calzetti2004, Dopita:2010fe, Hong:2011ch}, ultraviolet~\citep{Boissier2005}, and X-ray~\citep{Soria:2002dn}, reveal intense star--forming regions in its nucleus and spiral arms. In the past century, at least six supernova events have been detected in M83, making it one of the most active nearby starburst galaxies. Indeed, M83 has been found to be a molecular gas--rich galaxy compared to the Milky Way. It contains nearly the same molecular mass although it is 3 times less massive in its stellar mass~\citep{Young1991, Crosthwaite:2002cr}. Within the inner radius of $7.3\ \kpc$, the molecular mass is estimated to be more than twice that of atomic gas~\citep{Lundgren2004a}. Observational evidence also indicates that the emission of the lowest three transitions of CO in M83 is more concentrated in the nucleus, compared with M51~\citep{Kramer2005}. Compared to other nearby galaxies, the starburst nucleus of M83 shows relatively brighter CO emission for the same physical scale, as suggested by the PDR models that are fitted to the observed CO transitions up to $\mathrm{J}=6-5$~\citep{Bayet2006}. The near-infrared morphology in the nuclear region~($<\,20''$~in radius) of M83 shows a complex structure which contains a bright point source, coinciding with the optical nucleus, and a star--forming arc extending from the southeast to the northwest~\citep{Gallais:1991ti}. This morphological structure is also seen in the spatially resolved observations of CO transitions, $\mathrm{J}=4-3$ and $\mathrm{J}=3-2$~\citep{Petitpas1998} and in the age distribution of young star clusters~\citep{Harris2001}. Morphologically, observations of CO have demonstrated that the structure of molecular gas in M83 is quite complex. However, ground-based studies of physical conditions for molecular gas in M83 have been restricted by the difficulty of observing CO transitions in the higher--J states. The highest transition detected with ground--based telescopes is $\mathrm{J}=6-5$, which may be at or near the peak of the CO spectral line energy distribution~(SLED) of M83~\citep{Bayet2006}.
	}

	\par{
	In this paper, we investigate how the spatially resolved physical conditions of the molecular gas compares with the dust properties and star formation. We adopt a distance of $4.5\,\mpc$ based on the measurement of the Cepheid distance to M83~\citep{Thim2003}. The structure of this paper is as follows: in Section~\ref{data}, we describe in detail how the observed data is treated and sampled to generate the final products for the analysis; in Section~\ref{model}, we present the methodology for our estimation of molecular gas and dust physical conditions; we show the results and their discussion in Section~\ref{results}; a conclusion of this paper is given in Section~\ref{conclusion}.
	}

\section{Observation and data processing}\label{data}

	\begin{table*}
		\centering
		\caption{Emission lines observed by the \textsl{Herschel} SPIRE FTS and the ground--based telescopes used in this work\tablefootmark{a}.}
		\label{linetable}
		\begin{minipage*}{15cm}
			\centering
			\begin{tabular}{cccccccc}
				\hline\hline
				Transition & Frequency & E$_{upper}$ & g$_{upper}$&A&FWHM\tablefootmark{b}& Luminosity\tablefootmark{c}\\
						&	($\ghz$)&	    ($\kelvin$)    &&($10^{-6}\,\mathrm{s^{-1}}$)&($''$)&$\mathrm{(L_{\odot})}$\\
				\hline
				$\mathrm{CO\ J=1-0}$\tablefootmark{d}&$115.271$&$5.53$&$3$&$0.072$&$22$&$(5.94\pm0.62)\,\times\,10^{3}$\\
				$\mathrm{CO\ J=2-1}$\tablefootmark{d}&$230.538$&$16.60$&$5$&$0.691$&$14$&$(3.18\pm0.46)\,\times\,10^{4}$\\
				$\mathrm{CO\ J=3-2}$\tablefootmark{e}&$345.796$&$33.19$&$7$&$2.497$&$14$&$(8.72\pm0.18)\,\times\,10^{4}$\\
				\hline
				$\mathrm{CO\ J=4-3}$&$461.041$&$38.45$&$9$&$6.126$&$41.7$&$(1.78\pm0.27)\,\times\,10^{5}$\\
				$\mathrm{[C\textsc{I}\,]^{3}\mathrm{P}_{1}-^{3}\mathrm{P}_{0}}$&$492.161$&$16.42$&$3$&$0.079$&$38.1$&$(1.01\pm0.16)\,\times\,10^{5}$\\
				$\mathrm{CO\ J=5-4}$&$576.268$&$57.67$&$11$&$12.210$&$33.5$&$(2.08\pm0.27)\,\times\,10^{5}$\\
				$\mathrm{CO\ J=6-5}$&$691.473$&$80.74$&$13$&$21.370$&$29.3$&$(1.73\pm0.21)\,\times\,10^{5}$\\
				$\mathrm{CO\ J=7-6}$&$806.652$&$107.64$&$15$&$34.220$&$33.0$&$(1.24\pm0.15)\,\times\,10^{5}$\\
				$\mathrm{[C\textsc{I}\,]^{3}\mathrm{P}_{2}-^{3}\mathrm{P}_{1}}$&$809.342$&$43.41$&$5$&$0.265$&$33.0$&$(1.68\pm0.18)\,\times\,10^{5}$\\
				$\mathrm{CO\ J=8-7}$&$921.800$&$138.39$&$17$&$51.340$&$33.2$&$(9.48\pm1.50)\,\times\,10^{4}$\\
				$\mathrm{CO\ J=9-8}$&$1036.912$&$172.98$&$19$&$73.300$&$19.1$&$(7.88\pm3.28)\,\times\,10^{4}$\\
				$\mathrm{CO\ J=10-9}$&$1151.985$&$211.40$&$21$&$100.600$&$17.6$&$(6.83\pm3.99)\,\times\,10^{4}$\\
				$\mathrm{CO\ J=11-10}$&$1267.015$&$253.67$&$23$&$133.900$&$17.3$&$<\,(7.39)\,\times\,10^{4}$\tablefootmark{(f)}\\
				$\mathrm{CO\ J=12-11}$&$1381.995$&$299.77$&$25$&$173.500$&$16.9$&$<\,(3.18)\,\times\,10^{4}$\tablefootmark{(f)}\\
				$\mathrm{[N\textsc{II}\,]^{3}\mathrm{P}_{1}-^{3}\mathrm{P}_{0}}$&$1461.133$&$70.15$&$3$&$2.070$&$16.6$&$(1.62\pm0.17)\,\times\,10^{6}$\tablefootmark{(g)}\\
				$\mathrm{CO\ J=13-12}$&$1496.923$&$349.70$&$27$&$220.000$&$16.6$&$<\,(5.13)\,\times\,10^{4}$\tablefootmark{(f)}\\
				\hline\hline
			\end{tabular}
		\end{minipage*}\\
			\raggedright
			\tablefoottext{a}{References for spectroscopic parameters: \textsl{The Cologne Database for Molecular Spectroscopy} (http://www.astro.uni-koeln.de/cdms), \textsl{Leiden Atomic and Molecular Database} (http://home.strw.leidenuniv.nl/$\sim$moldata/) and \textsl{NIST Atomic Spectra Databse} (http://www.nist.gov/pml/data/asd.cfm)}
			\tablefoottext{b}{The equivalent full width at half maximum~(FWHM) of the maps. For the SPIRE FTS, the number is given as the width of a 2D Gaussian approximation to the beam profile.}
			\tablefoottext{c}{Integrated luminosity at the resolution of $42''$ over the un-masked pixels in Figure~\ref{coverage}}
			\tablefoottext{d}{Observation done at the \textsl{Swedish-ESO Submillimetre Telescope}~(SEST), see \citet{Lundgren2004a} for details.}
			\tablefoottext{e}{Observation done at the \textsl{Atacama Submillimeter Telescope Experiment}~(ASTE), see \citet{Muraoka2009} for details.}
			\tablefoottext{f}{$S/N\,<\,1$, the value given is the $1\,\sigma$ upper limit.}
			\tablefoottext{g}{The $\niitof$ luminosity is calculated from the map at a spatial resolution of $22''$.}
	\end{table*}

	\subsection{The \textsl{Herschel} SPIRE FTS}

		\begin{figure}[h!]
			\centering
			\includegraphics[width=0.5\textwidth]{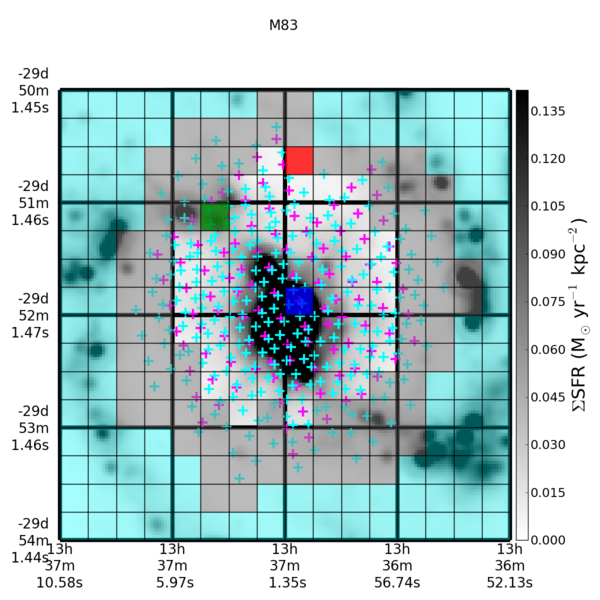}
			\caption{The common grid (indicated by the black lines) generated for the SPIRE FTS observation overlaid on the star formation rate~(SFR) map of M83. The displayed SFR map has an equivalent spatial resolution~(FWHM$=6.43''$), the same as the MIPS $24\,\mum$ broad-band image. Each pixel~(black box) has a size of $15''\times15''$, which is equivalent to $330\,\mathrm{pc}\,\times330\,\mathrm{pc}$ at the distance to M83. The SFR map is constructed from the far-ultraviolet~(FUV) and $24\ \mum$ photometry maps following the calibration given in \citet{Hao2011}~(see Section~\ref{ancillary} and \ref{sfr_nii} for more description). The magenta and cyan crosses indicate the pointings of the SLW and SSW detectors from all jiggle positions. The pixels masked by semi-transparent cyan color are outside the coverage of the SPIRE FTS maps presented in this work. The pixels masked by the semi-transparent gray color are truncated after the maps for observed lines are convolved to $42''$. The spectral energy distributions of three colored pixels, in blue, green, and red, indicating the nuclear, arm, and inter-arm regions, are displayed in Figure~\ref{fig:SEDs}~(see Section~\ref{dust_model}). The thick black lines are present as a guide to the labeled coordinates in the J2000 epoch.}
			\label{coverage}
		\end{figure}

		\indent\par{
		The spectra presented in this work are extracted from the observations conducted by \textsl{Herschel} on the operational day number 602~(2011-01-05). The data are observed as part of the \textsl{Herschel} key program, Very Nearby Galaxies Survey (VNGS) (PI: C. D. Wilson). The assigned observation identifier to this observation is 1342212345. The target is observed with the SPIRE FTS, which covers a bandwidth of $194\,<\,\lambda\,<\,671\ \mum$, in the high-resolution~($R=1000$ at 250$\ \mum$) single-pointing mode and is fully-sampled, that is, the beam is directed as 16-point jiggles to provide complete Nyquist sampling of the requested area. At each jiggle position, we request eight repetitions of scan, which, in turn, records data of sixteen scans in one jiggle position with each detector. Each scan spends 33.3~$\sec$~on the target. This gives an effective integration time of 533~$\sec$~on each jiggle position. We begin the data processing with the unapodized averaged point--source calibrated data, which is calibrated by the observations of Uranus. For each detector, the observed spectrum is calibrated with the calibration tree, spire\_cal\_11\_0.
			}
		
		\subsubsection{Observation}
		
			\indent\par
\begin{figure*}[htbp!]
			\centering
			\includegraphics[width=\textwidth]{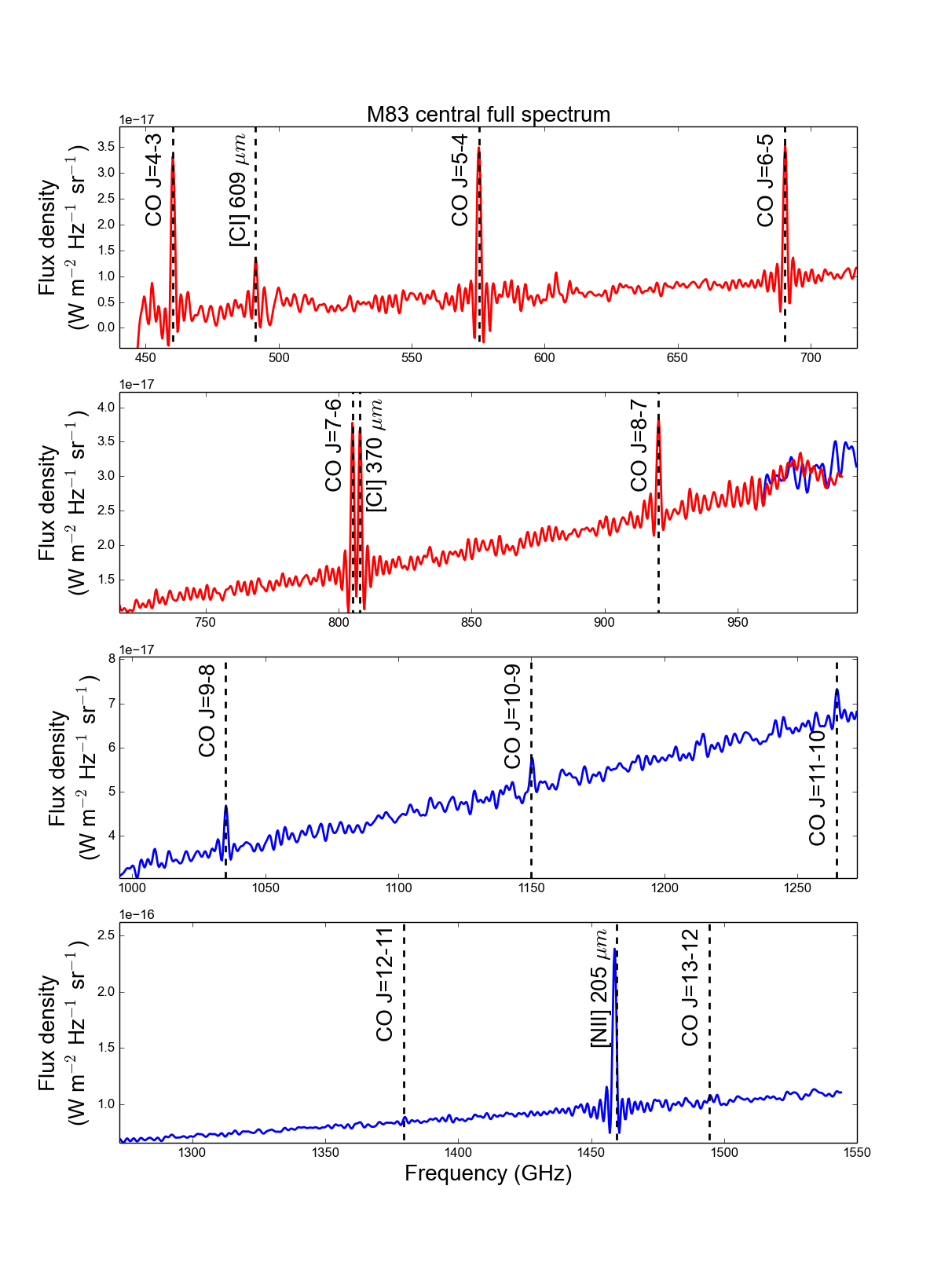}
			\caption{The spectrum taken with the two central bolometers, SLWC3~(red) and SSWD4~(blue), from the center~(blue pixel in Figure~\ref{coverage}) of the SPIRE FTS map of M83. This spectrum has been corrected for the source--beam coupling effect with the method introduced in~\citet{Wu2013}. The source distribution is assumed to be a Gaussian profile with FWHM$=18''$, estimated from the overlap bandwidth of the two bolometers. The expected locations of the CO, [C\textsc{I}], and [N\textsc{II}] transitions listed in Table~\ref{linetable} are noted with the black dashed lines.
			\label{spec}}
		\end{figure*}

{The two spectrometer arrays of the SPIRE FTS contain 7 (\textsl{SPIRE Long ($316-672\,\mum$) Wavelength Spectrometer Array}, SLW) and 19 (\textsl{SPIRE Short ($194-324\,\mum$) Wavelength Spectrometer Array}, SSW) hexagonally packed unvignetted detectors. In a complete Nyquist sampled map, the bolometer spacing is approximately $12.7''$~($\sim\,270\ \mathrm{pc}$) for SLW and $8.1''$~($\sim\,170\ \mathrm{pc}$) for SSW~\citep{observersmanual}. To assign the observed value from each detector to a uniformly gridded map, we create a common spatial pixel grid which covers a $5'\times5'$ area with each pixel covering an area of $15''\times15''$~($\sim\,330\times330\ \mathrm{pc^{2}}$). The map is centered on the pointing of the center bolometer of the SLW array, SLWC3, at the jiggle position labeled 6. Figure~\ref{coverage} shows the pointing of the detectors on the common grid overlaid on the star formation rate~(SFR, see Section~\ref{sfr_nii} for details) map. Figure~\ref{spec} gives an example of the spectrum taken by the SPIRE FTS. This spectrum is from the central bolometers, SLWC3 and SSWD4, at the central jiggle~(approximately from the blue pixel in Figure~\ref{coverage}) and has been corrected for the source--beam coupling effect with the method introduced in~\citet{Wu2013}. Due to the uncertainty of the beam profiles for the side bolometers, the data maps presented in this work are not corrected for the source--beam coupling effects. The spatial size of each pixel is chosen so that within the  field of view~(FOV) of the \textsl{Herschel Space Observatory}, each pixel contains at least one observed spectrum. Although the pixel size of the SPIRE FTS maps used in this work is less than half of $42''$, the equivalent beam full width at half maximum~(FWHM) at the instrument's longest wavelength, we choose to sample the maps this way based on two reasons: 1) the smallest FWHM of the instrument is approximately $16''$, so this choice of pixel size can better preserve the relatively higher spatial resolution at the short wavelengths; 2) even at the longest wavelength, the values assigned to any two adjacent pixels are not completely independent, since the average spacing of a given bolometer to its neighbors is smaller than $15''$. The data extracted from individual bolometers are co-added and sampled to the created grid following the method described below.
	}

		\subsubsection{Map-making procedure}\label{fts_maps}

		\indent\par
\begin{figure*}[htp]
			\centering
			\subfloat[][]{
				\includegraphics[width=0.62\textwidth]{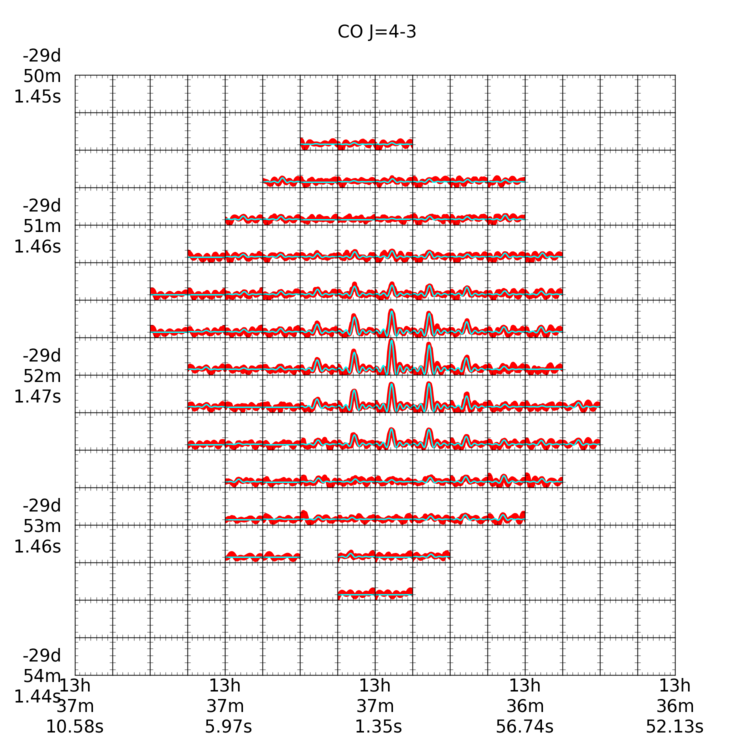}
				\label{co43_linestack}
			}
			\quad
			\subfloat[][]{
				\includegraphics[width=0.62\textwidth]{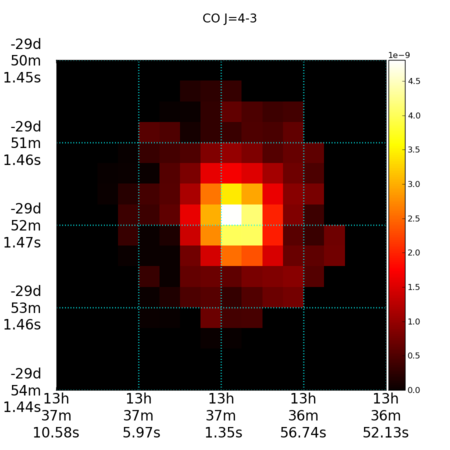}
				\label{co43_intmap}
			}
			\caption{The spatial distribution of the observed CO $J=4-3$ line. Figure~\ref{co43_linestack} shows the continuum-removed coadded spectrum on every pixel within a range of $454<\nu<468\ \mathrm{GHz}$. The vertical axis in each pixel ranges between $-9.8\,\times\,10^{-19}$ and $5.4\,\times\,10^{-19}\,\mathrm{W\ m^{-2}\ sr^{-1}\ Hz^{-1}}$. 
			 Figure~\ref{co43_intmap} shows the measured CO $J=4-3$ intensity distribution, in the unit of $\mathrm{W\,m^{-2}\,sr^{-1}}$, within the mapped region of M83. Each pixel has size $15''\,\times\,15''$~or $330\,\mathrm{pc}\,\times330\,\mathrm{pc}$. The spatial resolution of the map is $42''$.\label{co43_map}}
		\end{figure*}

\begin{figure*}[htp]
			\centering
			\subfloat[][]{
				\includegraphics[width=0.62\textwidth]{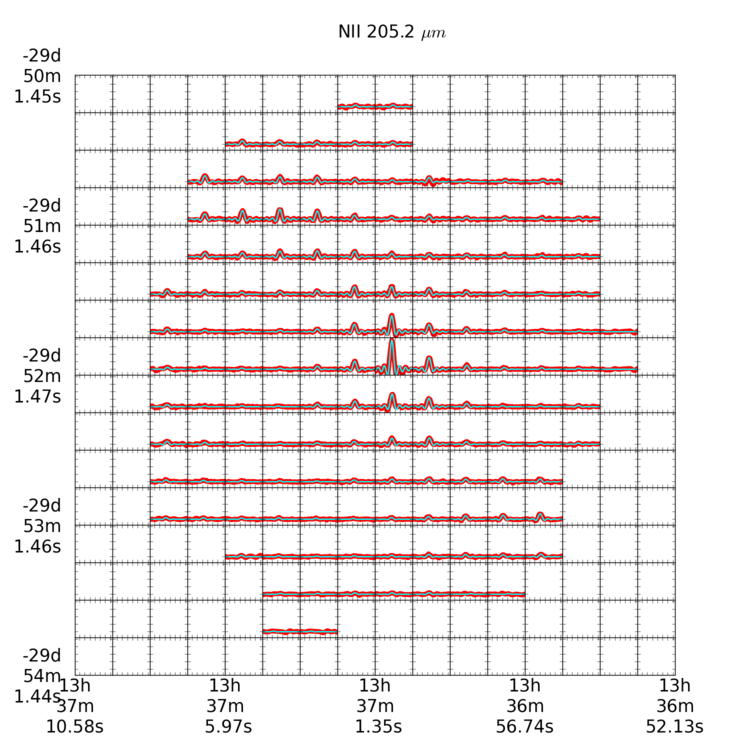}
				\label{nii_linestack}
			}
			\quad
			\subfloat[][]{
				\includegraphics[width=0.62\textwidth]{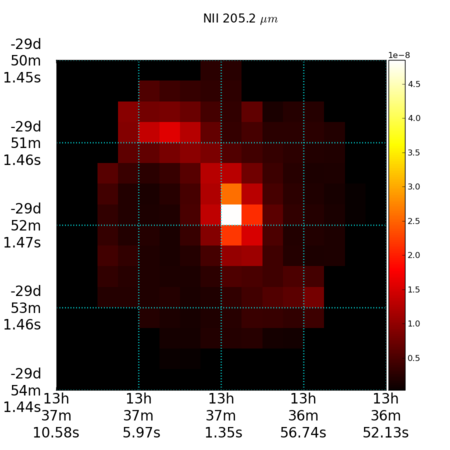}
				\label{nii_intmap}
			}
			\caption{The spatial distribution of the observed $\niitof$ line. Figure~\ref{nii_linestack} shows the continuum-removed coadded spectrum on every pixel within a range of $1452<\nu<1467\ \mathrm{GHz}$.  The vertical axis in each pixel ranges between $-1.7\,\times\,10^{-17}$ and $9.3\,\times\,10^{-17}\,\mathrm{W\ m^{-2}\ sr^{-1}\ Hz^{-1}}$.
			 Figure~\ref{nii_intmap} shows the measured $\niitof$ intensity distribution, in the unit of $\mathrm{W\,m^{-2}\,sr^{-1}}$, within the mapped region of M83. Each pixel has size $15''\,\times\,15''$~or $330\,\mathrm{pc}\,\times330\,\mathrm{pc}$. The spatial resolution of the map is $17''$.\label{nii_map}}
		\end{figure*}

{
		For each line observed by the \textsl{Herschel} FTS~(see Table~\ref{linetable}), a map is made based on the observed spectra in the frequency range, $\nu_{\mathrm{line}}-15\,\ghz<\nu<\nu_{\mathrm{line}}+15\,\ghz$, where $\nu_{\mathrm{line}}$ is the frequency of the emission line in the lab frame. For each detector, the observed spectrum is assumed to be uniform in a $15''\times15''$ area centered at the pointing of detector. The continuum is first removed from each observed spectrum prior to co-addition. The level of the continuum is simultaneously measured with a spectral model, which is a combination of a parabola~(continuum) and a sinc~(emission line) function, through the \texttt{MPFIT} procedure~\citep{Markwardt2009}. The spectrum assigned to each pixel is then computed as the error-weighted sum of each overlapping spectrum in proportion to its effective area in the pixel. Figures~\ref{co43_linestack} and~\ref{nii_linestack} demonstrate example maps of the spectra in the range of $\nu_{\mathrm{line}}-15<\nu<\nu_{\mathrm{line}}+15\,(\ghz)$ for the transition, CO $J=4-3$ and $\mathrm{[N\textsc{II}\,]^{3}\mathrm{P}_{0}-^{3}\mathrm{P}_{1}}$, assigned to the pixels. We then fit the spectrum from each pixel with the spectral model to obtain a map of line fluxes~(see Figure~\ref{co43_intmap} and \ref{nii_intmap} for an example).
		}

		\par{
		As to the uncertainty estimate, although the pipeline gives statistical uncertainties which are computed as the variance of observed spectra from the total 16 scans for each detector, we found that the signal-to-noise ratio~($S/N$) is often over-estimated when considering only the pipeline-statistical uncertainty, judging from the noise level observed in the spectral data cube. This is especially the case when the central frequency of the line is close to the edge of the bandwidths. Because of this, we derive the random uncertainty from the residual after subtracting the continuum for each detector in the desired frequency range before co-addition, with exclusion of the central 10$\,\ghz$ around the emission line so that the result of line measurement does not affect the uncertainty estimation. For each detector, the random uncertainty, which is to be propagated through the map-making, is generated based on the root-mean-square~(RMS) of the residuals. At each frequency grid, the random uncertainty is assigned by a random number generator assuming a normal distribution with the RMS of the residuals as the 1-$\mathrm{\sigma}$. The uncertainties estimated with the RMS of the residuals are termed ``RMS uncertainties'' from here on.
		}
		
		\par{
		The estimated absolute calibration uncertainty for the instrument is about 10\% of the observed flux~\citep{Swinyard2014}. The uncertainty assigned to each pixel is the summation in quadrature of the RMS and calibration uncertainties. These uncertainties are propagated by the Monte-Carlo~(MC) method with 300 iterations. Considering that the calibration uncertainty of each line and bolometer should not be independent, during one MC iteration, the percentage of calibration uncertainty is fixed for all the lines, and the RMS uncertainty for each bolometer at each jiggle position is set to vary on every frequency grid. The final maps used in our analysis are presented in Figure~\ref{co43_map}, \ref{nii_map}, and Appendix~\ref{appmaps} in the observed beam width~(see Table~\ref{linetable}). Interested readers should note that, toward the edge of the maps, the level of uncertainty increases~(see Fig. A.X (a)), which sometimes makes the pixels on the edge appear falsely bright in Fig. A.X (b).
		}

		\par{
		At the central pixel~(marked by blue color in Figure~\ref{coverage}), the measured line intensities for the CO $\mathrm{J}=4-3$ and $\mathrm{J}=6-5$ transitions are $(4.81\pm0.66)\times10^{-9}$ and $(7.13\pm0.33)\times10^{-9}\ \mathrm{W\ m^{-2}\ sr^{-1}}$ at the observed resolution, $42''$ and $29.3''$, respectively. At the same position, these two lines have been observed by the \textsl{Caltech Submillimeter Observatory~(CSO)}, and the measured intensities~(converted from the main-beam temperature) for the CO $\mathrm{J}=4-3$ and $\mathrm{J}=6-5$ transitions are $(1.2\pm0.2)\times10^{-8}$ and $(2.8\pm0.2)\times10^{-8}$ at the spatial resolutions of $21.9''$~\citep{Bayet2006}. Following \citet[][Equation~A.4 and Table~2]{Bayet2006}, the intensities for the CO $\mathrm{J}=4-3$ and $\mathrm{J}=6-5$ transitions measured by \textsl{CSO} can be convolved to the SPIRE FTS resolution, resulting in $(3.9\pm0.6)\times10^{-9}$ and $(1.69\pm0.4)\times10^{-9}\ \mathrm{W\ m^{-2}\ sr^{-1}}$, respectively. The intensity for CO $\mathrm{J}=4-3$ compares well between the two instruments, implying that the spatial distribution of CO $\mathrm{J}=4-3$ transition can be regarded as uniform within $42''$ in the nucleus. The SPIRE FTS measured intensity is smaller than the CSO measured one for CO $\mathrm{J}=6-5$ by a factor of 2.4, implying that the spatial distribution of the CO $\mathrm{J}=6-5$ transition should have a FWHM less than $29.3''$ and larger than $21.9''$.
		}
	
	\par{
	In order to compare all the CO molecular and [C\textsc{I}] lines under a common spatial resolution, we convolve these maps to the largest beam size, $\sim\,42''$~($\sim\,910\ \mathrm{pc}$), which is the beam FWHM of the CO $\mathrm{J}=4-3$ transition. The beam shape for the SSW can be well approximated by a pure Gaussian profile. However, the distribution of the beam profiles in the SLW is generally more complex and cannot be well described by a pure Gaussian except at the lowest frequency in its bandwidth. In order to construct kernels for convolving all the SPIRE FTS CO and [C\textsc{I}] maps, we adopt the parameters from fitting a 2D Hermite-Gaussian function to the beam profiles of the center detectors, SLWC3 and SSWD4. The off-center detectors are found to have similar beam profiles as the center ones~\citep{Makiwa2013}, so we apply the same kernel to all bolometers on each map. The kernels used in the convolution process are constructed using the method described in~\citet{Gordon2008} with the cutoff frequency in the Hanning function chosen so that the inclusion of the high spatial frequency is minimal. The pixels on the edge of each convolved map are truncated, so the effective FOV for the observed CO transitions at a common spatial resolution is as indicated by the unmasked pixels in Figure~\ref{coverage}.
		}

		\subsection{Ancillary data}\label{ancillary}
	
			\indent\par{
			We compare the FTS data to the available ground-based CO observations, which include the maps of the $\mathrm{J}=1-0$ and $\mathrm{J}=2-1$ transitions observed with the Swedish-ESO Submillimetre Telescope~(SEST) from \citet{Lundgren2004a} and the $\mathrm{J}=3-2$ transition observed with the the \textsl{Atacama Submillimeter Telescope Experiment}~(ASTE) from \citet{Muraoka2009}. Since our analysis is based on the point--source calibrated FTS data, we use the main--beam temperature for the CO transitions from the ground-based observations~(see Table~\ref{linetable} for the information of the ground--based lines). We also calculate the star formation rate based on the far-ultraviolet~(FUV) observation by the \textsl{Galaxy Evolution Explorer}~(GALEX, \citealt{Martin2005}) and the 24$\,\mum$ observation by the \textsl{Multiband Imaging Photometer for Spitzer}~(MIPS,~\citealt{Rieke2004}). We derive dust properties based on the broad-band images from $3$ to $500\ \mum$ observed by \textsl{Spitzer} and \textsl{Herschel}. One of the objectives of this work is to compare the atomic and molecular gas with dust properties. The atomic gas properties are derived from the data taken as part of The H\textsc{I} Nearby Galaxy Survey~(THINGS,~\citealt{Walter2008}). In this section, we provide a short summary of how the supplementary data are treated prior to comparison.
			}
			\subsubsection{CO $\mathrm{J}=1-0$ and $\mathrm{J}=2-1$}\label{co10_section}
		
				\indent\par{
				The CO $\mathrm{J}=1-0$ and $\mathrm{J}=2-1$ observations of M83 were done using the 15m SEST during two epochs, 1989--1994 and 1997--2001 and presented in \citet{Lundgren2004a}. During the observations, the receivers were centered on the CO $\mathrm{J}=1-0$ and CO $\mathrm{J}=2-1$ lines ($115.271$ and $230.538\,\ghz$, respectively), where the FWHM of the observed maps are $45''$ and $23''$, respectively. The CO $\mathrm{J}=1-0$ spectra were taken with $11''$ spacing, and the coverage was complete out to a radius of $4'20''$. The CO $\mathrm{J}=2-1$ spectra were taken with $7''$ spacing in the inner region of approximately $5'\,\times\,3'$ around the nucleus. The absolute uncertainty of the observations is about 10\%. \citet{Lundgren2004a} adopted relatively dense grid spacing, compared to the beamwidth, so that the spatial resolution of the data can be increased by using a \texttt{MEM-DECONVOLUTION} routine (maximum entropy method). The angular resolutions of the \texttt{MEM-DECONVOLVED} maps, which are adopted in this work, are $\sim\,22''$ and $\sim\,14''$ for the CO $\mathrm{J}=1-0$ and $\mathrm{J}=2-1$ maps, respectively. By convolving the \texttt{MEM-DECONVOLVED} data cube and comparing with the observed spectra, it has been verified that the \texttt{MEM-DECONVOLVED} results are reliable. Details of the data calibration and map-making for these two lines are described in~\citet{Lundgren2004a}.
				}
				\par{
				The pixel size of the CO $\mathrm{J}=1-0$ and $\mathrm{J}=2-1$ maps is $11''\times11''$. We first re-project these maps to the same world coordinate system~(WCS) as the FTS maps, and convolve the re-projected maps to the equivalent spatial resolution of $42''$ to match the spatial resolution of the largest beam FWHM~($42''$) of the SPIRE FTS. After convolution, the maps are resampled to the pixel size of the FTS maps~($15''\times15''$) with the bicubic interpolation.
				}

			\subsubsection{CO $J=3-2$}
				\indent\par{
				The CO $\mathrm{J}=3-2$ observations of M83 were made using the ASTE from 2008 May 15--21. The size of the CO $\mathrm{J}=3-2$ map is about $8'\times8'$ ($10.5\times10.5\,\kpc$), including the whole optical disk. The half-power beam width~(HPBW) of the ASTE $10\,\mathrm{m}$ dish is $22''$ at $345\,\ghz$. The raw data were gridded to $7.5''$ per pixel, giving an effective spatial resolution of approximately $25''$. The absolute uncertainty of the observation is about 20\%. Details of the data calibration and map-making can be found in~\citet{Muraoka2009}. We follow the same procedure described in Section~\ref{co10_section} to reproject and convolve the map to the same pixel size and spatial resolution as the SPIRE FTS maps~($15''\times15''$, $42''$).
				}

			\subsubsection{FUV map}
				\indent\par{
				The FUV map of M83 is made available as part of the GR7 (ops-v7\_2\_1) of \textsl{GALEX}. We obtain the FUV map through the Barbara A. Mikulski Archive for Space Telescopes (MAST\footnote[1]{http://archive.stsci.edu/}). The wavelength bandpass of these detectors in the FUV covers a range of 1400 to 1800 \AA. The M83 FUV map used in this work originates from the Nearby Galaxy Survey (NGS), which is a survey targeting nearby galaxies, including M83, with a nominal exposure time of 1000 to 1500s. In the case of M83, the exposure time is 1349 seconds. The observed intensity and sky-background map is recorded in units of (photon) counts per second~(CPS), and the conversion factor to units of $\mathrm{erg\ s^{-1}\ cm^{-2}\ \AA^{-1}}$ is $1.40\times10^{-15}\,\mathrm{erg\ s^{-1}\ cm^{-2}\ \AA^{-1}\ CPS^{-1}}$~\citep{Morrissey2007}. 
				}
				\par{
			The pixel size of the FUV map is $1.5''\times1.5''$ with a point spread function~(PSF) of $4.2''$ FWHM. We first re-project the FUV map to the same WCS as the FTS maps. For the purpose of analysis, we convolve the re-projected map to the equivalent spatial resolution of $22''$ to match the spatial resolution of the CO $\mathrm{J}=1-0$ map~(see Section~\ref{sfr_nii}), and of $42''$ to match the spatial resolution of the CO $\mathrm{J}=4-3$ map~(see Section~\ref{gas_depletion_section}). After that, the FUV map is resampled to the pixel size of the FTS maps~($15''\times15''$) with the bicubic interpolation.
			}
			
			\subsubsection{H\textsc{I} map}
				\indent\par{
				In order to derive the total gas mass in this work, we estimate the mass of atomic gas with the observations from the THINGS sample. All the galaxies in the THINGS sample are observed with the Very Large Array~(VLA) in its B array configuration (baselines: 210 m to 11.4 km) with an addition of the D array (35 m to 1.03 km) and C array (35 m to 3.4 km) data to recover extended emission in the objects. The H\textsc{I} map of M83 covers an area of $34'\,\times\,34'$. The final angular resolution of the H\textsc{I} map is $6''$. A detailed discussion of the data reduction procedure can be found in \citet{Walter2008}.}
				\par{We convert the observed H\textsc{I} map, from which the fluxes are calculated with a ``natural weighting'' scheme, with a conversion factor, $1.823\times10^{18}\ \mathrm{cm^{-2}\ (Jy\ Beam^{-1}\ km\ s^{-1})^{-1}}$, to derive the total H\textsc{I} column density, $N(\mathrm{H\textsc{I}})$. The map of $N(\mathrm{H\textsc{I}})$ is re-projected and convolved to match the pixel size~($15''\times15''$) and angular resolution~($42''$) of the FTS maps.
			\begin{table*}
				\centering
				\caption{Broad-band image information}
				\label{photometrytable}
				\begin{minipage*}{15cm}
					\centering
					\begin{tabular}{cccccc}
						\hline\hline
						&	& Beam &	Calibration &  Background&	Background\\
								Broadband&Wavelength			& FWHM  & Uncertainty & Center\tablefootmark{a}  &  Radius \\
						& $(\mum)$& $('')$ & $(\%)$ &	($\alpha$,\ $\delta$ in J2000)&	$('')$\\
						\hline
						IRAC1\tablefootmark{b}&	$3.6$&	$1.90$&	10&	$13:36:50.7\ -29:37:22$&	$200$\\
						IRAC2\tablefootmark{b}&	$4.5$&	$1.81$&	10&	$13:36:43.5\ -29:38:35$&	$200$\\
						IRAC3\tablefootmark{b}&	$5.8$&	$2.11$&	10&	$13:36:51.5\ -29:39:48$&	$200$\\
						IRAC4\tablefootmark{b}&	$8.0$&	$2.82$&	10&	$13:36:42.8\ -29:38:57$&	$200$\\
						MIPS1\tablefootmark{c}&	$24$&	$6.43$&	4&	$13:36:46.6\ -29:40:51$&	$130$\\
						PACS1\tablefootmark{d}&	$70$&	$5.67$&	5&	$13:36:22.2\ -29:46:42$&	$180$\\
						PACS3\tablefootmark{d}&	$160$&	$11.18$&	 5&	$13:36:22.2\ -29:46:42$&	$180$\\
						SPIRE1\tablefootmark{d}&	$250$&	$18.15$&	 7&	$13:36:24.8\ -29:38:33$&	$230$\\
						SPIRE2\tablefootmark{d}&	$350$&	$24.88$&	 7&	$13:36:24.8\ -29:37:59$&	$230$\\
						SPIRE3\tablefootmark{d}&	$500$&	$36.09$&	 7&	$13:36:30.7\ -29:39:28$&	$230$\\
						\hline\hline
					\end{tabular}
					\raggedright
					\tablefoottext{a}{The FTS maps are centered at ($\alpha$, $\delta$)=($13:37:0.78$, $-29:51:53.97$) in J2000.}\\
					\tablefoottext{b}{http://vizier.u-strasbg.fr/}\\
					\tablefoottext{c}{http://www.jb.man.ac.uk/~gbendo/exchange/SpitzerData/spitzerdata\_vngs.html}\\
					\tablefoottext{d}{http://hedam.lam.fr; Data are processed with the version 5 of the Herschel Interactive Processing Environment~(HIPE).}\\
				\end{minipage*}
			\end{table*}
				}
		\subsubsection{Broad-band images} \label{sec:photoMC}
			\indent\par{
			In order to derive the dust parameters, we collect the fully reduced \textsl{Spitzer} and \textsl{Herschel} broadband maps of M83 previously presented by different studies. The 4 IRAC bands are those presented by \citet{Dong:2008ho} and are public from the VizieR Catalogue Service. The MIPS 24$\,\mum$ image was presented by \citet{Bendo2012a}. Information for the broad-band images used in this work is listed in Table~\ref{photometrytable}. 
			\par{
			The PACS and SPIRE data are presented by \citet{Foyle2012} and are directly downloaded from the \textsl{HeDaM} website~\citep{Oliver2012}. First, a median local background is subtracted from each map~(see Table~\ref{photometrytable} for the aperture information). We then degraded the resolution to $42''$, corresponding to the SPIRE/FTS spatial resolution at 650$\,\mum$, which is the wavelength of the CO $\mathrm{J}=4-3$ transition. This step was performed using the method and the kernels provided by \citet{Aniano2011}. Each map was then reprojected on the FTS common grid. Since the previous steps induce significant noise correlation between pixels, the statistical uncertainties for each photometry measurement are propagated through the data processing procedure. The uncertainty propagation is similar to the MC procedure described in Section~\ref{fts_maps}. We perturb the value from each pixel of the photometry map~(perturbed MC maps) by a normal distribution with the correspondent statistical uncertainty as the $1\sigma$~value. The final statistical uncertainty for each map is the standard deviation derived from the 300 perturbed MC maps. These uncertainties are then propagated through the SED fitting. The error on the dust SED parameters are the standard deviation of the distribution of the MC SED parameters.
			}

\section{Gas and dust modeling}\label{model}
		\begin{figure*}[htbp!]
			\centering
			\includegraphics[width=\textwidth]{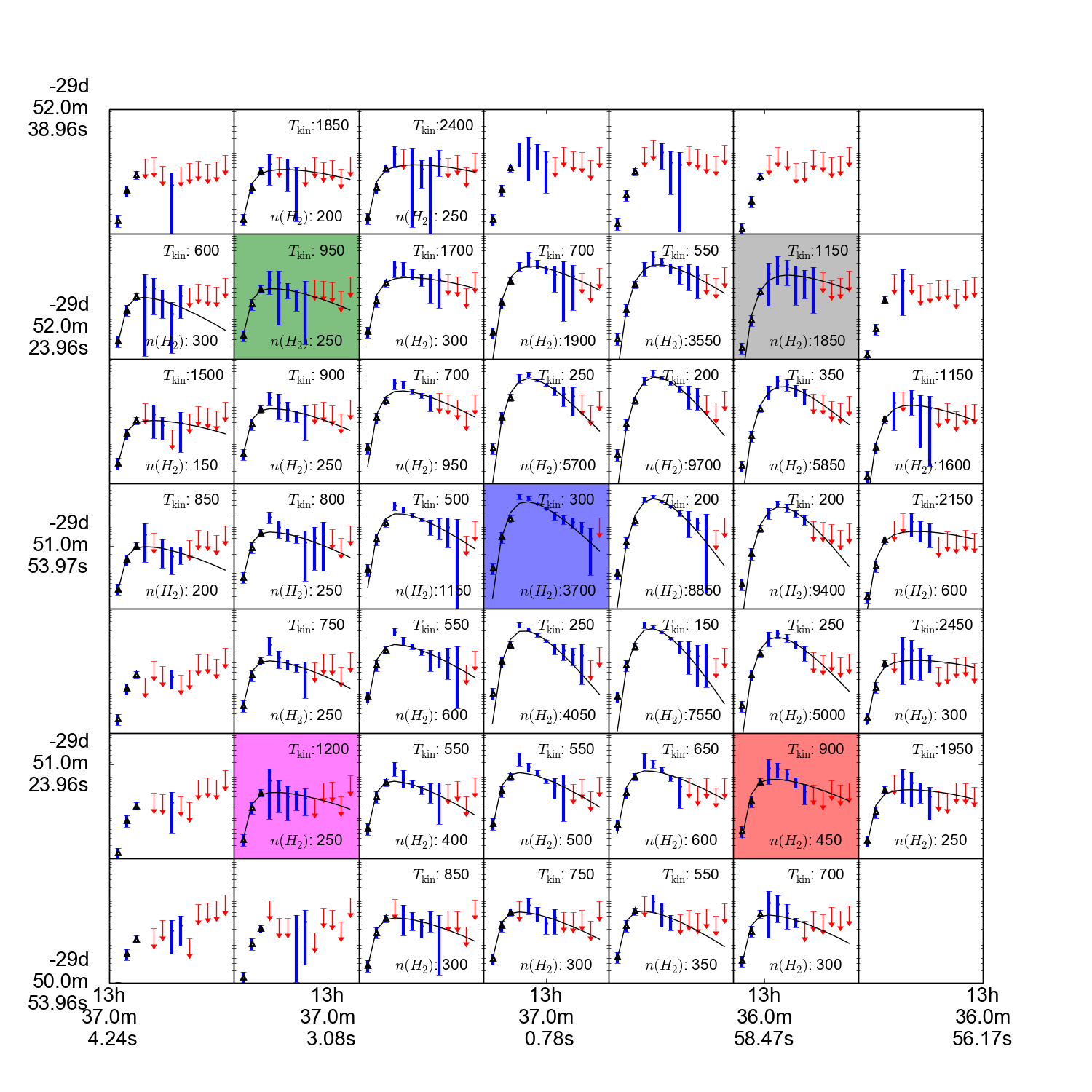}
			\caption{A map of the CO SLEDs used in this study. Each pixel in this map is equal in its physical size to one pixel in Figure~\ref{coverage}. The CO transitions~(in $\mathrm{W\ m^{-2}\ sr^{-1}}$) observed by the SPIRE FTS from unmasked pixels in Figure~\ref{coverage} are shown here in blue~(detected, $\mathrm{S/N}>1$) or red (upper limits). The dynamical range for each plotted SLED is the same as in Figure~\ref{sled_compare}. CO transitions of $\mathrm{J}=1-0, 2-1, 3-2$ observed by ground-based telescopes are shown as black triangles. All SLEDs that contain at least 2 detections with the SPIRE FTS are fitted with \texttt{RADEX} to obtain the best-fit parameters. The black solid line shows the SLED that is described by the best-fit parameters in \texttt{RADEX} with its best-fit $T_{\mathrm{kin}}$ and $n(\mathrm{H_{2}})$ labeled in each pixel, in units of $\mathrm{K}$ and $\mathrm{cm^{-3}}$ respectively. The data contained in the five color--masked pixels are highlighted in Figure~\ref{sled_compare} and discussed in Section~\ref{sec_co_sled}.}
			\label{co_sled}
		\end{figure*}
			
	\subsection{Using \texttt{RADEX} to interpret the CO SLEDs}\label{radex}

		\indent\par{
		Our interpretation of the CO spectral line energy distribution~(SLED) is based on results computed by \texttt{RADEX}, a non-local thermal equilibrium~(non-LTE) radiative transfer model with the assumption of local statistical equilibrium~\citep{vanderTak2007}. The on-line manual\footnote[2]{http://home.strw.leidenuniv.nl/\textasciitilde moldata/radex\_manual.pdf} of \texttt{RADEX} includes detailed discussions and derivation of the radiative transfer process in the model. We adopt the uniform-sphere approximation for the photon escape probability in our calculation and assume the cosmic microwave background~(CMB) as the only background radiation field. The collision partners taken into account include H$_{2}$ and $e^{-}$, with the electron density set to $1\,\mathrm{cm^{-3}}$. We use the theoretical values of collisional cross-sections for CO, which were valid between $1$ and $3000\,\mathrm{K}$ and calculated by \citet{Yang2010} based on the potential energy surfaces reported by \citet{Jankowski2005}. In the computation, the line-width is assumed to be $50\,\mathrm{km\,s^{-1}}$, which is the average velocity width of the CO $\mathrm{J}=3-2$ map convolved to a spatial resolution of $42''$~within the central $1.5'$ around the M83 nucleus. The best-fit parameters are derived by comparing directly our intensities of CO transitions with the results from \texttt{RADEX}. In this section, we describe our approach of using \texttt{RADEX} as our data interpretation tool.
		}
		
		\par{
		We first create a grid of \texttt{RADEX}-computed CO rotational line intensities based on the three main parameters: kinetic temperature~($T_{\mathrm{kin}}: \mathrm{K}$), H$_{2}$ number density~($n(\mathrm{H_{2}}): \mathrm{cm^{-3}}$), and CO column density per velocity width~($N(\mathrm{CO})/\Delta v: \mathrm{cm^{-2}\,(km\,s^{-1}})^{-1}$). For each parameter, the input value varies along 101 uniformly spaced~(in natural logarithmic scale) points within the range reported in Table~\ref{parameter_table}. This gives 1 million SLEDs generated in the three dimensional parameter space.
		}
			\begin{table}
				\caption{RADEX Parameters}
				\label{parameter_table}
				\begin{minipage*}{8.8cm}
					\begin{center}
						\begin{tabular}{cccc}
							\hline\\[-1.5ex]
							Parameter & Range & Unit & Array Size\\
							\hline\\[-1.5ex]
							$\mathrm{N}(\mathrm{CO})/\Delta v$  & $10^{12}$ -- $10^{19}$ & $\mathrm{cm^{-2}\,(km\,s^{-1})^{-1}}$ & 101\\
							n(H$_{2}$) & $10$ -- $10^{8}$ & $\mathrm{cm^{-3}}$ & 101\\
							T$_{kin}$ & $10$ -- $10^{4}$ & $\mathrm{K}$ & 101\\
							\hline
						\end{tabular}
					\end{center}
				\end{minipage*}	
			\end{table}
		\par{
		In order to efficiently search for the best-fit values in the parameter space, we create a function for each transition by linearly interpolating its values~($I_{k}\,(\mathrm{J}=i-j)$) between points in a three--dimensional space, which is labeled $T_{\mathrm{kin}}$, $n(\mathrm{H_{2}})$ and $N(\mathrm{CO})/\Delta\,v$, with \texttt{InterpolatingFunction} of the \texttt{ScientificPython} library (v2.8). These thirteen functions~(one for each line) are then called by \texttt{MPFIT} with the same set of parameters, $T_{\mathrm{kin}}$, $n\,(\mathrm{H_{2}})$ and $N(\mathrm{CO})/\Delta\,v$, to compute a SLED that best fits the observed values on a given pixel. The SLED generated by this method compares well with the results output by \texttt{RADEX} at each parameter grid. To ensure the best-fit parameters returned by \texttt{MPFIT} are indeed at the absolute minimum $\chi^{2}$ in the parameter space and not biased toward a local minimum due to our choice of initial parameters, we initiate the fitting process with parameters randomly selected within the range given in Table~\ref{parameter_table} and output one thousand sets of best-fit parameters and $\chi^{2}$ values given by \texttt{MPFIT}. We do this test with two scenarios such that the molecular cloud in the observed region (1) can be approximated by one temperature component (hereafter \emph{one-component fit}), or (2) can be decomposed into two components of different temperatures (hereafter \emph{two-component fit}). In reality, neither proposed scenario can describe all the observed molecular clouds, which might be tens or even hundreds in number, in detail. Given the limited spatial resolution of the instrument, the derived physical conditions should be treated as average values within the beam. We perform a one-component fit only at the pixels where there are more than 4 detections of CO transitions~(S/N$\,>\,1$) and a two-component fit where there are more than 7 detections, including the ground-based observations. Since this work includes multiple observed CO transitions, the $S/N$ threshold is so chosen that the model can be the most constrained by as many observed lines simultaneously. The average $S/N$ of the observed SLED is propagated with the MC method and reflected by the uncertainties of the derived physical parameters.
		}
		\par{
		Based on the previous observational results from the nucleus of M83, we expect the CO transitions of the SLED at the central pixel~(marked by blue) of Figure~\ref{co_sled} to be dominated, on average, by a cold~($T_{\mathrm{kin}}\,\lesssim\,100\,\mathrm{K}$) and dense~(n(H$_{2}$)$\,>1000\,\cmcube$) component~\citep{Israel2001, Kramer2005, Bayet2006}. We have observed that the one-component fits can often return best-fit parameters that are comparable to the values in the literature at its absolute minimum $\chi^{2}$. On the other hand, the results from two-component fits often are in two combinations: (1) a cold~($T_{\mathrm{kin}}\,<\,100\,\kelvin$) but rather diffuse~(n(H$_{2}$)$\,<100\,\cmcube$) component with a column density higher than a warm~($100\,<\,$T$_{kin}\,<1000\,\kelvin$) component which is of a reasonable density~(n(H$_{2}$)$\,\sim3000\,\cmcube$), or (2) a cold and dense~(n(H$_{2}$)$\,>\,10^{5}\,\cmcube$) component with lower column density than a warm and diffuse~(n(H$_{2}$)$\,<100\,\cmcube$) component.  The molecular gas density for the cold component predicted in the first scenario is too diffuse, while the cold-to-warm gas ratio predicted in the second scenario is too low. Neither scenario appears physical. To quantitatively test whether a second component is statistically necessary based on our data, we use the output $\chi^{2}$ and the corresponding degrees of freedom for both scenarios to perform an F-test, in which a ratio $F_{\chi}$ is computed to determine whether adding $n$ terms of parameters to the original $m$ terms can significantly improve the fit~\citep{bevington2003data} over the SLED. The ratio $F_{\chi}$ is defined as follows
		\begin{equation}
			F_{\chi}=\displaystyle{\frac{\chi^{2}(m)-\chi^{2}(m+n)}{\chi^{2}(m+n)/(N-m-n)}}.
			\label{f_test}
		\end{equation}
		For pixels with a sufficient number of detected lines for two-component fits, $F_{\chi}$ is generally less than 20, which corresponds to a $95\%$ confidence level for comparing a model of 5 degrees of freedom with another of 2 degrees of freedom, and accordingly implies that the additional parameters contributed by adding a second component does not statistically improve the fit. Therefore, we derive our parameters using exclusively a single component model.
		}
		\par{
		In order to better constrain the physical parameters, we attempt to include the two neutral carbon lines, $\cison$ and $\cits$, observed by the SPIRE FTS in the fitting. Galactic observations with the Antarctic Submillimeter Telescope and Remote Observatory~(AST/RO) have shown that the spatial distribution of $\cison$ is similar to that of CO $\mathrm{J}=1-0$ within Milky Way~\citep{Martin2004}. Assuming that the distributions of $\cison$ and $\cits$ are co-spatial with CO transitions on the $\sim\,330\,\mathrm{pc}$ scale, we assume the same kinetic temperature and molecular gas density in \texttt{RADEX} for [C\textsc{I}] as for CO but introduce the column density of [C\textsc{I}]~($N(\mathrm{C})$) as an additional parameter. With and without the inclusion of $\cison$ and $\cits$, the derived values of $n(\mathrm{H}_{2})$ and $T_{\mathrm{kin}}$ are in agreement. However, similar to what \citet{Kamenetzky2012} have found in their study of the SPIRE FTS observation of M82, $N(\mathrm{C})$ is generally larger than $N(\mathrm{CO})$ by a factor between 4 and 10 throughout the map. This result differs from the result derived from observations of Large Magellanic Cloud that $N(\mathrm{C})$ is found to be 25\% of $N(\mathrm{CO})$ in 30 Doradus and $\approx\,N(\mathrm{CO})$ in N159W~\citep{Pineda2012}, hinting that the beam-filling factor for atomic carbon might be larger than CO transitions in our observations. Another possible explanation for the large $N(\mathrm{C})/N(\mathrm{CO})$ ratio may be the high kinetic temperatures, which lead to a high value of the partition function, hence a large column density. On the other hand, recent studies of CO $\mathrm{J}=1-0$ in M51 suggest that more than half of the CO $\mathrm{J}=1-0$~emission might originate on scales larger than $1.3\,\mathrm{kpc}$~\citep{Pety2013}. Assuming that the two lowest transitions of CO, $\mathrm{J}=1-0$ and $\mathrm{J}=2-1$ originate from more diffuse regions than other transitions, we have excluded these two lines when searching for the best-fit parameters with \texttt{RADEX}. In this scenario, the derived physical conditions remain consistent with the results derived with all available CO transitions in this study, which suggests that the dataset used in this work may not be sufficient to differentiate the origins of each transition spatially. We would also like to note that when the analysis in this work is based on the extended-source calibrated FTS data and the corrected antenna temperatures from the ground-based telescope, the derived physical conditions remain the same in general. However, since the extended-source calibrated data is smaller than the point-source calibrated data, the derived $N(\mathrm{CO})$ is generally smaller when using the extended-source calibrated data.
		}
		
		\par{
		Based on these results, we opt for the one-component fit scenario in our analysis with the inclusion of only the CO transitions, based on the point-source calibrated FTS data. Figure~\ref{co_sled} shows the available SLEDs extracted from the SPIRE FTS observation, with their best-fit \texttt{RADEX} SLEDs indicated by solid black lines. The reduced $\chi^{2}$ for each fitted SLED is labeled on the individual pixel. The median reduced $\chi^{2}$ value is 0.76. The uncertainties estimated for the observed values are propagated through the MC method by randomly varying the SLEDs within a normal distribution and using the standard deviation of the best-fit parameters given by \texttt{MPFIT} as the 1-$\sigma$ uncertainty of the evaluated parameters. Further discussion of the CO SLEDs is presented in Section~\ref{sec_co_sled}.
		}
		
	\subsection{Dust properties}\label{dust_model}
	
		\begin{figure}[htbp!]
			\includegraphics[width=0.45\textwidth]{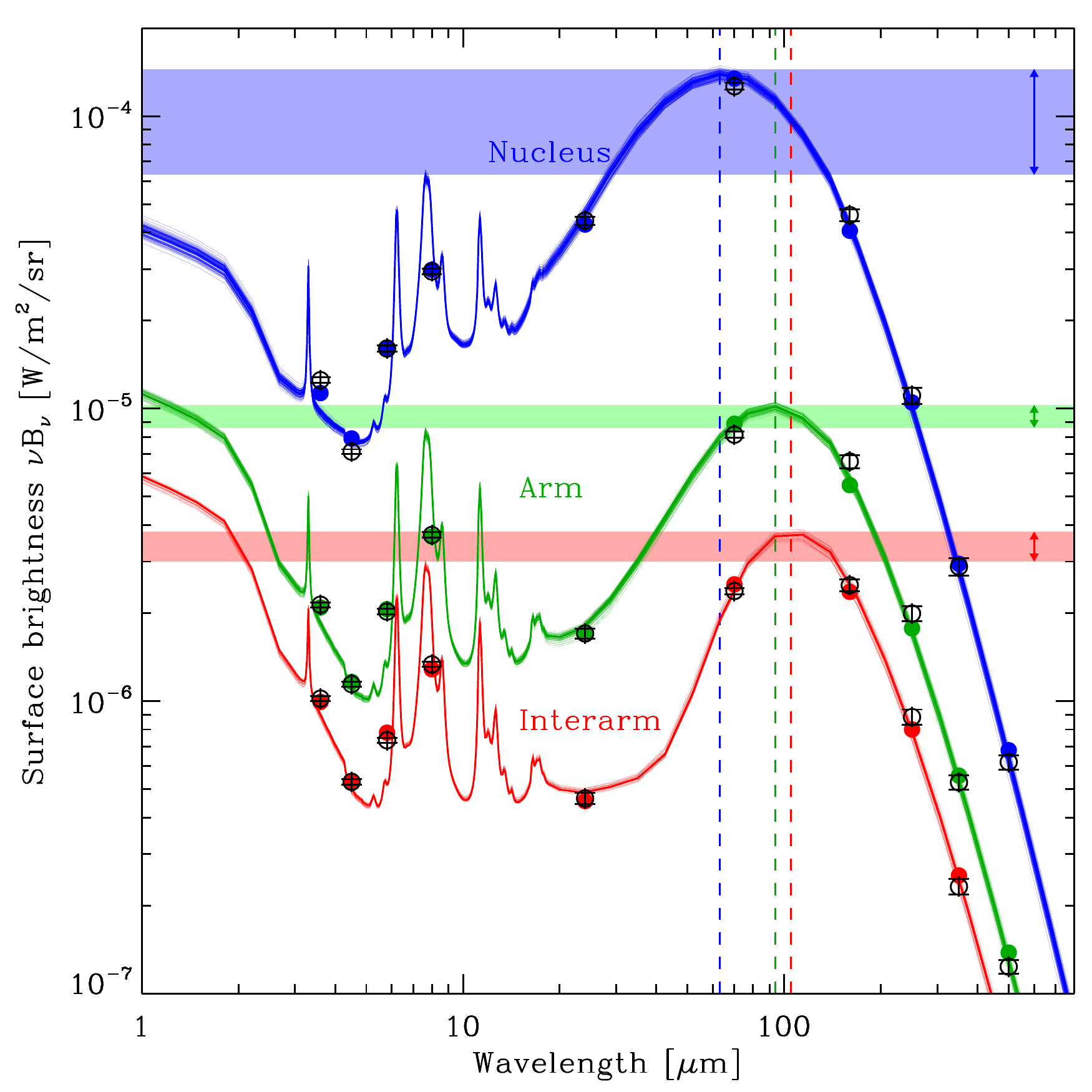}
			\caption{Selected SEDs of M83. They correspond to three pixels: one in the central region ({\it Nucleus}; blue); one on the spiral arm ({\it Arm}; green); one in the interarm region ({\it Interarm}; red). These three pixels are masked in the same colors in Figure~\ref{coverage}. The open circles with error bars are the observed photometry. The solid lines are the \citet{Galliano2011} model fit to these fluxes. The widths of the line indicate the levels of uncertainty of the model fit. The solid circles are the synthetic photometry computed from the model. The vertical dashed line shows the peak wavelength of the SED, which is an indication of the dust temperature. The horizontal stripes show the difference between the peak fluxes of the 7.7$\:\mu m$ PAH feature and the FIR. They demonstrate that the PAH-to-FIR peak ratio, which can be translated to the PAH-to-dust mass ratio~($f_{\mathrm{PAH}}$), is similar in the arm and interarm regions, but is lower in the nucleus.\label{fig:SEDs}}		
		\end{figure}
	
		\indent\par{
		One of our main objectives is to compare the physical properties derived from the CO SLEDs with the dust properties. In order to derive physical conditions for the dust, we fit the SED of each pixel of the regridded maps, using the dust model presented by \citet{Galliano2011}. In brief, this model assumes that the distribution of starlight intensities heating the dust follows a power-law \citep{Dale2001,Galliano2011}. The parameters of this power-law are derived by the fitter, and are constrained by the actual shape of the infrared SED. It therefore accounts for the fact that several physical conditions are mixed within each pixel. This model uses the dust model and the Galactic grain composition presented by \citet{Zubko2004}. We arbitrarily fix the charge fraction of polycyclic aromatic hydrocarbon~(PAH) to $1/2$, due to lack of constraints. The main parameters derived by the fitter include
		\begin{enumerate}
  			\item $M_{\mathrm{dust}}$: the dust mass;
	  		\item $\langle U\rangle$: the average intensity of the interstellar radiation field~(ISRF), normalized by the value in the solar neighborhood, $2.2\times 10^{-5}\;\rm W\,m^{-2}\,sr^{-1}$;
			\item $f_{\mathrm{PAH}}$: the PAH-to-dust mass fraction, normalized by the Galactic value~(4.6\%, \citealt{Zubko2004}).
		\end{enumerate}
		To derive the uncertainties on the fitted parameters, we perform the SED fitting on each one of the perturbed MC maps~(see Section~\ref{sec:photoMC}) with the addition of the calibration uncertainties~(see Table~\ref{photometrytable}), following the method outlined in \citet[][Sect.~3.4]{Galliano2011}.
		}
		
		\begin{figure*}[htbp!]
  			\begin{tabular}{cc}
    				\includegraphics[width=0.45\textwidth]{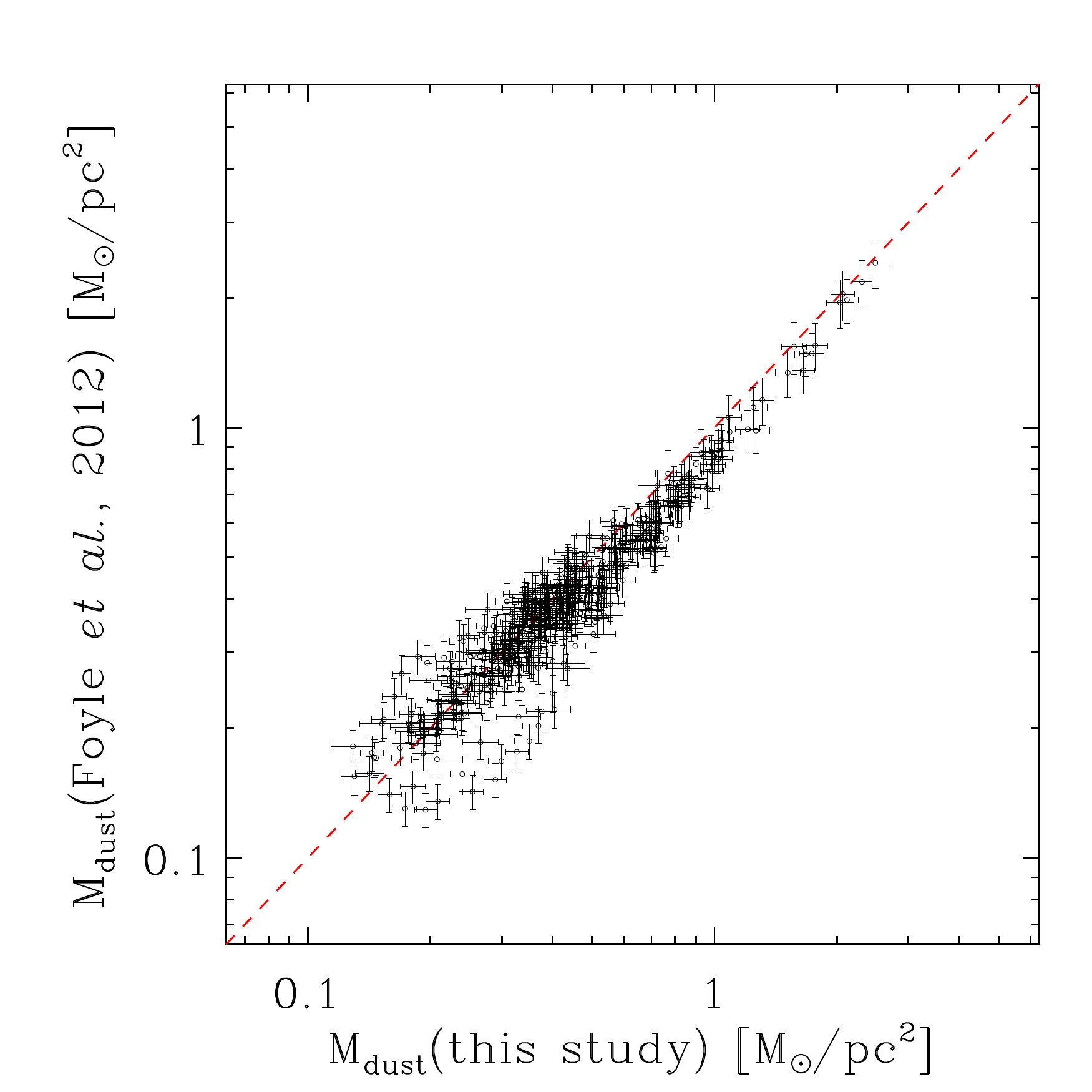} &
   			 	\includegraphics[width=0.45\textwidth]{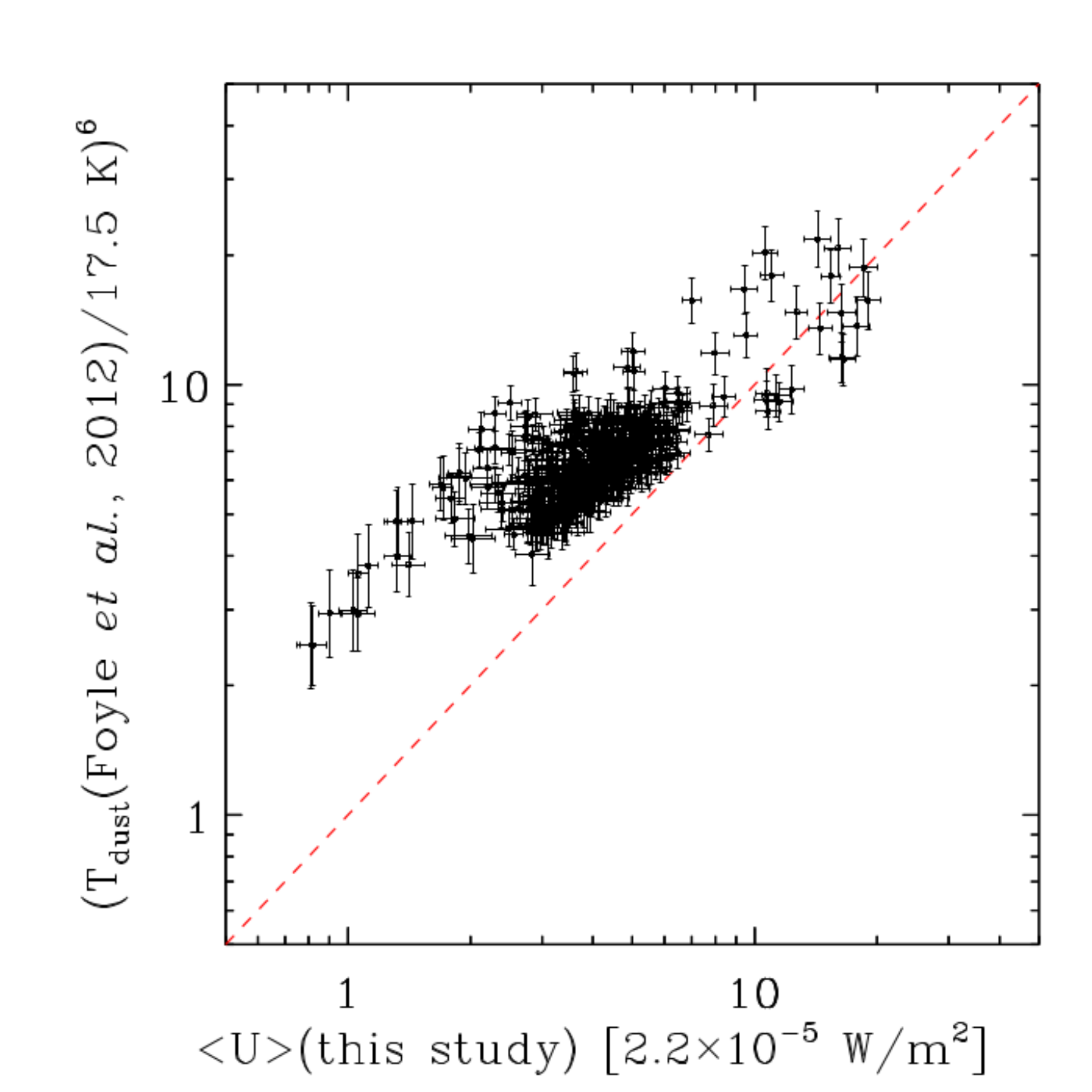}
  			\end{tabular}
  			\caption{Comparison between the results from this work and those from \citet{Foyle2012} after reprojecting the dust properties in \citet{Foyle2012} to the common grid used in this work. The left panel compares the dust masses from the two studies, pixel by pixel. The right panel compares the average intensity of the interstellar radiation field~(\uav) with the dust temperature~($T_{\mathrm{dust}}$) derived by \citet{Foyle2012}. $T_{\mathrm{dust}}$ has been related to \uav~with the relationship $U\propto T_{\mathrm{dust}}^{6}$~(see text). The red dashed lines indicate the relationship when the quantities in horizontal and vertical axes are equivalent.\label{fig:comparison}}
		\end{figure*}
		
		\par{
		Three SEDs with best-fit parameters for three pixels selected from the nucleus~(blue), arm~(green), and interarm~(red) are displayed in Figure~\ref{fig:SEDs}. The locations of these three pixels are marked by blue, green, and red colors in Figure~\ref{coverage}. The SEDs show difference in that the central region is on average hotter than the other regions, as the far-IR peaks at shorter wavelengths~(indicated by vertical dashed lines). The far-IR component is also broader at the nucleus, as a consequence of the large variety of grain excitation conditions within the pixel. It indicates that the region likely contains one or several dense phases. For instance, if a region contains a molecular cloud, which is illuminated by a star cluster on one side, then the UV-edge of the cloud will be hot, and the temperature will decrease into the cloud, until it reaches a relatively cold temperature in the core. On the other hand, in a pixel containing mainly a diffuse phase, the far-IR SED will look isothermal, as the optical depth across the pixel will be moderate. Finally, the peak of the PAH features is much lower than the peak of the far-IR intensity at the nucleus, compared with that at the arm and interarm regions~(indicated by the color stripes in Figure~\ref{fig:SEDs}). It indicates that $f_{\mathrm{PAH}}$ is lower in the nucleus while it is close to the Galactic value in the rest of the map. This latter point is likely a sign that the nucleus hosts harder radiation fields, where PAHs are destroyed~\citep{Galliano2003, Galliano2005, Madden2006, Gordon2008, Wu2011}. 
		}

		\par{
		The FIR broad-band images from the FOV modeled here have been previously studied by \citet{Foyle2012}. In \citet{Foyle2012}, five broad-band images, including the PACS and SPIRE photometry data, are fitted with a single modified blackbody function, with an assumption that the emissivity~($\beta$) is equal to 2, to derive the dust mass and the equilibrium grain temperature~($T_{\mathrm{dust}}$). In this work, we perform a more complex and realistic modeling, by accounting for the temperature mixing within each pixel, while \citet{Foyle2012} assumed an isothermal distribution of equilibrium grains. The model used in this work allows us to constrain the PAH mass fraction, which is interesting to be compared with the density of molecular gas~($n(\mathrm{H_{2}}$) derived from the CO SLEDs). To check the consistency between the results given in \citet{Foyle2012} and this work, the derived dust properties are compared in Figure~\ref{fig:comparison}. The dust mass compares well throughout the FOV. However, comparison with $T_{\mathrm{dust}}$ is less straightforward, as we derive the intensity of ISRF, \uav, within each pixel, instead. In the FIR/sub-mm regime, the relationship between \uav~and $T_{\mathrm{dust}}$ can be written as $U\propto T_{\mathrm{dust}}^{4+\beta}$~($\beta=2$). Moreover, $T_{\mathrm{dust}}$ is approximately $17.5\;\rm K$ in the solar neighborhood~\citep{Boulanger1996}, implying \uav$\simeq\left(T_{\mathrm{dust}}/17.5\:\rm K\right)^6$. The two quantities are in general agreement, as shown on the right panel of Fig.~\ref{fig:comparison}. However, the relation is more scattered than with that for the dust masses, as~\uav~and $T_{\mathrm{dust}}$~are not exactly equivalent.
		}
					
\section{Results and discussion}\label{results}

	\indent\par{
	The presentation of results is organized as follows. We first discuss the observed star formation rate~(SFR) in a global scale~($\sim\,3.5'$ in diameter) within our FOV~(all pixels in Figure~\ref{coverage}, except the cyan-masked ones), and whether it is spatially related to the ionized gas tracers $\niitof$, to $\cits$, and to $\scoone$ in M83. We then zoom in to a smaller region~($\sim\,2.3'$ in diameter) of the disk to discuss what we find with the observed CO transitions from the unmasked pixels in Figure~\ref{coverage}. We only derive physical parameters with \texttt{RADEX} from the pixels where more than four CO transitions are detected~(see also Section~\ref{radex}) in the CO spectral line energy distribution~(SLED), including the available ground-based observations, so the derived physical parameters are from an area approximately $1.3'$ around the nucleus
	}
	
	\subsection{Star formation rate and the fine--structure lines}\label{sfr_nii}

		\begin{figure}[htbp]
			\centering
			\includegraphics[width=0.5\textwidth]{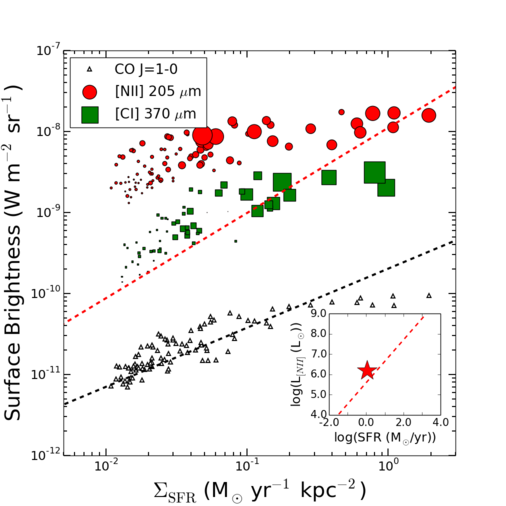}
			\caption{Relationships between $\sfrd$ and $\niitof$~(red dots), between $\sfrd$ and $\cits$~(green squares), and between $\sfrd$ and $\scoone$~(black triangles). The sizes of the red dots and green squares are proportional to the $S/N$ of the measured intensities. For $\niitof$, the $S/N$ ranges between 0.3 and 10.0 with a median value of 2.4. For $\cits$,  the $S/N$ ranges between 0.1 and 10.0 with a median value of 2.3.  $\niitof$ and $\scoone$ are compared with $\sfrd$ at a spatial resolution of $22''$, and $\cits$ is compared with $\sfrd$ at a spatial resolution of $38''$. The red and black dashed lines indicate the SFR calibration converted from that given by \citet{Zhao2013} for $\niitof$ and by \citet{Kennicutt2007} for $\scoone$, respectively. The box in the bottom right shows the integrated quantities of SFR and $L_{\niitof}$ of M83 from this work, with the dynamical range of both axes set to be the same as that in \citet[][Figure~2]{Zhao2013}.\label{sfr_nii_co}}			
		\end{figure}

		\indent\par{
		The SPIRE FTS is by far the most sensitive instrument that can spatially resolve the $\niitof$ emission from an extragalactic object. In the bandwidth covered by the SPIRE FTS, the $\niitof$ is the most widely detected emission line from the observed region of M83. From the spatial distribution, the emission of $\niitof$ peaks at the nucleus and extends from northeast to southwest in the FOV of the SPIRE FTS, where local maxima of $\sfrd$ can be identified in several spots.  The fine--structure line of the atomic carbon, the $\cits$ line~(see Figure~\ref{ci370_linestack}), is also well detected over our FOV. Compared with the $\niitof$ emission, the spatial distribution of the $\cits$ emission appears more concentrated around the nuclear region.
		}
		
		\par{
		Tracing star formation with $\niitof$ has recently been proposed. \citet{Zhao2013} used a subsample of 70 spatially--unresolved galaxies from the \textsl{Herschel} open time project, \textsl{Herschel} Spectroscopic Survey of Warm Molecular Gas in Local Luminous Infrared Galaxies (LIRGs) (PI: N. Lu), and found a relationship between the measured luminosity of $\niitof$ and the SFR, where the SFR in \citet{Zhao2013} was calibrated from the total infrared luminosity by using the relationship given in~\citet{Kennicutt2012}. The calibration they found also well describes 30 star--forming galaxies, which are also included in their analysis. The star--forming galaxies in their analysis are taken from the sample in \citet{Brauher2008}, where $\niitof$ is scaled from the $\mathrm{[N\textsc{II}]}\ 122\ \mum$ with a theoretical emission ratio of 2.6~(see \citet{Zhao2013} for more details). As to the atomic carbon emission, studies of nearby galaxies have revealed that the contribution of atomic carbon~(traced by $\cison$) to the gas cooling stays comparable with the emission of CO $\mathrm{J}=1-0$ in a variety of environments, and it has been suggested that the atomic carbon emissions can be a good molecular gas tracer~\citep{Gerin2000, Israel2002}.
		}
	
		\par{
		We investigate how the $\niitof$ and $\cits$ emissions relate to $\sfrd$ and how their relationships compare with the existing one between $\sfrd$ and the CO $\mathrm{J}=1-0$ transition~\citep{Kennicutt2007}, where $\sfrd$ is calibrated with the H$\alpha$ and $24\,\mum$ emission. Because $\cits$ is observed at a spatial resolution of $\sim\,38''$ while the spatial resolutions of the $\niitof$ and $\scoone$ maps are comparable~($\sim\,17''$ and $22''$, respectively), we compare the pixel-by-pixel values of the $\niitof$ surface brightness with the $\sfrd$, and of the $\scoone$ with $\sfrd$ at a spatial resolution of $22''$ and of the $\cits$ surface brightness with the $\sfrd$ at a spatial resolution of $38''$. The $\sfrd$ is calculated from the FUV and $24\ \mum$ photometry maps of M83 following the calibration given in \citet{Hao2011}. Figure~\ref{sfr_nii_co} shows that a general relationship exists between the values of $\niitof$ and the $\sfrd$. Compared with the $\scoone$, the surface brightness of $\niitof$ is around two orders of magnitudes higher, and the relationship between $\niitof$ and $\sfrd$ appears to have a shallower slope in Figure~\ref{sfr_nii_co}. It is interesting to note that the values of $\niitof$ appear to reach a maximum around $2\times10^{-8}\ \mathrm{W\ m^{-2}\ sr^{-1}}$~(at a spatial resolution of $22''$) and range only over one order of magnitude while the values of $\sfrd$ range over about two orders of magnitude in the FOV.
		}		
		
		\par{
		\citet{Zhao2013} have shown that the $\niitof$ luminosity holds the following relationship with the total SFR~(calibrated with the total infrared luminosity, $L_{\mathrm{IR}}$) among a sample of 70 luminous infrared galaxies~(LIRGs) and 30 star--forming galaxies
			\begin{equation}
				\centering
				\displaystyle{\log\left(\frac{\mathrm{SFR}}{\mathrm{M_{\mathrm{\odot}}}\,\mathrm{yr}^{-1}}\right)=-5.31+0.95\,\log\left(\frac{L_{\niitof}}{L_{\mathrm{\odot}}}\right)}.
				\label{zhao_sfr}
			\end{equation}
After converting the total SFR to $\sfrd$, and the $\niitof$ luminosity to surface brightness by using the pixel size~($15''$, $330\ \mathrm{pc}$) of the FTS maps, Equation~\ref{zhao_sfr} is compared with our spatial results from M83 as the red dashed line in Figure~\ref{sfr_nii_co}. Equation~\ref{zhao_sfr} intersects with the data only at the high $\sfrd$~($\sfrd\,>\,1\,\sfrdunit$) end, but for regions with $\sfrd\,<\,1\,\sfrdunit$, Equation~\ref{zhao_sfr} does not fit well to the resolved data from M83. This suggests that the relationship between $\sfrd$ and $\niitof$ described by Equation~\ref{zhao_sfr} is dominated by active star--forming regions. In order to compare M83 directly with the sample included in making the calibration in Equation~\ref{zhao_sfr}, we integrate the $\niitof$ surface brightness and $\sfrd$ from the observed region to obtain the $L_{\niitof}$ and total SFR. The integrated value from M83 is shown in the embedded plot in Figure~\ref{sfr_nii_co} where Equation~\ref{zhao_sfr}~(indicated by the red dashed line) well matches the values from M83. This result shows that Equation~\ref{zhao_sfr} can possibly be applied to galaxies in its global scale but cannot well describe the emission averaged from regions on $\sim\,300\ \mathrm{pc}$ scale. This is probably caused by the fact that $\niitof$ can also originate from more diffuse regions than the FUV and $24\ \mum$ emission due to its low critical density~($\sim\,50\,\cmcube$, main collisional partner: $e^{-}$). This implies that the emission of $\niitof$ is more uniformly distributed within the galaxy than the adopted SFR indicator, even though its emission is still related to the star--forming activity. Indeed, Galactic observations with the \textsl{Far-InfraRed Absolute Spetrophotometer~(FIRAS)} on the \textsl{COsmic Background Explorer~(COBE)} have suggested that most of the Galactic $\niitof$ emission arises from from diffuse~($n_{e}\,<\,100\,\mathrm{cm^{-3}}$) regions~\citep{Wright1991, Bennett1994}. On the other hand, the relationship between the $\cits$ emission and $\sfrd$ appears more linear. Within our FOV, the $\cits$ surface brightness ranges over 1.5 orders of magnitude, which is comparable to the range of the values of $\sfrd$. $\cits$ has a critical density of $\sim\,1200\,\cmcube$~(main collisional partners: H and H$_{2}$), which is closer to the critical density of $\scoone$~($\sim\,3000\,\cmcube$, main collisional partners: H and H$_{2}$), and its emission is brighter than $\scoone$ by at least one order of magnitude. The $\cits$ emission might originate from regions of similar physical conditions as $\scoone$, and can potentially be a more reliable star formation tracer than the $\niitof$ emission.
		}
		
		\par{
		Figure~\ref{sfr_nii_co} also compares the $\scoone$ with $\sfrd$. The calibration of $\sfrd$ as a function of $\htwod$~(estimated from the CO $\mathrm{J}=1-0$ intensity, $I_{\mathrm{CO}}$, \citealt{Kennicutt2007}) is indicated by the black dashed line~(Equation~\ref{kennicutt_sfr}).
		
			\begin{equation}
				\centering
				\displaystyle{\log\left(\frac{\sfrd}{\sfrdunit}\right)=-3.78+1.37\,\log\left(\frac{\htwod}{\mathrm{M_{\odot}\,pc^{-2}}}\right)}
				\label{kennicutt_sfr}
			\end{equation}
			
		\noindent
		We use $N(\mathrm{H_{2}})\,=\,2.8\times10^{20}\ I_{\mathrm{CO}}\ \mathrm{cm^{-2}\,(K\,km\,s^{-1})^{-1}}$, the same conversion adopted by \citet{Kennicutt2007}, to convert $\htwod$ to the $\scoone$ in Equation~\ref{kennicutt_sfr}. The relationship between the $\sfrd$ and $\scoone$ observed in M83 compares well with the spatially resolved study in M51.
		}

	\subsection{CO spectral line energy distribution}\label{sec_co_sled}
		\begin{figure}[h]
			\centering
			\includegraphics[width=0.5\textwidth]{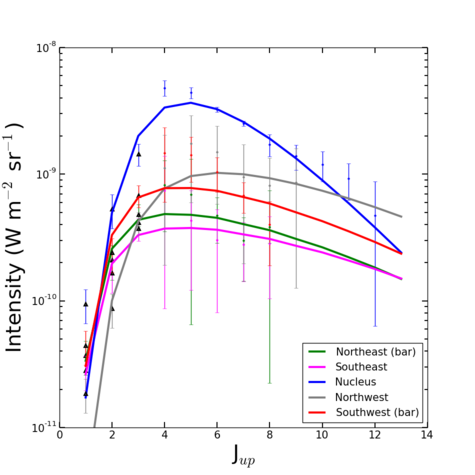}
			\caption{Comparison of the CO SLEDs from the five color--masked pixels in Figure~\ref{co_sled}. The five pixels are chosen to be the nucleus~(blue) and four pixels that have equal distance~($\sim\,1\,\kpc$) from the nucleus at the northeast~(green), southeast~(magenta), northwest~(gray), and southwest~(red). The SLEDs and their best-fit from \texttt{RADEX} are shown in the same color.}
			\label{sled_compare}
		\end{figure}

		\indent\par{
		Before translating the observed quantities into physical parameters through \texttt{RADEX}, we first investigate how the intensity contributed by CO varies in different excitation states. Figure~\ref{sled_compare} shows the SLEDs from five pixels chosen from Figure~\ref{co_sled} at the nucleus~(blue), and at the northeast~(green), southeast~(magenta), northwest~(gray), and southwest~(red), that have equal distance~($\sim42''$, $\sim\,1\,\kpc$; approximately the FWHM of the beam at the CO $\mathrm{J}=4-3$ transition) from the nucleus. From the nucleus, where the peaks of intensity of the CO transitions are found~(see images in Section~\ref{appmaps}), the average $S/N$ for the detected CO transitions is $\sim\,10$. The shape of the CO SLED from the nucleus of M83 resembles the CO SLED observed from the nucleus of M82, but with the peak intensity of CO transitions found at $\mathrm{J=4-3}$~\citep{Panuzzo2010, Kamenetzky2012}, hinting possibly softer radiation fields in this region, compared with the nucleus of M82. The average $S/N$ for the detected CO transitions from the four pixels around the nucleus is $\sim\,3$, on average. The SLEDs along the bar~(the northeast and southwest pixels) and from the southeast pixel appear to have similar shapes in that the highest intensity is found at $\mathrm{J}=4-3$. The SLED from the northwest pixel, on the other hand, shows higher intensities for CO transitions beyond $\mathrm{J}=4-3$, although it shows lower intensity at the CO $\mathrm{J}=1-0$ and $\mathrm{J}=2-1$ transitions than the three pixels from northeast, southeast and southwest. At the nucleus~(the blue pixel in Figure~\ref{co_sled}), where CO transitions are detected up to $\mathrm{J}=12-11$, the proportion of $\scowarm$ to the total emitted intensity by CO is approximately 80\%, assuming that the emission at $\mathrm{J_{up}}>13$, beyond the bandwidth of the SPIRE FTS, does not make a significant contribution. At the same pixel, the contribution of $\scoone$ to the total emitted intensity is only $\sim1\%$. From Figures~\ref{co_sled} and ~\ref{sled_compare}, one can observe that, toward the nucleus, the intensities of higher excitation states ($4\leq\mathrm{J_{up}}\leq13$) increase. Moreover, the fact that the peak intensity of the CO SLED from the northwest~(gray pixel in Figure~\ref{co_sled}) of the nucleus is found at higher-J transition~($\mathrm{J}=6-5$) implies that the radiation field may be harder toward this region.
		}
		\begin{figure}[htbp!]
			\centering
			\includegraphics[width=0.45\textwidth]{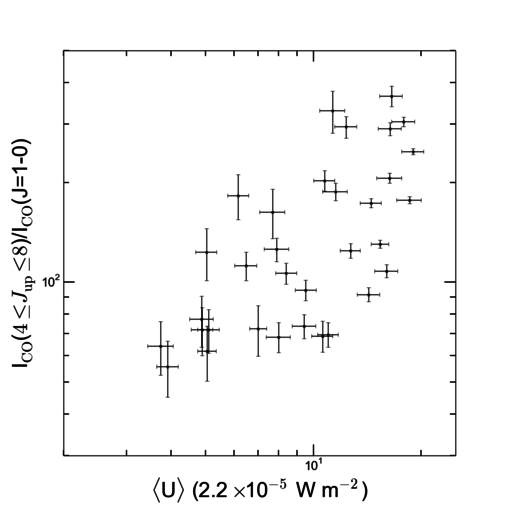}
			\caption{The relationship between the ratio of intensities~(in $\mathrm{W\ m^{-2}\ sr^{-1}}$) contributed by CO in the higher excitation states~($\mathrm{J}=4-3$ to $\mathrm{J}=8-7$) to that of CO at $\mathrm{J}=1-0$ versus \uav.\label{warmtocold_uav}}
		\end{figure}

		\begin{figure*}[htpb]
			\centering
				\subfloat[][]{				
					\includegraphics[width=0.45\textwidth]{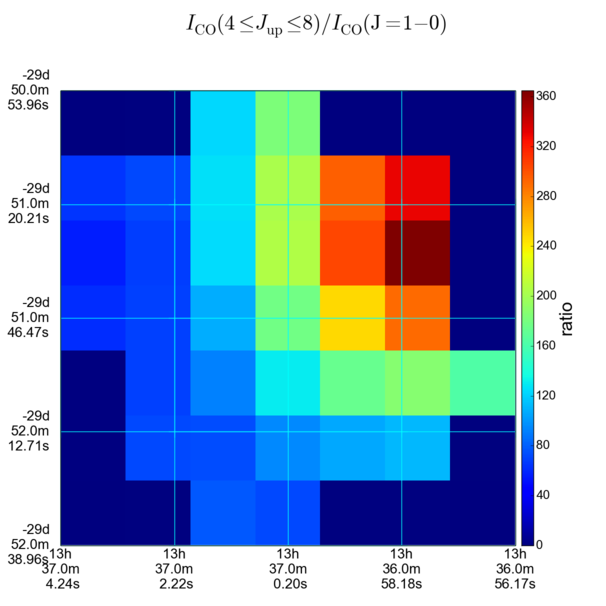}
					\label{warm_to_cold}
				}
				\quad
				\subfloat[][]{
					\includegraphics[width=0.45\textwidth]{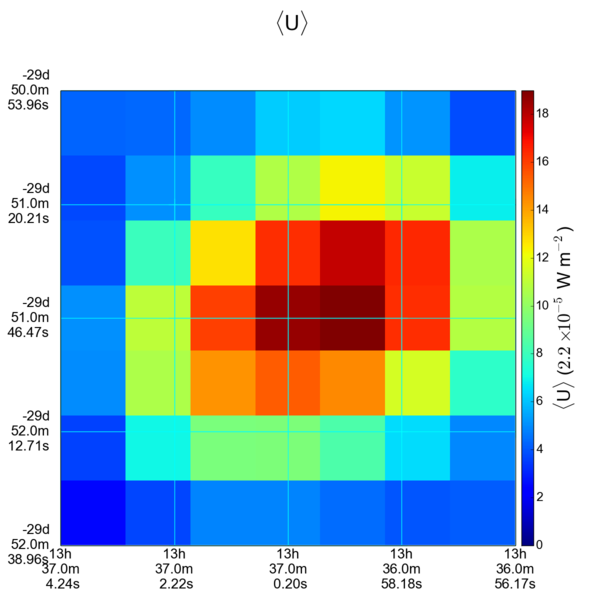}
					\label{uav}
				}
			\caption{{\it Left:} The spatial distribution of the intensity ratio plotted in the y-axis of Figure~\ref{warmtocold_uav}. {\it Right:} The spatial distribution of \uav, estimated by fitting the dust model to the SEDs over the FOV~\label{warm_to_cold_uav}. }
			
		\end{figure*}

		\par{
		To investigate how the emission of different CO transitions is affected by} the local radiation field, we express the trend described above quantitatively in Figure~\ref{warmtocold_uav}, in which we compare the ratio of total surface brightness contributed by $\mathrm{J}=4-3$ to $\mathrm{J}=8-7$~($\scowarm$, mid-J transitions) and $\scoone$ with~\uav, which is a measure of average strength of ISRF. $\scowarm$ covers most of the CO emission beyond the peak of the CO SLEDs within our FOV~(see Figure~\ref{co_sled}). Because the transitions with $\mathrm{J_{up}}\geq9$ are detected only at the few central pixels~(see Section~\ref{appmaps}), we exclude them from $\scowarm$ in our comparison. Figure~\ref{warmtocold_uav} shows a clear increase of total emitted intensity contributed by transitions from higher excitation states when the total starlight intensity increases. This increase is in qualitative, although not quantitative, agreement to what is predicted by the PDR models~\citep{Kaufman1999, Wolfire2010, LePetit2006}. The spatial distribution of $\scowarm/\scoone$ appears to increase toward the northwest of nucleus~(see Figure \ref{warm_to_cold_uav}). This implies that the $\mathrm{J}=1-0$ transition alone may not be sufficient to represent the entire population of CO molecules. Emission of CO $\mathrm{J}=1-0$ is widely employed as an estimate of the total H$_{2}$ column density~($N(\mathrm{H_{2}}$)). With spatially resolved CO transitions of higher excitation states, we can examine how the derived masses of molecular gas with \texttt{RADEX} compare to those with the intensities of CO $\mathrm{J}=1-0$, and how the physical parameters derived from Figure~\ref{co_sled} compare with dust properties and star formation rate.
		}
				
		\begin{figure*}[htbp!]
			\centering
			\subfloat[][]{
				\includegraphics[width=0.45\textwidth]{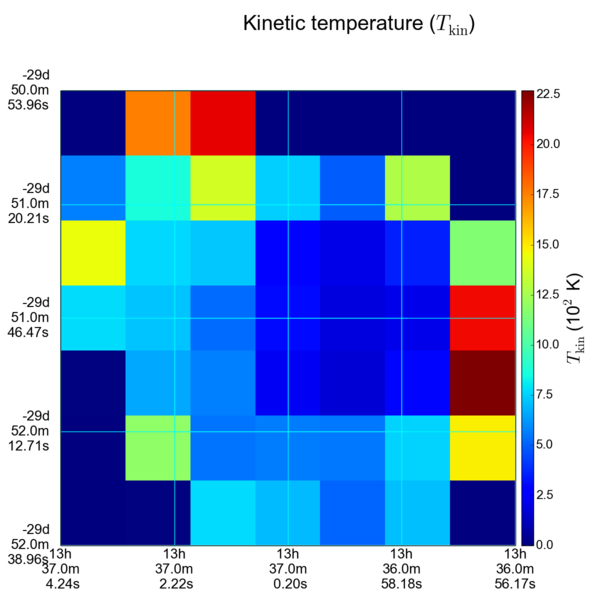}
				\label{tkin}
				}
			\quad
			\subfloat[][]{
				\includegraphics[width=0.45\textwidth]{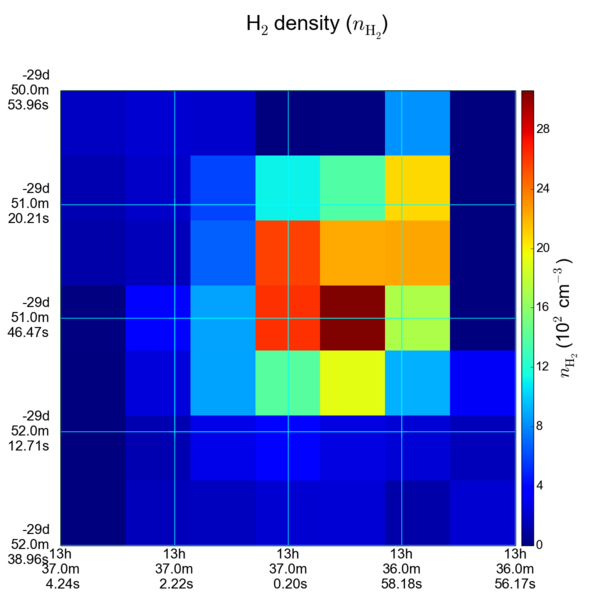}
				\label{nh2}
			}
			\quad
			\subfloat[][]{
				\includegraphics[width=0.45\textwidth]{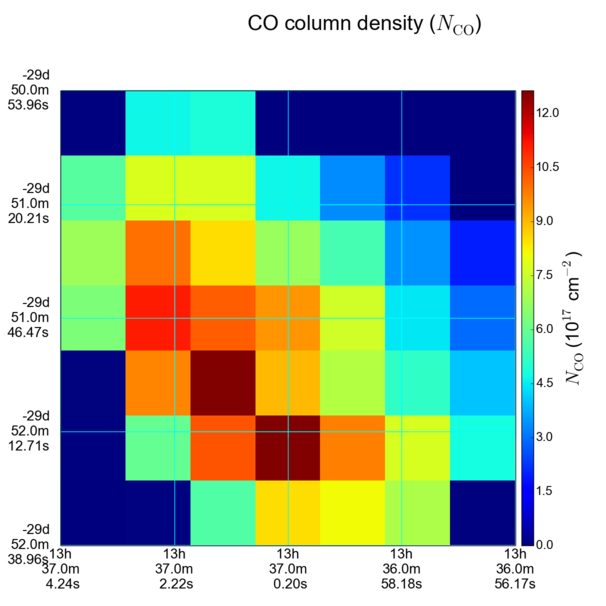}
				\label{nco}
			}
			\quad
			\subfloat[][]{
				\includegraphics[width=0.45\textwidth]{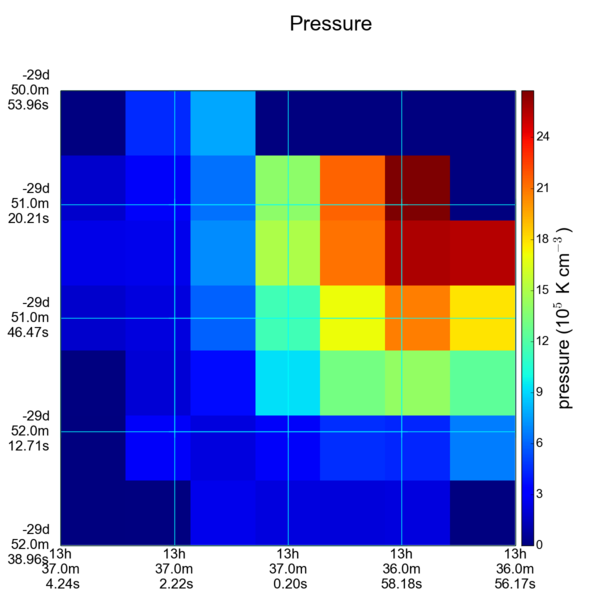}
				\label{pressure}
			}
			\caption{The best-fit physical parameters, T$_{kin}$~(top left), n(H$_{2}$)~(top right), N(CO) (bottom left), and pressure~(bottom right) derived from Figure~\ref{co_sled} by \texttt{RADEX}. Figure~\ref{pressure} is a multiplication of Figure~\ref{tkin} and Figure~\ref{nh2}.}
			\label{radex_result}
		\end{figure*}
				
		\indent\par{
		Figure~\ref{radex_result} shows the beam-averaged physical parameters derived from the observed CO SLEDs in Figure~\ref{co_sled}. We would like to emphasize that the parameter $N(\mathrm{CO})$ reported in this work actually includes an uncertainty due to our ignorance of the beam--filling factor at each transition. Constraining the beam-filling factor in the modeling introduces an additional parameter that is bound to be degenerate with $N(\mathrm{CO})$, therefore the derived $N(\mathrm{CO})$ should be viewed as the average value within the beam, which has a size of $\sim\,1\ \kpc$.  At the nucleus of Figure~\ref{co_sled} (the pixel highlighted in blue), the estimated $T_{\mathrm{kin}}$, $n(\mathrm{H_{2}})$, and, $N(\mathrm{CO})$, are $306\pm63\ \kelvin$, $2630\pm960\ \cmcube$, and $(9.49\pm1.63)\times10^{17}\ \mathrm{cm^{-2}}$, respectively. The CO transitions in this region have been observed by various ground-based telescopes. \citet{Israel2001} report $n(\mathrm{H_{2}})\,=\,1000\ \cmcube$ and $N(\mathrm{CO})/\Delta v\,=\,1\times10^{17}\,\mathrm{cm^{-2}}$. \citet{Kramer2005} report $n(\mathrm{H_{2}})\,=\,3000\ \cmcube$ and $N(\mathrm{CO})/\Delta v\,=\,3.2\times10^{16}\,\mathrm{cm^{-2}}$. \citet{Bayet2006} report $n(\mathrm{H_{2}})\,=\,6.5\times10^{5}\ \cmcube$ and $N(\mathrm{CO})/\Delta v\,=\,6\times10^{16}\,\mathrm{cm^{-2}}$. Our derived n(H$_{2}$) and N(CO) are in general agreement with previously derived values in~\citet{Israel2001, Kramer2005} but different from the values reported in \citet{Bayet2006}. The range of kinetic temperatures, T$_{kin}$, derived in this work, however, are higher than the values derived from previous observations~($30-150\,\kelvin$,~\citealt{Israel2001}; $15\,\kelvin$,~\citealt{Kramer2005}); and $40\,\kelvin$,~\citealt{Bayet2006}). This discrepancy is largely due to our inclusion of transitions from higher excitation states, while previously reported values are based on CO SLEDs that include transitions between excitation states lower than or equal to $\mathrm{J}=6-5$. The remaining part of this Section is dedicated to the presentation of the physical parameters derived with \texttt{RADEX} from Figure~\ref{co_sled}. Due to the inherent degeneracy of $T_{\mathrm{kin}}$ and $n(\mathrm{H_{2}}$) from \texttt{RADEX}, we quantitatively present them in this work as one parameter, the thermal pressure of molecular gas~($P_{\mathrm{th}}=T_{\mathrm{kin}}\cdot n(\mathrm{H_{2}})$). However, the spatial distributions of $T_{\mathrm{kin}}$ and $n(\mathrm{H_{2}})$ individually are discussed and compared with other physical parameters.
		}
		
		\subsubsection{Column density of CO: emissivity of CO $\mathrm{J}=1-0$}
			
		\indent\par{
		The result shown in Figure~\ref{warmtocold_uav} implies that the emission of CO in the mid-J transitions, compared with that in the $\mathrm{J}=1-0$ transition, shows a steep increase with \uav. This can be seen when one compares Figure~\ref{warm_to_cold} with Figure~\ref{uav}. We define the emissivity of CO as the CO emission per molecule. The emissivity of CO in the $\mathrm{J}=1-0$ transition can be then expressed as $\emico=\scoone/N(\mathrm{CO})$. We investigate how $\emico$ varies spatially, using $N(\mathrm{CO})$ derived  with \texttt{RADEX}. Since $\scoone$ is widely used as an estimate of total mass of molecular gas through a conversion factor, $X_{CO}$, and $N(\mathrm{CO})$ can be related to $N(\mathrm{H_{2}}$) through the abundance ratio of CO and H$_{2}$~([CO/H$_{2}$]), $\emico$ can be approximated as $X_{CO}^{-1}\ \mathrm{[CO/H_{2}]}^{-1}$. The spatial variation of $\emico$ can therefore be approximately regarded as a combined variation of $X_{\mathrm{CO}}$ and $\mathrm{[CO/H_{2}]}$. Figure~\ref{ncoico_uav} shows the relationship between \uav\ and $\emico$. We derive an empirical relationship between $\emico$ and \uav\ as follows:
		\begin{equation}
			\displaystyle{\emico=(3.08\pm0.89)+(0.36\pm0.08)\cdot\langle U\rangle}
			\label{emico_uav}
		\end{equation}
		
		\begin{figure}[htbp]
			\centering
			\includegraphics[width=0.45\textwidth]{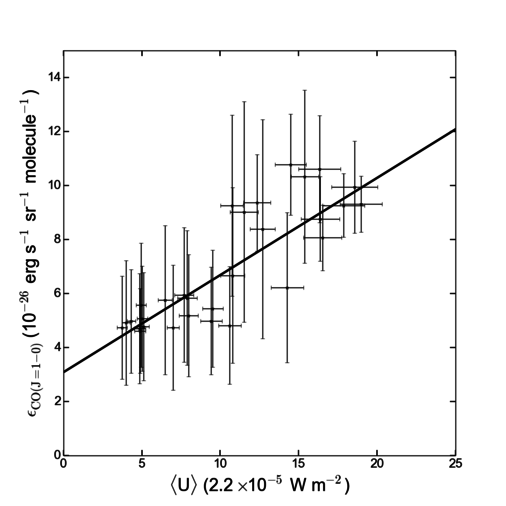}
			\caption{A relationship between the emissivity of CO $\mathrm{J}=1-0$ and \uav, derived by fitting the dust model to the SEDs.\label{ncoico_uav}}
		\end{figure}

		\noindent
		where $\emico$ is in units of $10^{-26}\,\mathrm{erg\ s^{-1}\ sr^{-1}\ molecule^{-1}}$, and \uav\ in units of $2.2\times10^{-5}\,\mathrm{W\ m^{-2}}$. The relation given in Equation~\ref{emico_uav} is fitted to Figure~\ref{ncoico_uav} with a reduced $\chi^{2}$ of 1.07.
		}
		
		\par{
		M83 is known to have a shallow radial metallicity gradient~\citep{Dufour:1980ja}. A measurement from 11 targeted H\textsc{II} regions in M83 show that the oxygen abundance, expressed as $12+\log\mathrm{(O/H)}$, in the central $\sim\,80''$ of diameter~(approximately the area shown in Figure~\ref{radex_result}) is around 9.15 and varies between 9.07 and 9.25, which corresponds to a metallicity of $\sim\,2\,Z_{\odot}$\footnote[3]{$12+\log(\mathrm{O/H})_{\mathrm{\odot}}=8.69$~\citep{Asplund2009}}~\citep{Bresolin2002}. Although $[\mathrm{CO}/\mathrm{H_{2}}]$ depends not only on the metallicity but also on the relative spatial distribution of CO and H$_{2}$, the variation of the mass fraction of ``CO--free'' H$_{2}$ might not be significant given that the metallicity within our FOV is super solar and has a shallow radial gradient~\citep{Wolfire2010}. If one assumes that the relative abundance of CO to H$_{2}$ is uniform within our FOV, the relationship observed in Figure~\ref{ncoico_uav} implies that $X_{CO}$ decreases as \uav\ increases. This relationship is similar to the existing relationship between $\alpha_{\mathrm{CO}}$, the molecular--mass-to-intensity ratio, and luminosity of CO $\mathrm{J}=1-0$ transition~($L_{\mathrm{CO}}$), in which a decreasing trend in $\alpha_{\mathrm{CO}}$ is found when the $L_{\mathrm{CO}}$ of giant molecular clouds~(GMC) in the Milky Way increases~\citep{Solomon1987, Bolatto2013}. With a sample of 26 nearby galaxies, \citet{Sandstrom2013} have found that $\alpha_{\mathrm{CO}}$ decreases when \uav\ increases when comparing the dust SEDs with $\scoone$ at the $\sim\,1\kpc$ scale, assuming a metallicity--dependent gas-to-dust mass ratio. Lower values of $X_{\mathrm{CO}}$ have also been observed from local ULIRGs~\citep{Papadopoulos2012} and mergers~\citep{Narayanan2011}. 
		}
		
		\begin{figure}[htbp]
			\centering
			\includegraphics[width=0.45\textwidth]{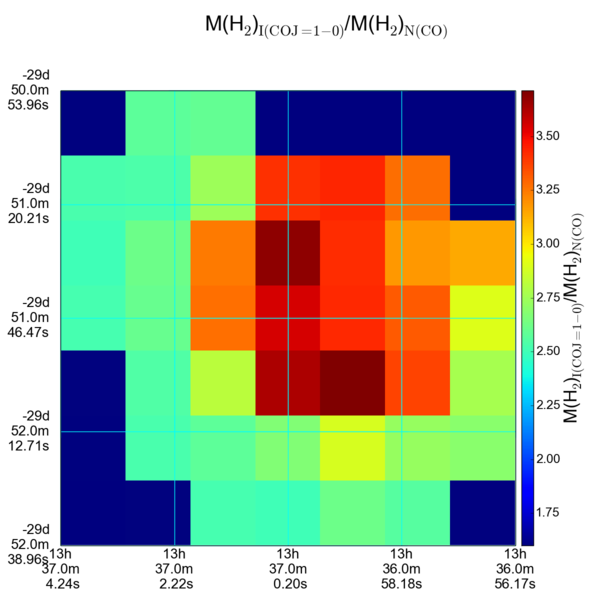}
			\caption{The spatial distribution of $\mhtwoico/\mhtwonco$\ from the area where CO SLEDs are modeled with \texttt{RADEX}.\label{mh2_ratio}}
		\end{figure}
		
		\par{
		The spatial variation of $\emico$ can also be regarded as the variation of the mass of H$_{2}$ derived from two different approaches. We define $\mhtwoico$ as the mass of H$_{2}$ derived from $\scoone$ with the Galactic $X_{\mathrm{CO}}$ value, $2\times10^{20}\ \mathrm{cm^{-2}\ (K\ km\ s^{-1})^{-1}}$~\citep{Strong1996}, and $\mhtwonco$ as the total H$_{2}$ mass converted from the estimated $N(\mathrm{CO})$ with \texttt{RADEX} in this work, assuming $[\mathrm{CO}/\mathrm{H_{2}}]=2.7\times10^{-4}$, which is comparable to the solar abundance and is measured from the nearby Flame Nebula, NGC\ 2024~\citep{Lacy1994}. Figure~\ref{uav} and Figure~\ref{mh2_ratio} spatially compare the distribution of \uav~and $\mhtwoico/\mhtwonco\propto\emico$. As suggested by Figure~\ref{ncoico_uav}, \uav\ and $\mhtwoico/\mhtwonco$ generally resemble each other in their spatial distribution. Values of $\mhtwoico/\mhtwonco$ in Figure~\ref{mh2_ratio} have an average of $2.4$\ and range between $1.6$ and $3.7$. Given that $\mhtwoico$ and $\mhtwonco$ are estimated from two different approaches, the adopted conversion factors generally agree with each other. However, our result suggests that, on the local scale, the ISRF might have a role in determining both $X_{\mathrm{CO}}$ and $[\mathrm{CO}/\mathrm{H_{2}}]$, whether through direct or indirect effect, and that one needs to be cautious when deriving the $X_{CO}$ factor based on the CO transitions studied in this work.
		}

		\subsubsection{Column density of CO: gas-to-dust mass ratio}

		\begin{figure}[h!]
			\centering
			\includegraphics[width=0.5\textwidth]{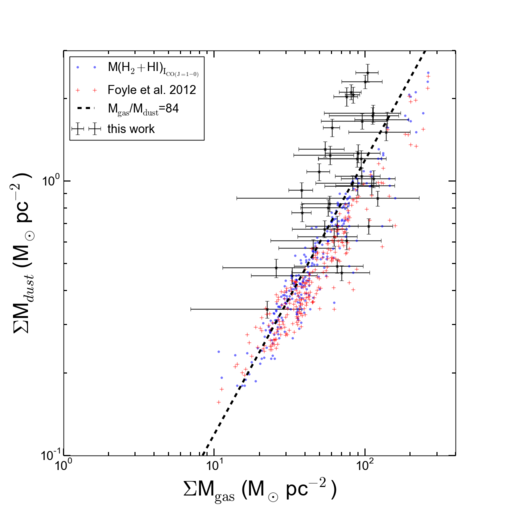}
			\caption{A comparison of the dust molecular gas masses estimated in M83. Points with error bars are the values derived using the \textsl{Herschel} SPIRE FTS data. Blue points indicate the values estimated with $\scoone$~($X_{\mathrm{CO}}\,=\,2\times10^{20}\ \mathrm{cm^{-2}\ (K\ km\ s^{-1})^{-1}}$) when there is no available data from FTS observation. Red crosses indicate the values used in~\citet{Foyle2012}, in which the $G/D$ is estimated to be $84\pm4$~(dashed line) within M83.}
			\label{mdust_mgas}
		\end{figure}

		\indent\par{
		The derived $N(\mathrm{CO})$ also provides an alternative way to estimate the gas-to-dust mass ratio~($G/D$) in M83. Figure~\ref{mdust_mgas} shows the relationship between the dust and total gas mass surface density in M83 from the entire area of Figure~\ref{coverage}. \citet{Foyle2012} has estimated an average $G/D=84\pm4$~(the dashed line in Figure~\ref{mdust_mgas}) within an area of $12'\times12'$ centered at the M83 nucleus, using the same $X_{\mathrm{CO}}$ factor as used in this work. They derive the dust mass with a modified blackbody fit to the two \textsl{Herschel} PACS and three SPIRE broad band images at $70,\ 160,\ 250,\ 350$, and, $500\ \mum$. For the H$_{2}$ mass, they scale the observed CO $\mathrm{J}=3-2$ map to match the CO $\mathrm{J}=1-0$ map and derive the H$_{2}$ mass from the scaled $\mathrm{J}=3-2$ observation. The H\textsc{I} gas mass in~\citet{Foyle2012} is derived from the M83 observation in the THINGS program, which is the same map used in this study. The data points from \citet{Foyle2012} are shown as red crosses in Figure~\ref{mdust_mgas}. We add the H\textsc{I} gas mass to $\mhtwonco$ for the points with $S/N>1$ for $N(\mathrm{CO})$ to derive the total hydrogen gas mass. The total gas mass, including the helium contribution, is then calculated as 1.36 times the total hydrogen gas mass\footnote[4]{$\mathrm{[He}/\mathrm{H]}\,\sim\,10\%$}. The $G/D$ estimated from the CO SLEDs is $106\pm47$~(the black points with error bars in Figure~\ref{mdust_mgas}), which is similar to the values found within central $1\ \mathrm{kpc}$ of radius in \citet{Foyle2012}. Because the available pixels in Figure~\ref{co_sled} only cover a $\sim\ 1.5'\times1.5'$ area around the M83 nucleus, for pixels that do not have sufficient $S/N$ data observed by the SPIRE FTS, we estimate the total gas mass with $\mhtwoico$ (blue points in Figure~\ref{mdust_mgas}). The total $G/D$ we estimated from the entire area, including the cyan-masked pixels in Figure~\ref{coverage}, is $93\pm19$, which is in general agreement with the value estimated in \citet{Foyle2012} and is about half of the $G/D$ estimated in the diffuse ISM of the Milky Way~\citep{Zubko2004}. This result is consistent with the metallicity of the central region of M83 being twice solar, as it implies that the dust-to-metal mass ratio is similar to the Galactic value.
		}
	
		\subsubsection{Column density of CO: gas depletion time}\label{gas_depletion_section}
		
		\begin{figure}[h]
			\centering
			\includegraphics[width=0.45\textwidth]{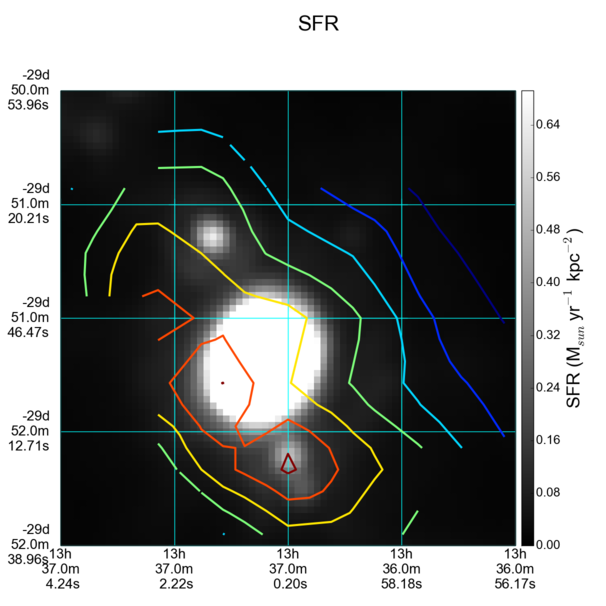}
			\caption{Contours of N(CO), derived in this work, overlaid on the SFR map. The contours are spaced by $1.5\,\times\,10^{17}\ \mathrm{cm^{-2}}$. The red and blue contours correspond to the highest~($1.2\,\times\,10^{18}\ \mathrm{cm^{-2}}$) and lowest~($3\,\times\,10^{17}\ \mathrm{cm^{-2}}$) values of N(CO)~(see also Figure~\ref{nco}).}
			\label{sfr_nco_contour}
		\end{figure}
		
		\indent\par{
		The analysis of the relationship between gas mass and SFR has given indirect evidence that nearby spiral galaxies are forming stars with a constant gas depletion time~\citep{Bigiel2008, Leroy2008, Bigiel2011, Leroy2013a}. Using a sample of 30 disk, non-edge-on~(inclination$\ \lesssim75^{\circ}$) and spatially resolved~(down to $\sim1\ \kpc$) galaxies, \citet{Bigiel2011} found a linear relationship between $\htwod$ and $\sfrd$, which gives a constant molecular gas depletion time~($\ghtwodep$) of $2.35\pm0.24\ \mathrm{Gyr}$ with $1\sigma$ scatter $0.24\ \mathrm{dex}$. This sample spans an oxygen abundance range of $8.36\ \lesssim\ 12+\log(\mathrm{O/H})\ \lesssim\ 8.93$ and a mass range of $8.9\ \lesssim\ \log\,(M_{*}/\mathrm{M_{\odot}})\ \lesssim\ 11.0$, which are compatible with the global properties of M83, which has a stellar mass of $\log\,(M_{*}/\mathrm{M_{\odot}})\,\sim\,10.9$~\citep{Jarrett2013}. As already discussed in Section~\ref{sfr_nii}, we have found that the $\htwod$~(derived with $\scoone$) and the $\sfrd$ are related spatially within M83. Figure~\ref{sfr_nco_contour} shows this relationship spatially with $N(\mathrm{CO})$ derived with \texttt{RADEX}. It is clear that $N(\mathrm{CO})$ has a very similar spatial distribution to that of the SFR map. We now revisit the quantitative relationship found in \citet{Bigiel2011} but with the addition of $\mhtwonco$, in Figure~\ref{gas_depletion}. 
		
		\begin{figure}[h]
			\centering
			\includegraphics[width=0.5\textwidth]{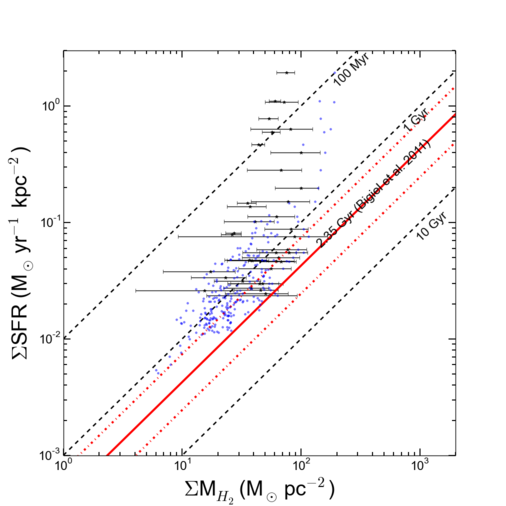}
			\caption{A comparison of the molecular gas mass and star formation rate surface density estimated in M83. Points with error bars are the values derived using the \textsl{Herschel} SPIRE FTS data. Blue points indicate the values estimated with $\scoone$~($X_{\mathrm{CO}}\,=\,2\times10^{20}\ \mathrm{cm^{-2}\ (K\ km\ s^{-1})^{-1}}$) when there is no available data from FTS observation. The $2.35\ \mathrm{Gyr}$, which is the gas depletion time found within 30 nearby spatially resolved spiral galaxies in \citet{Bigiel2011}, is shown as the solid line, with its $1\sigma$ scatter marked by the red dash-dotted lines.}
			\label{gas_depletion}
		\end{figure}
		
		Within the area of $3'\times3'$ around the nucleus, we found an average (molecular) gas depletion time of $1.13\pm0.6\ \mathrm{Gyr}$, estimated from the combination of $\mhtwonco$ and $\mhtwoico$. This number is smaller than the result previously found~($\sim2.35\ \mathrm{Gyr}$, shown as the red solid line with the $1\sigma$ values indicated by the red dotted lines in Figure~\ref{gas_depletion}). The smaller gas depletion time found in this work can be explained by the fact that the map covered in Figure~\ref{sfr_nco_contour} is dominated by M83's starburst nucleus region, within which strong starburst activity is found accompanied by a concentration of X-ray sources~\citep{Soria:2002dn}. It can be that the depletion of molecular gas by star formation is more efficient in this region, or that the molecular gas, traced by CO, is dissociated through other mechanisms. Figure~\ref{gas_depletion} also shows that $\htwod$ appears to reach a maximum at $\sim\,100\ \mathrm{M_{\odot}\ pc^{-2}}$. The appearance of the $\htwod$ saturation at $\sim\,100\ \mathrm{M_{\odot}\ pc^{-2}}$ can be caused by two reasons. First, the $N(\mathrm{CO})$ derived in this work may underestimate the total CO column density, due to the fact that the $N(\mathrm{CO})$ is an averaged property of cold and warm CO molecular gas together over a large scale~($\sim\,1\ \kpc$). However, we would like to point out that in Figure~\ref{gas_depletion}, $\mhtwoico$~(blue points) also shows the saturation toward the higher SFR regions. Second, whether through direct or indirect effects, when the SFR increases, self-shielding of CO might become less efficient so that the molecules are dissociated while H$_{2}$ molecules can still remain in the molecular state due to more efficient self-shielding~\citep{Dishoeck1988, Liszt1998, Wolfire2010, Levrier2012}. Based on these two reasons, the observed ``saturation'' of molecular mass at $\sim\,100\ \mathrm{M_{\odot}\ pc^{-2}}$ does not directly imply a maximum value of $\htwod$ when $\sfrd$ increases.
		}
				
		\subsubsection{Pressure: excitation of molecular CO}
			
			\begin{figure}[h!]
				\centering
				\includegraphics[width=0.5\textwidth]{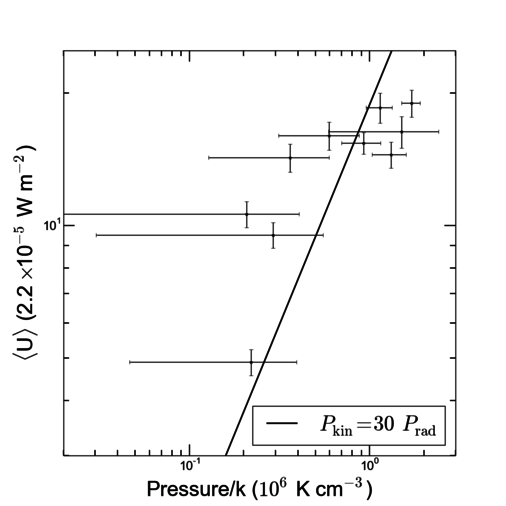}
				\caption{A relationship between $P_{\mathrm{th}}$, derived from CO SLEDs, and \uav, derived from the SEDs. The plotted points all have $S/N>1$ for the derived $P_{\mathrm{th}}$. A linear relationship is found between $P_{\mathrm{th}}$ and radiation pressure~($P_{\mathrm{rad}}$). This relationship is shown in the box on the lower right of the graph.}
				\label{u_pressure}
			\end{figure}
			
			\indent\par{
			Similar to the spatial distribution of $\scowarm/\scoone$, the spatial distribution of $P_{\mathrm{th}}$ appears to increase toward the northwest of nucleus, where peak of \uav ~is found. Figure~\ref{u_pressure} compares $P_{\mathrm{th}}$ and \uav~pixel-by-pixel, wherever the derived $P_{\mathrm{th}}$ has $S/N>1$. A general relationship between $P_{\mathrm{th}}$ and \uav~appears that, if expressing \uav~as radiation pressure~($P_{\mathrm{rad}}=\langle U\rangle/3\,c$, where $c$ is the speed of light), the pressure of the molecular gas is approximately thirty times the average radiation pressure generated by the ISRF. This means that the energy density of the ISRF is less than 5\% of the kinetic energy density of molecular gas, so that it is unlikely that \uav~can supply the energy to maintain the observed CO transitions in M83. The transitions of CO molecules observed by the SPIRE FTS may be tracing other radiation sources than the ISRF. Similarly, studies of the nearby galaxies, M82 and Arp220, also suggest that the CO transitions observed by the SPIRE FTS cannot be explained with photo-dissociation regions~(PDR) alone and still be consistent with solutions from \texttt{RADEX}~\citep{Rangwala2011, Kamenetzky2012}. However, we would also like to point out that, by studying the CO transitions in the nearby interacting system, the Antennae, with the SPIRE FTS, \citet{Schirm2013} conclude that the ISRF and the radiation generated by turbulence due to the ongoing merger or supernova are both likely responsible for the observed CO transitions. The energy source traced by the observed CO transitions should also depend on the radiation environments.}
			
			\begin{figure}[h!]
				\centering
				\includegraphics[width=0.5\textwidth]{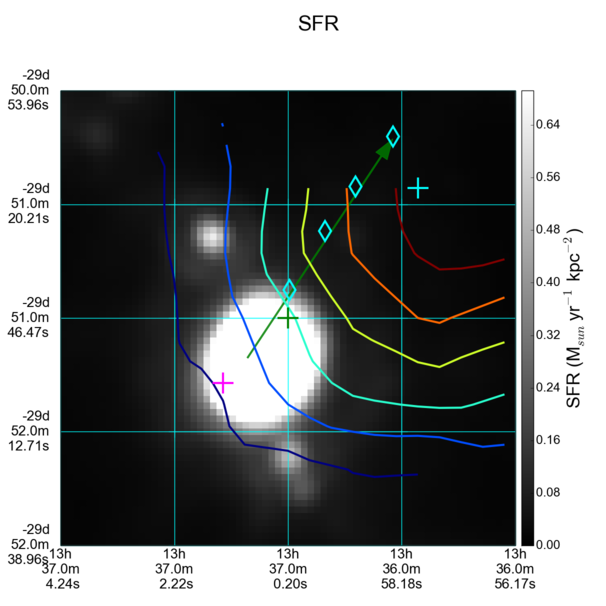}
				\caption{A graph showing the contours of $P_{\mathrm{th}}$ on the SFR map. The gray scale of SFR is recorded in the colorbar to the right. The contours of $P_{\mathrm{th}}$ ranges between $4\,\times\,10^{5}$~(blue) and $2.4\times10^{6}\,\mathrm{cm^{-3}\ K}~(red)$ and have spacings equal to $4\,\times\,10^{5}\,\mathrm{cm^{-3}\ K}$. The crosses mark the locations of three pixels chosen from Figure~\ref{co_sled}~(see text for more details). The four cyan diamonds mark the locations of four radio sources observed in \citet{Maddox:2006ey}.}
				\label{sfr_pressure}
			\end{figure}
			
			\par{
			The pressure of the molecular gas derived in this work is $1.63\,\times\,10^{5}\,<\,P_{\mathrm{th}}\,<\,2.12\,\times10^{6}\,\mathrm{cm^{-3}\ K}$, which is consistent with the estimated pressure of the warm neutral medium~(WNM) associated with supernova remnants~(SNR),  based on the shock models of \citet{Dopita1996} and the radiative shock theory. Assuming a shock velocity of~$200\,\mathrm{km s^{-1}}$ and a typical temperature of the WNM~($\sim\,5000\mathrm{K}$), \citet{Dopita:2010fe} derive from the observed H$\alpha$ luminosity of SNR in the radiative phase that the range of ISM pressures in M83 is of the order $5\,\times\,10^{3}\,<\,P\,<\,1.5\,\times10^{6}\,\mathrm{cm^{-3}\ K}$. Although the estimated pressure in \citet{Dopita:2010fe} spans a larger range than the values derived from the CO SLED, it shows that radiative shocks associated with SNRs cannot be excluded from the possible energy sources traced by the observed CO transitions. Following similar analysis as in \citet[][Equation~14]{Maloney1999}, we estimate the energy released by supernovae to be $L_{\mathrm{SN}}=1.5\,\times\,10^{8}\,L_{\mathrm{\odot}}$ in the nucleus of M83, with a supernova rate of $0.02\,\mathrm{yr^{-1}}$ estimated at the M83 nucleus~\citep{Dopita:2010fe}. The total CO luminosity of the M83 nucleus~(blue pixel in Figure~\ref{co_sled}), calculated from $\scowarm$, is approximately $4.9\,\times\,10^{4}\,L_{\mathrm{\odot}}$. This represents less than 0.5\% of $L_{\mathrm{SN}}$ and makes radiative shock produced by supernovae a possible energy source to supply the necessary energy budget for the observed CO transitions.
			}
			\par{
			Another interesting feature we observe from the map of $P_{\mathrm{th}}$ is shown in Figure~\ref{sfr_pressure} where the contours of $P_{\mathrm{th}}$ are overlaid on the SFR map. There is a clear gradient of $P_{\mathrm{th}}$ pointing from the southeast of the SFR nucleus~($\sim\ (3.5\pm2.0)\,\times\,10^{5}\ \mathrm{cm^{-3}\ K}$, magenta cross in Figure~\ref{sfr_pressure}) toward the northwest~($\sim\ (9.9\pm1.7)\,\times\,10^{5}$ and $\sim\ (2.1\pm1.5)\,\times\,10^{6}\ \mathrm{cm^{-3}\ K}$, green and cyan crosses in Figure~\ref{sfr_pressure}) within a physical scale of $\sim\,1\ \kpc$. As already pointed out in \citet{Harris2001}, based on the determined age of young star clusters with EW(H$\alpha$), they conclude that the starburst in M83 generally propagates from the southern end~(aged around or older than $10\ \mathrm{Myr}$) toward the northern end~(aged around or less than $5\ \mathrm{Myr}$) of the nucleus, within a range of $\sim\,300\ \mathrm{pc}$. The increase of pressure from the magenta toward green crosses in Figure~\ref{sfr_pressure} is in agreement with this picture. The increase of pressure from magenta toward cyan crosses also traces the alignment of the observed radio sources, which are distributed along the green arrow in Figure~\ref{sfr_pressure}, from the optical nucleus toward sources \#32, 30, 29, and 28 listed in \citet{Maddox:2006ey}~(here after R32, etc.). The locations of these sources are marked by the cyan diamonds in Figure~\ref{sfr_pressure} from the nucleus along the green arrow, respectively. Although \citet{Maddox:2006ey} suggest that the source R28 in their observation is likely the nucleus of a background radio galaxy with its radio lobes traced by the radio sources R29 and R27, \citet{Dottori:2010cz} propose that R28, which coincides with an X-ray source observed with the ACIS-S3 chip of the \textsl{Chandra X-ray Observatory}~(labeled \#39 in \citealt{Soria:2003cx}), might be a local emission from M83, based on the low redshift~($z\ll1$) derived from the marginally detected Fe--K line by \textsl{Chandra}. There is a chance that the pressure gradient is linked to the observed radio-jet. 
			}
			\par{
			Although due to the large beam size of the SPIRE FTS the derived values of pressure along the green arrow in Figure~\ref{sfr_pressure} cannot be regarded as fully independent of each other, the general observed trend and its coinciding with previous findings in the literature are encouraging. Similar studies using spatial distributions of pressure with higher spatial resolution will potentially reveal the effects of stellar feedback on molecular gas. 
			}
		
			\begin{figure*}[htbp!]
				\centering
				\subfloat[][]{
					\includegraphics[width=0.45\textwidth]{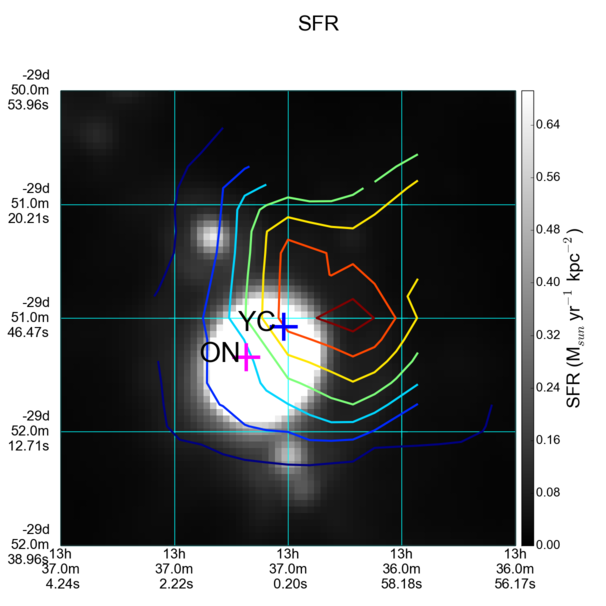}
					\label{sfr_nh2_contour}
				}
				\quad
				\subfloat[][]{
					\includegraphics[width=0.45\textwidth]{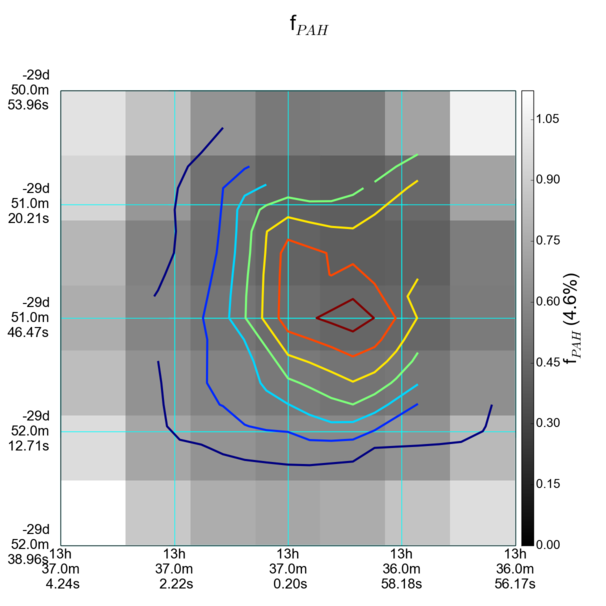}
					\label{fpah_nh2_contour}
				}
				\caption{Comparison of derived $n(\mathrm{H_{2}})$ contours with the SFR~(left) and $f_{\mathrm{PAH}}$~(right) images. In both figures, the contours for $n(\mathrm{H_{2}})$ are spaced by $400\ \cmcube$, with the red and dark blue colors correspond to the highest~($2800\ \cmcube$) and lowest~($400\ \cmcube$) values. The two crosses in the left figure indicate the locations for the optical nucleus~(magenta) and the young~(age$\,\lesssim\,3\ \mathrm{Myr}$) star clusters~(blue). The values of $f_{\mathrm{PAH}}$ in the right figure are normalized by the Galactic value, which is $4.6\%$~(see Section~\ref{dust_model}).}
			\end{figure*}
			
			\subsubsection{Off-centered $n(\mathrm{H_{2}})$~density peak}
			\indent\par{
			Comparing with the SFR map, Figure~\ref{sfr_nh2_contour} shows that the peak of the distribution of $n(\mathrm{H_{2}})$ is slightly offset toward the northwest of the SFR nucleus. Although the offset from the optical nucleus~(the magenta cross in Figure~\ref{sfr_nh2_contour}, \citealt{Thatte:2000uo}) observed in our work is $\sim25''$, which is smaller than the beam FWHM at CO $\mathrm{J}=4-3$~($\sim\,42''$), considering the bolometer spacing of SLW~($\sim\,12.7''$) from a complete Nyquist sampled map, this offset is likely to be physical. We have noted that, in \citet{Crosthwaite:2002cr}, a small offset of the peak of CO $\mathrm{J}=1-0$ emission, observed with the former NRAO $12\ \mathrm{m}$ telescope at Kitt Peak, from M83's optical nucleus~(from a $B$-band image) is also present. \citet{Sakamoto2004} also observe from the CO $\mathrm{J}=2-1$ map that its redshifted peak is separated from the blueshifted peak by approximately $10''$. Using the \textsl{Hubble Space Telescope} Wide Field Planetary Camera 2, \citet{Harris2001} has determined the age of 45 star clusters within the $300\ \mathrm{pc}$ of M83's optical nucleus based on the comparison of measured optical colors with predictions made by stellar population synthesis models generated with \texttt{Starburst99}~\citep{Leitherer1999}. Within their sample, they found four heavily extincted objects, whose ages are difficult to determine from the model prediction. Two of these four clusters, situated close to each other, aged $3.07$ and $2.22\ \mathrm{Myr}$~(\#5 and \#7 in \citealt{Harris2001}, shown as a blue cross in Figure~\ref{sfr_nh2_contour}) based on the measured equivalent widths of H$\alpha$, are found close to the peak of $n(\mathrm{H_{2}})$, derived with the SPIRE FTS. As pointed out in Section~\ref{dust_model}, the result from dust SED modeling reveals that $f_{\mathrm{PAH}}$ is lower near the nuclear region than elsewhere in M83, within our FOV, hinting at the possibility of the destruction of PAH in the nucleus. The derived $f_{\mathrm{PAH}}$ from this work is shown as the background image in Figure~\ref{fpah_nh2_contour}. At the pixel where the highest $n(\mathrm{H_{2}})$ is derived~($3000\pm2000\,\cmcube$), the value of $f_{\mathrm{PAH}}$ is $\sim0.45$, which is equivalent to a mass fraction of $\sim\,2\%$ with respect to dust, while $f_{\mathrm{PAH}}$ is close to one elsewhere. This is also one of the pixels where CO transitions are detected up to $\mathrm{J}=10-9$. The lowest value of $f_{\mathrm{PAH}}$ within our map is $\sim0.41$ and is found at the north adjacent pixel from the $n(\mathrm{H_{2}})$ peak. Since $n(\mathrm{H_{2}})$ and $f_{\mathrm{PAH}}$ are derived independently of two different methods and datasets, the observed northwestern offsets of the lowest value of $f_{\mathrm{PAH}}$ and the peak of CO emission from the optical nucleus are likely tracing very recent~($\lesssim3\ \mathrm{Myr}$) starbursts, which are heavily extincted in FUV and optical. It hints that the CO transitions more energetic than $\mathrm{J}=9-8$, which are only detected in the nucleus of M83, may be tracing radiative mechanisms directly or indirectly associated with very recent starbursts, such as shocks produced by supernovae or stellar winds.
			}
			
\section{Concluding remarks and summary}\label{conclusion}
	\indent\par{
	The \textsl{Herschel} SPIRE FTS uses a pioneering design to simultaneously observe a broad range of CO transitions in spectral imaging mode. Such a design enables us to derive the physical properties of molecular gas in a macroscopic field of view. Although the available spatial and spectral resolutions of the instrument are insufficient to differentiate the dynamics of individual molecular clouds in galaxies beyond the Milky Way, the main strength of such observations lies in its ability to provide a broad and average view of the physical conditions of the molecular gas properties traced by CO. Up to this point, our understanding about interstellar molecular gas properties is still limited. There are still many open questions, such as: what are the limitations of using observed CO to trace the states of molecular gas? What is the relationship between the formations of molecular clouds and stars? Results derived from observations by the SPIRE FTS help pave the way for the current and future studies of gas and dust, whether using ground-based telescopes, such as ALMA and JCMT, airborne observatories, such as SOFIA, or space observatories, such as JWST and SPICA. We summarize our findings from this work in the following six points:
	 \begin{enumerate}
	 	\item{
	 	Comparing the star formation surface density~($\sfrd$) with $\niitof$, $\cits$, and surface brightness of CO $\mathrm{J}=1-0$~($\scoone$), we find that the previously established relationship between $\sfrd$ and $\scoone$ can also well describe the observations in M83. On a global scale, the integrated SFR and $L_{\mathrm{\niitof}}$ from M83 agree with the calibration proposed by \citet{Zhao2013}. However, the relationship between $\sfrd$ and $\niitof$ appears to be affected by local conditions within M83 such that the regions with lower $\sfrd$ show higher $\niitof$ than that given by the calibration. On the other hand, the observed spatially resolvedlinear relationship between $\cits$ and $\sfrd$ implies that, compared with the results from $\niitof$, $\cits$ can potentially be a better SFR tracer.
	 	}
	 	\item{
	 	Based on the derived $N(\mathrm{CO})$ from the CO SLED map, we find a relationship between emissivity of CO molecules in the $\mathrm{J}=1-0$ state~($\emico=\scoone/N(\mathrm{CO})$) and \uav, the strength of the interstellar radiation field, determined from fitting the dust model to the SEDs. The increasing relationship of $\emico$ and \uav~implies a decreasing relationship between $X_{\mathrm{CO}}$ and \uav.
	 	}
	 	\item{
	 	We convert $N(\mathrm{CO})$ to mass of H$_{2}$ and derive gas-to-dust ratios mass of $77\pm33$ within the central $1\ \mathrm{kpc}$~(in radius) of M83 and $93\pm19$ from the region enclosed by the inner spiral arms, which are consistent with the results from \citet{Foyle2012}. The gas depletion time in M83 is estimated to be $1.13\pm0.6\ \mathrm{Gyr}$, which is smaller than that estimated statistically from 30 nearby galaxies~($2.35\ \mathrm{Gyr}$, \citealt{Bigiel2011}). This difference can be caused by either the limitation of the derived $N(\mathrm{CO})$ to trace mass of H$_{2}$ or the strong starburst within the M83 nucleus.
	 	}
	 	\item{
	 	Overall, the derived molecular gas pressure is consistent in its values with the estimates from H$\alpha$ photometry based on the shock models of \citet{Dopita1996}. We find that the molecular gas pressure~($P_{\mathrm{th}}$) is generally linearly related to the radiation pressure of the average starlight intensity~($P_{\mathrm{rad}}$). This relationship can be expressed as $P_{\mathrm{th}}\,=\,30\,P_{\mathrm{rad}}$, which indicates that the ISRF in M83 alone is insufficient to supply the energy budget for} the observed CO transitions in the SPIRE FTS.
	 	\item{
	 	Our calculation shows that it requires less than $0.5\%$ of supernovae luminosity~($L_{\mathrm{SN}}$) to power the observed luminosity of CO in M83. This makes the CO transitions observed with the SPIRE FTS possible tracers for radiative shocks of supernova remnants. We also observe a gradient of molecular gas pressure pointing from the M83 nucleus toward $1\ \mathrm{kpc}$ northwest of it. This direction is generally consistent with the observed propagation of the starburst pointed out by \citet{Harris2001}. The gradient of molecular gas pressure also follows the observed ``jet-like'' distribution of radio sources~\citep{Maddox:2006ey}, whose location, whether they are local or in the background, is still under debate~\citep{Dottori:2010cz}. Because our results are restricted by the spatial resolution of the instrument, we are not able to pinpoint the association of the gradient with its cause(s). However, the coincidence of these results demonstrates the strong potential in studying stellar feedback with spatial distributions of pressure.
	 	}
	 	\item{
		From the derived map of $n(\mathrm{H_{2}})$, we find an offset of the molecular density peak from the optical nucleus by $\sim\,25''$. This offset traces well the decrease in PAH-to-dust mass fraction~($f_{\mathrm{PAH}}$) as well as two heavily extincted young clusters in its location. We conclude that the observed CO transitions are possibly tracing the radiative mechanisms directly or indirectly associated with very recent starbursts.
	 	}
	 \end{enumerate}
	}

%
%
%
%
			

\begin{acknowledgements}
We would like to thank the anonymous referee for the constructive comments. RW would like to thank Dr. Estelle Bayet, Dr. Kelly Foyle, Dr. Andreas A. Lundgren, Dr. Kazuyuki Muraoka and Dr. Glenn R. Petitpas for generously sharing their data, and Dr. George Bendo, Dr. Hor\'acio Dottori, and Dr. Eric Pellegrini for fruitful discussions. Special thanks go to the members of the \textsl{Herschel} SPIRE Instrument Control Centre~(ICC) for their continuous support throughout the project. This research was supported in part by the Grant-in-Aid for Scientific Research for the Japan Society of Promotion of Science (140500000638). This research was also supported by grants from the Canadian Space Agency and the Natural Sciences and Engineering Research Council of Canada (PI: C.D. Wilson).\\
-- The FUV data presented in this paper were obtained from the Mikulski Archive for Space Telescopes (MAST). STScI is operated by the Association of Universities for Research in Astronomy, Inc., under NASA contract NAS5-26555. Support for MAST for non-HST data is provided by the NASA Office of Space Science via grant NNX13AC07G and by other grants and contracts.\\
-- This research has made use of data from HerMES project (http://hermes.sussex.ac.uk/). HerMES is a Herschel Key Programme utilising Guaranteed Time from the SPIRE instrument team, ESAC scientists and a mission scientist. The HerMES data was accessed through the Herschel Database in Marseille (HeDaM: http://hedam.lam.fr) operated by CeSAM and hosted by the Laboratoire d'Astrophysique de Marseille.\\
-- This research made use of Astropy, a community-developed core Python package for Astronomy~\citep{Robitaille2013}.\\
-- SPIRE has been developed by a consortium of institutes led by Cardiff University (UK) and including Univ. Lethbridge (Canada); NAOC (China); CEA, LAM (France); IFSI, Univ. Padua (Italy); IAC (Spain); Stockholm Observatory (Sweden); Imperial College London, RAL, UCL-MSSL, UKATC, Univ. Sussex (UK); and Caltech, JPL, NHSC, Univ. Colorado (USA). This development has been supported by national funding agencies: CSA (Canada); NAOC (China); CEA, CNES, CNRS (France); ASI (Italy); MCINN (Spain); SNSB (Sweden); STFC, UKSA (UK); and NASA (USA).\\
-- The Herschel spacecraft was designed, built, tested, and launched under a contract to ESA managed by the Herschel/Planck Project team by an industrial consortium under the overall responsibility of the prime contractor Thales Alenia Space (Cannes), and including Astrium (Friedrichshafen) responsible for the payload module and for system testing at spacecraft level, Thales Alenia Space (Turin) responsible for the service module, and Astrium (Toulouse) responsible for the telescope, with in excess of a hundred subcontractors.
\end{acknowledgements}

\bibliography{mendeley}

\begin{thebibliography}{112}
\expandafter\ifx\csname natexlab\endcsname\relax\def\natexlab#1{#1}\fi

\bibitem[{Abdo {et~al.}(2010)Abdo, Ackermann, Ajello, Baldini, Ballet,
  Barbiellini, Bastieri, Baughman, Bechtol, Bellazzini, Berenji, Bloom,
  Bonamente, Borgland, Bregeon, Brez, Brigida, Bruel, Burnett, Buson,
  Caliandro, Cameron, Caraveo, Casandjian, Cecchi, \c{C}elik, Chekhtman,
  Cheung, Chiang, Ciprini, Claus, Cohen-Tanugi, Cominsky, Conrad, Dermer,
  de~Palma, Digel, {do Couto e Silva}, Drell, Dubois, Dumora, Farnier, Favuzzi,
  Fegan, Focke, Fortin, Frailis, Fukazawa, Funk, Fusco, Gargano, Gehrels,
  Germani, Giavitto, Giebels, Giglietto, Giordano, Glanzman, Godfrey, Grenier,
  Grondin, Grove, Guillemot, Guiriec, Harding, Hayashida, Horan, Hughes,
  Jackson, J\'{o}hannesson, Johnson, Johnson, Kamae, Katagiri, Kataoka, Kawai,
  Kerr, Kn\"{o}dlseder, Kuss, Lande, Latronico, Lemoine-Goumard, Longo,
  Loparco, Lott, Lovellette, Lubrano, Makeev, Mazziotta, McEnery, Meurer,
  Michelson, Mitthumsiri, Mizuno, Monte, Monzani, Morselli, Moskalenko, Murgia,
  Nolan, Norris, Nuss, Ohsugi, Okumura, Omodei, Orlando, Ormes, Paneque,
  Pelassa, Pepe, Pesce-Rollins, Piron, Porter, Rain\`{o}, Rando, Razzano,
  Reimer, Reimer, Reposeur, Rodriguez, Ryde, Sadrozinski, Sanchez, Sander,
  Parkinson, Sgr\`{o}, Siskind, Smith, Spandre, Spinelli, Starck, Strickman,
  Strong, Suson, Takahashi, Tanaka, Thayer, Thayer, Thompson, Tibaldo, Torres,
  Tosti, Tramacere, Uchiyama, Usher, Vasileiou, Vilchez, Vitale, Waite, Wang,
  Winer, Wood, Ylinen, \& Ziegler}]{Abdo2010}
Abdo, A.~A., Ackermann, M., Ajello, M., {et~al.} 2010, The Astrophysical
  Journal, 710, 133

\bibitem[{Aniano {et~al.}(2011)Aniano, Draine, Gordon, \&
  Sandstrom}]{Aniano2011}
Aniano, G., Draine, B.~T., Gordon, K.~D., \& Sandstrom, K. 2011, Publications
  of the Astronomical Society of the Pacific, 123, 1218

\bibitem[{Asplund {et~al.}(2009)Asplund, Grevesse, Sauval, \&
  Scott}]{Asplund2009}
Asplund, M., Grevesse, N., Sauval, A.~J., \& Scott, P. 2009, Annual Review of
  Astronomy and Astrophysics, 47, 481

\bibitem[{Bayet {et~al.}(2006)Bayet, Gerin, Phillips, \& Contursi}]{Bayet2006}
Bayet, E., Gerin, M., Phillips, T.~G., \& Contursi, A. 2006, Astronomy and
  Astrophysics, 460, 467

\bibitem[{Bendo {et~al.}(2012)Bendo, Galliano, \& Madden}]{Bendo2012a}
Bendo, G.~J., Galliano, F., \& Madden, S.~C. 2012, Monthly Notices of the Royal
  Astronomical Society, 423, 197

\bibitem[{Bennett {et~al.}(1994)Bennett, Fixsen, Hinshaw, Mather, Moseley,
  Wright, Eplee, Gales, Hewagama, Isaacman, Shafer, Turpie, \& {Eplee, R.
  E.}}]{Bennett1994}
Bennett, C.~L., Fixsen, D.~J., Hinshaw, G., {et~al.} 1994, The Astrophysical
  Journal, 434, 587

\bibitem[{Bevington \& Robinson(2003)}]{bevington2003data}
Bevington, P.~R. \& Robinson, D.~K. 2003, {Data reduction and error analysis
  for the physical sciences}, McGraw-Hill Higher Education (McGraw-Hill)

\bibitem[{Bigiel {et~al.}(2008)Bigiel, Leroy, Walter, Brinks, de~Blok, Madore,
  \& Thornley}]{Bigiel2008}
Bigiel, F., Leroy, A., Walter, F., {et~al.} 2008, The Astronomical Journal,
  136, 2846

\bibitem[{Bigiel {et~al.}(2011)Bigiel, Leroy, Walter, Brinks, de~Blok, Kramer,
  Rix, Schruba, Schuster, Usero, \& Wiesemeyer}]{Bigiel2011}
Bigiel, F., Leroy, a.~K., Walter, F., {et~al.} 2011, The Astrophysical Journal,
  730, L13

\bibitem[{Blitz {et~al.}(1982)Blitz, Fich, \& Stark}]{Blitz1982}
Blitz, L., Fich, M., \& Stark, A.~A. 1982, The Astrophysical Journal Supplement
  Series, 49, 183

\bibitem[{Boissier {et~al.}(2005)Boissier, de~Paz, Madore, Boselli, Buat,
  Burgarella, Friedman, Barlow, Bianchi, Byun, Donas, Forster, Heckman,
  Jelinsky, Lee, Malina, Martin, Milliard, Morrissey, Neff, Rich, Schiminovich,
  Siegmund, Small, Szalay, Welsh, \& Wyder}]{Boissier2005}
Boissier, S., de~Paz, A.~G., Madore, B.~F., {et~al.} 2005, The Astrophysical
  Journal, 619, L83

\bibitem[{Bolatto {et~al.}(2013)Bolatto, Wolfire, \& Leroy}]{Bolatto2013}
Bolatto, A.~D., Wolfire, M., \& Leroy, A.~K. 2013, Annual Review of Astronomy
  and Astrophysics, 51, 207

\bibitem[{Boselli {et~al.}(2002)Boselli, Lequeux, \& Gavazzi}]{Boselli2002}
Boselli, A., Lequeux, J., \& Gavazzi, G. 2002, Astronomy and Astrophysics, 384,
  33

\bibitem[{Boulanger {et~al.}(1996)Boulanger, Abergel, Bernard, Burton, Desert,
  Hartmann, Lagache, \& Puget}]{Boulanger1996}
Boulanger, F., Abergel, A., Bernard, J.-P., {et~al.} 1996, Astronomy \&
  Astrophysics1, 312, 256

\bibitem[{Brauher {et~al.}(2008)Brauher, Dale, \& Helou}]{Brauher2008}
Brauher, J.~R., Dale, D.~A., \& Helou, G. 2008, The Astrophysical Journal
  Supplement Series, 178, 280

\bibitem[{Bresolin {et~al.}(2002)Bresolin, Kennicutt, \& {Kennicutt,
  Jr.}}]{Bresolin2002}
Bresolin, F., Kennicutt, R. C.~J., \& {Kennicutt, Jr.}, R.~C. 2002, The
  Astrophysical Journal, 572, 838

\bibitem[{Calzetti {et~al.}(2004)Calzetti, Harris, {Gallagher III}, Smith,
  Conselice, Homeier, \& Kewley}]{Calzetti2004}
Calzetti, D., Harris, J., {Gallagher III}, J.~S., {et~al.} 2004, The
  Astronomical Journal, 127, 1405

\bibitem[{Comte(1981)}]{Comte:1981tw}
Comte, G. 1981, Astronomy and Astrophysics Supplement, 44, 441

\bibitem[{Crosthwaite {et~al.}(2002)Crosthwaite, Turner, Buchholz, Ho, \&
  Martin}]{Crosthwaite:2002cr}
Crosthwaite, L.~P., Turner, J.~L., Buchholz, L., Ho, P. T.~P., \& Martin, R.~N.
  2002, The Astronomical Journal, 123, 1892

\bibitem[{Dale {et~al.}(2001)Dale, Helou, Neugebauer, Soifer, Frayer, \&
  Condon}]{Dale2001}
Dale, D.~A., Helou, G., Neugebauer, G., {et~al.} 2001, The Astronomical
  Journal, 122, 1736

\bibitem[{Dong {et~al.}(2008)Dong, Calzetti, Regan, Thilker, Bianchi, Meurer,
  \& Walter}]{Dong:2008ho}
Dong, H., Calzetti, D., Regan, M., {et~al.} 2008, The Astronomical Journal,
  136, 479

\bibitem[{Dopita {et~al.}(2010)Dopita, Blair, Long, Mutchler, Whitmore, Kuntz,
  Balick, Bond, Calzetti, Carollo, Disney, Frogel, O’Connell, Hall, Holtzman,
  Kimble, MacKenty, McCarthy, Paresce, Saha, Silk, Sirianni, Trauger, Walker,
  Windhorst, \& Young}]{Dopita:2010fe}
Dopita, M.~A., Blair, W.~P., Long, K.~S., {et~al.} 2010, The Astrophysical
  Journal, 710, 964

\bibitem[{Dopita \& Sutherland(1996)}]{Dopita1996}
Dopita, M.~A. \& Sutherland, R.~S. 1996, The Astrophysical Journal Supplement
  Series, 102, 161

\bibitem[{Dottori {et~al.}(2010)Dottori, D\'{\i}az, {Facundo Albacete-Colombo},
  \& Mast}]{Dottori:2010cz}
Dottori, H., D\'{\i}az, R.~J., {Facundo Albacete-Colombo}, J., \& Mast, D.
  2010, The Astrophysical Journal, 717, L42

\bibitem[{Dufour {et~al.}(1980)Dufour, {Talbort, R. J.}, Jensen, \&
  Shields}]{Dufour:1980ja}
Dufour, R.~J., {Talbort, R. J.}, J., Jensen, E.~B., \& Shields, G.~A. 1980, The
  Astrophysical Journal, 236, 119

\bibitem[{Elmegreen(2007)}]{Elmegreen2007}
Elmegreen, B.~G. 2007, The Astrophysical Journal, 668, 1064

\bibitem[{Elmegreen {et~al.}(1998)Elmegreen, Chromey, \&
  Warren}]{Elmegreen1998}
Elmegreen, D.~M., Chromey, F.~R., \& Warren, A.~R. 1998, The Astronomical
  Journal, 116, 2834

\bibitem[{Foyle {et~al.}(2012)Foyle, Wilson, Mentuch, Bendo, Dariush, Parkin,
  Pohlen, Sauvage, Smith, Roussel, Baes, Boquien, Boselli, Clements, Cooray,
  Davies, Eales, Madden, Page, \& Spinoglio}]{Foyle2012}
Foyle, K., Wilson, C.~D., Mentuch, E., {et~al.} 2012, Monthly Notices of the
  Royal Astronomical Society, 421, 2917

\bibitem[{Gallais {et~al.}(1991)Gallais, Rouan, Lacombe, Tiphene, \&
  Vauglin}]{Gallais:1991ti}
Gallais, P., Rouan, D., Lacombe, F., Tiphene, D., \& Vauglin, I. 1991,
  Astronomy and Astrophysics, 243, 309

\bibitem[{Galliano {et~al.}(2011)Galliano, Hony, Bernard, Bot, Madden,
  Roman-Duval, Galametz, Li, Meixner, Engelbracht, Lebouteiller, Misselt,
  Montiel, Panuzzo, Reach, \& Skibba}]{Galliano2011}
Galliano, F., Hony, S., Bernard, J., {et~al.} 2011, Astronomy \& Astrophysics,
  536, A88

\bibitem[{Galliano {et~al.}(2003)Galliano, Madden, Jones, Wilson, Bernard, \&
  {Le Peintre}}]{Galliano2003}
Galliano, F., Madden, S.~C., Jones, A.~P., {et~al.} 2003, Astronomy and
  Astrophysics, 407, 159

\bibitem[{Galliano {et~al.}(2005)Galliano, Madden, Jones, Wilson, \&
  Bernard}]{Galliano2005}
Galliano, F., Madden, S.~C., Jones, A.~P., Wilson, C.~D., \& Bernard, J.-P.~P.
  2005, Astronomy and Astrophysics, 434, 19

\bibitem[{Gerin \& Phillips(2000)}]{Gerin2000}
Gerin, M. \& Phillips, T.~G. 2000, The Astrophysical Journal, 537, 644

\bibitem[{Glover \& Clark(2012)}]{Glover2012}
Glover, S. C.~O. \& Clark, P.~C. 2012, Monthly Notices of the Royal
  Astronomical Society, 421, 9

\bibitem[{Gordon {et~al.}(2008)Gordon, Engelbracht, Rieke, Misselt, Smith,
  {Kennicutt, Jr.}, \& Kennicutt}]{Gordon2008}
Gordon, K.~D., Engelbracht, C.~W., Rieke, G.~H., {et~al.} 2008, The
  Astrophysical Journal, 682, 336

\bibitem[{Grenier {et~al.}(2005)Grenier, Casandjian, \& Terrier}]{Grenier2005}
Grenier, I.~A., Casandjian, J.-M., \& Terrier, R. 2005, Science (New York,
  N.Y.), 307, 1292

\bibitem[{Griffin {et~al.}(2010)Griffin, Abergel, Abreu, Ade, Andr\'{e},
  Augueres, Babbedge, Bae, Baillie, Baluteau, Barlow, Bendo, Benielli, Bock,
  Bonhomme, Brisbin, Brockley-Blatt, Caldwell, Cara, Castro-Rodr\'{\i}guez,
  Cerulli, Chanial, Chen, Clark, Clements, Clerc, Coker, Communal, Conversi,
  Cox, Crumb, Cunningham, Daly, Davis, {De Antoni}, Delderfield, Devin,
  di~Giorgio, Didschuns, Dohlen, Donati, Dowell, Dowell, Duband, Dumaye, Emery,
  Ferlet, Ferrand, Fontignie, Fox, Franceschini, Frerking, Fulton, Garcia,
  Gastaud, Gear, Glenn, Goizel, Griffin, Grundy, Guest, Guillemet, {Hargrave,
  P. C.}, Harwit, Hastings, Hatziminaoglou, Herman, Hinde, Hristov, Huang,
  Imhof, Isaak, Israelsson, Ivison, Jennings, Kiernan, King, Lange, Latter,
  Laurent, Laurent, Leeks, Lellouch, Levenson, Li, Li, Lilienthal, Lim, Liu,
  Lu, Madden, Mainetti, Marliani, McKay, Mercier, Molinari, Morris, Moseley,
  Mulder, Mur, Naylor, Nguyen, O'Halloran, Oliver, Olofsson, Olofsson, Orfei,
  Page, Pain, Panuzzo, Papageorgiou, Parks, Parr-Burman, Pearce, Pearson,
  P\'{e}rez-Fournon, Pinsard, Pisano, Podosek, Pohlen, Polehampton, Pouliquen,
  Rigopoulou, Rizzo, Roseboom, Roussel, Rowan-Robinson, Rownd, Saraceno,
  Sauvage, Savage, Savini, Sawyer, Scharmberg, Schmitt, Schneider, Schulz,
  Schwartz, Shafer, Shupe, Sibthorpe, Sidher, Smith, Smith, Spencer, Stobie,
  Sudiwala, Sukhatme, Surace, Stevens, Swinyard, Trichas, Tourette, Triou,
  Tseng, Tucker, Turner, Vaccari, Valtchanov, Vigroux, Virique, Voellmer,
  Walker, Ward, Waskett, Weilert, Wesson, White, Whitehouse, Wilson, Winter,
  Woodcraft, Wright, Xu, Zavagno, Zemcov, Zhang, Zonca, Castro-Rodriguez, \&
  Hargrave}]{Griffin2010}
Griffin, M.~J., Abergel, A., Abreu, A., {et~al.} 2010, Astronomy and
  Astrophysics, 518, L3

\bibitem[{Hao {et~al.}(2011)Hao, {Kennicutt, Jr.}, Johnson, Calzetti, Dale, \&
  Moustakas}]{Hao2011}
Hao, C.-N., {Kennicutt, Jr.}, R.~C., Johnson, B.~D., {et~al.} 2011, The
  Astrophysical Journal, 741, 124

\bibitem[{Harris {et~al.}(2001)Harris, Calzetti, {Gallagher III}, Conselice,
  Smith, \& Gallagher}]{Harris2001}
Harris, J., Calzetti, D., {Gallagher III}, J.~S., {et~al.} 2001, The
  Astronomical Journal, 122, 3046

\bibitem[{Hong {et~al.}(2011)Hong, Calzetti, Dopita, Blair, Whitmore, Balick,
  Bond, Carollo, Disney, Frogel, Hall, Holtzman, Kimble, McCarthy, O'Connell,
  Paresce, Saha, Silk, Trauger, Walker, Windhorst, Young, \&
  Mutchler}]{Hong:2011ch}
Hong, S., Calzetti, D., Dopita, M.~A., {et~al.} 2011, The Astrophysical
  Journal, 731, 21

\bibitem[{Israel \& Baas(2001)}]{Israel2001}
Israel, F.~P. \& Baas, F. 2001, Astronomy and Astrophysics, 371, 433

\bibitem[{Israel \& Baas(2002)}]{Israel2002}
Israel, F.~P. \& Baas, F. 2002, Astronomy and Astrophysics, 383, 82

\bibitem[{Jankowski \& Szalewicz(2005)}]{Jankowski2005}
Jankowski, P. \& Szalewicz, K. 2005, The Journal of chemical physics, 123,
  104301

\bibitem[{Jarrett {et~al.}(2013)Jarrett, Masci, Tsai, Petty, Cluver, Assef,
  Benford, Blain, Bridge, Donoso, Eisenhardt, Koribalski, Lake, Neill, Seibert,
  Sheth, Stanford, \& Wright}]{Jarrett2013}
Jarrett, T.~H., Masci, F., Tsai, C.~W., {et~al.} 2013, The Astronomical
  Journal, 145, 6

\bibitem[{Kamenetzky {et~al.}(2012)Kamenetzky, Glenn, Rangwala, Maloney,
  Bradford, Wilson, Bendo, Baes, Boselli, Cooray, Isaak, Lebouteiller, Madden,
  Panuzzo, Schirm, Spinoglio, Wu, \& Survey}]{Kamenetzky2012}
Kamenetzky, J.~R., Glenn, J., Rangwala, N., {et~al.} 2012, The Astrophysical
  Journal, 753, 70

\bibitem[{Kaufman {et~al.}(1999)Kaufman, Wolfire, Hollenbach, \&
  Luhman}]{Kaufman1999}
Kaufman, M.~J., Wolfire, M.~G., Hollenbach, D.~J., \& Luhman, M.~L. 1999, The
  Astrophysical Journal, 527, 795

\bibitem[{Kennicutt {et~al.}(2007)Kennicutt, Calzetti, Walter, Helou,
  Hollenbach, Armus, Bendo, Dale, Draine, Engelbracht, Gordon, Prescott, Regan,
  Thornley, Bot, Brinks, de~Blok, de~Mello, Meyer, Moustakas, Murphy, Sheth, \&
  Smith}]{Kennicutt2007}
Kennicutt, R.~C., Calzetti, D., Walter, F., {et~al.} 2007, The Astrophysical
  Journal, 671, 333

\bibitem[{Kennicutt \& Evans(2012)}]{Kennicutt2012}
Kennicutt, R.~C. \& Evans, N.~J. 2012, Annual Review of Astronomy and
  Astrophysics, 50, 531

\bibitem[{Koyama \& Inutsuka(1999)}]{Koyama1999}
Koyama, H. \& Inutsuka, S. 1999, The Astrophysical Journal, 532, 21

\bibitem[{Kramer {et~al.}(2005)Kramer, Mookerjea, Bayet, Garcia-Burillo, Gerin,
  Israel, Stutzki, \& Wouterloot}]{Kramer2005}
Kramer, C., Mookerjea, B., Bayet, E., {et~al.} 2005, Astronomy \& Astrophysics,
  441, 961

\bibitem[{Lacy {et~al.}(1994)Lacy, Knacke, Geballe, \& Tokunaga}]{Lacy1994}
Lacy, J.~H., Knacke, R., Geballe, T.~R., \& Tokunaga, A.~T. 1994, The
  Astrophysical Journal, 428, L69

\bibitem[{{Le Petit} {et~al.}(2006){Le Petit}, Nehm\'{e}, {Le Bourlot}, Roueff,
  \& Nehme}]{LePetit2006}
{Le Petit}, F., Nehm\'{e}, C., {Le Bourlot}, J., Roueff, E., \& Nehme, C. 2006,
  The Astrophysical Journal Supplement Series, 164, 506

\bibitem[{Leitherer {et~al.}(1999)Leitherer, Schaerer, Goldader, Delgado,
  Robert, Kune, de~Mello, Devost, Heckman, \& {Gonz\'{a}lez
  Delgado}}]{Leitherer1999}
Leitherer, C., Schaerer, D., Goldader, J.~D., {et~al.} 1999, The Astrophysical
  Journal Supplement Series, 123, 3

\bibitem[{Lequeux {et~al.}(1994)Lequeux, {Le Bourlot}, des Forets, Roueff,
  Boulanger, Rubio, \& {Pineau des Forets}}]{Lequeux1994}
Lequeux, J., {Le Bourlot}, J., des Forets, G., {et~al.} 1994, Astronomy and
  Astrophysics, 292, 371

\bibitem[{Leroy {et~al.}(2011)Leroy, Bolatto, Gordon, Sandstrom, Gratier,
  Rosolowsky, Engelbracht, Mizuno, Corbelli, Fukui, \& Kawamura}]{Leroy2011}
Leroy, A.~K., Bolatto, A., Gordon, K., {et~al.} 2011, The Astrophysical
  Journal, 737, 12

\bibitem[{Leroy {et~al.}(2008)Leroy, Walter, Brinks, Bigiel, de~Blok, Madore,
  \& Thornley}]{Leroy2008}
Leroy, A.~K., Walter, F., Brinks, E., {et~al.} 2008, The Astronomical Journal,
  136, 2782

\bibitem[{Leroy {et~al.}(2013)Leroy, Walter, Sandstrom, Schruba, Munoz-Mateos,
  Bigiel, Bolatto, Brinks, de~Blok, Meidt, Rix, Rosolowsky, Schinnerer,
  Schuster, \& Usero}]{Leroy2013a}
Leroy, A.~K., Walter, F., Sandstrom, K., {et~al.} 2013, The Astronomical
  Journal, 146, 19

\bibitem[{Levrier {et~al.}(2012)Levrier, {Le Petit}, Hennebelle, Lesaffre,
  Gerin, \& Falgarone}]{Levrier2012}
Levrier, F., {Le Petit}, F., Hennebelle, P., {et~al.} 2012, Astronomy \&
  Astrophysics, 544, A22

\bibitem[{Liszt \& Lucas(1998)}]{Liszt1998}
Liszt, H.~S. \& Lucas, R. 1998, 339, 561

\bibitem[{Loren \& Wootten(1978)}]{Loren1978}
Loren, R.~B. \& Wootten, H.~A. 1978, The Astrophysical Journal, 225, L81

\bibitem[{Lu {et~al.}(2014)Lu, Zhao, Xu, Gao, Armus, Mazzarella, Isaak, Petric,
  Charmandaris, D\'{\i}az-Santos, Evans, Howell, Appleton, Inami, Iwasawa,
  Leech, Lord, Sanders, Schulz, Surace, \& van~der Werf}]{Lu2014}
Lu, N., Zhao, Y., Xu, C.~K., {et~al.} 2014, The Astrophysical Journal, 787, L23

\bibitem[{Lundgren {et~al.}(2004)Lundgren, Wiklind, Olofsson, \&
  Rydbeck}]{Lundgren2004a}
Lundgren, A.~A., Wiklind, T., Olofsson, H., \& Rydbeck, G. 2004, Astronomy and
  Astrophysics, 413, 505

\bibitem[{{Mac Low} \& Klessen(2004)}]{MacLow2004}
{Mac Low}, M.-M. \& Klessen, R. 2004, Reviews of Modern Physics, 76, 125

\bibitem[{Madden {et~al.}(2006)Madden, Galliano, Jones, \&
  Sauvage}]{Madden2006}
Madden, S.~C., Galliano, F., Jones, A.~P., \& Sauvage, M. 2006, Astronomy and
  Astrophysics, 446, 877

\bibitem[{Maddox {et~al.}(2006)Maddox, Cowan, Kilgard, Lacey, Prestwich,
  Stockdale, \& Wolfing}]{Maddox:2006ey}
Maddox, L.~A., Cowan, J.~J., Kilgard, R.~E., {et~al.} 2006, The Astronomical
  Journal, 132, 32

\bibitem[{Makiwa {et~al.}(2013)Makiwa, Naylor, Ferlet, Salji, Swinyard,
  Polehampton, \& van~der Wiel}]{Makiwa2013}
Makiwa, G., Naylor, D.~a., Ferlet, M., {et~al.} 2013, Applied Optics, 52, 3864

\bibitem[{Maloney(1999)}]{Maloney1999}
Maloney, P.~R. 1999, Astrophysics and Space Science, 266, 207

\bibitem[{Markwardt(2009)}]{Markwardt2009}
Markwardt, C.~B. 2009, Astronomical Data Analysis Software and Systems XVIII
  ASP Conference Series, 411, 251

\bibitem[{Martin {et~al.}(2004)Martin, Walsh, Xiao, Lane, Walker, \&
  Stark}]{Martin2004}
Martin, C.~L., Walsh, W.~M., Xiao, K., {et~al.} 2004, The Astrophysical Journal
  Supplement Series, 150, 239

\bibitem[{Martin {et~al.}(2005)Martin, Fanson, Schiminovich, Morrissey,
  Friedman, Barlow, Conrow, Grange, Jelinsky, Milliard, Siegmund, Bianchi,
  Byun, Donas, Forster, Heckman, Lee, Madore, Malina, Neff, Rich, Small,
  Surber, Szalay, Welsh, \& Wyder}]{Martin2005}
Martin, D.~C., Fanson, J., Schiminovich, D., {et~al.} 2005, The Astrophysical
  Journal, 619, L1

\bibitem[{Morrissey {et~al.}(2007)Morrissey, Conrow, Barlow, Small, Seibert,
  Wyder, Budavari, Arnouts, Friedman, Forster, Martin, Neff, Schiminovich,
  Bianchi, Donas, Heckman, Lee, Madore, Milliard, Rich, Szalay, Welsh, \&
  Yi}]{Morrissey2007}
Morrissey, P., Conrow, T., Barlow, T.~A., {et~al.} 2007, The Astrophysical
  Journal Supplement Series, 173, 682

\bibitem[{Muraoka {et~al.}(2009)Muraoka, Kohno, Tosaki, Kuno, Nakanishi, Sorai,
  Okuda, Sakamoto, Endo, Hatsukade, Kamegai, Tanaka, Cortes, Ezawa, Yamaguchi,
  Sakai, Kawabe, Sawada, Handa, \& Fukuhara}]{Muraoka2009}
Muraoka, K., Kohno, K., Tosaki, T., {et~al.} 2009, The Astrophysical Journal,
  706, 1213

\bibitem[{Narayanan {et~al.}(2011)Narayanan, Krumholz, Ostriker, \&
  Hernquist}]{Narayanan2011}
Narayanan, D., Krumholz, M., Ostriker, E.~C., \& Hernquist, L. 2011, Monthly
  Notices of the Royal Astronomical Society, 418, 664

\bibitem[{Oliver {et~al.}(2012)Oliver, Bock, Altieri, Amblard, Arumugam,
  Aussel, Babbedge, Beelen, B\'{e}thermin, Blain, Boselli, Bridge, Brisbin,
  Buat, Burgarella, Castro-Rodr\'{\i}guez, Cava, Chanial, Cirasuolo, Clements,
  Conley, Conversi, Cooray, Dowell, Dubois, Dwek, Dye, Eales, Elbaz, Farrah,
  Feltre, Ferrero, Fiolet, Fox, Franceschini, Gear, Giovannoli, Glenn, Gong,
  {Gonz\'{a}lez Solares}, Griffin, Halpern, Harwit, Hatziminaoglou, Heinis,
  Hurley, Hwang, Hyde, Ibar, Ilbert, Isaak, Ivison, Lagache, {Le Floc'h},
  Levenson, Faro, Lu, Madden, Maffei, Magdis, Mainetti, Marchetti, Marsden,
  Marshall, Mortier, Nguyen, O'Halloran, Omont, Page, Panuzzo, Papageorgiou,
  Patel, Pearson, P\'{e}rez-Fournon, Pohlen, Rawlings, Raymond, Rigopoulou,
  Riguccini, Rizzo, Rodighiero, Roseboom, Rowan-Robinson, {S\'{a}nchez Portal},
  Schulz, Scott, Seymour, Shupe, Smith, Stevens, Symeonidis, Trichas, Tugwell,
  Vaccari, Valtchanov, Vieira, Viero, Vigroux, Wang, Ward, Wardlow, Wright, Xu,
  \& Zemcov}]{Oliver2012}
Oliver, S.~J., Bock, J., Altieri, B., {et~al.} 2012, Monthly Notices of the
  Royal Astronomical Society, 424, 1614

\bibitem[{Panuzzo {et~al.}(2010)Panuzzo, Rangwala, Rykala, Isaak, Glenn,
  Wilson, Auld, Baes, Barlow, Bendo, Bock, Boselli, Bradford, Buat,
  Castro-Rodr\'{\i}guez, Chanial, Charlot, Ciesla, Clements, Cooray, Cormier,
  Cortese, Davies, Dwek, Eales, Elbaz, Fulton, Galametz, Galliano, Gear, Gomez,
  Griffin, Hony, Levenson, Lu, Madden, O'Halloran, Okumura, Oliver, Page,
  Papageorgiou, Parkin, P\'{e}rez-Fournon, Pohlen, Polehampton, Rigby, Roussel,
  Sacchi, Sauvage, Schulz, Schirm, Smith, Spinoglio, Stevens, Srinivasan,
  Symeonidis, Swinyard, Trichas, Vaccari, Vigroux, Wozniak, Wright, \&
  Zeilinger}]{Panuzzo2010}
Panuzzo, P., Rangwala, N., Rykala, A., {et~al.} 2010, Astronomy and
  Astrophysics, 518, L37

\bibitem[{Papadopoulos {et~al.}(2012)Papadopoulos, van~der Werf, Xilouris,
  Isaak, \& Gao}]{Papadopoulos2012}
Papadopoulos, P.~P., van~der Werf, P., Xilouris, E., Isaak, K.~G., \& Gao, Y.
  2012, The Astrophysical Journal, 751, 10

\bibitem[{Pellegrini {et~al.}(2013)Pellegrini, {Smith (PI)}, Wolfire, Draine,
  Crocker, Croxall, van~der Werf, Dale, Rigopoulou, Wilson, Schinnerer, Groves,
  Kreckel, Sandstrom, Armus, Calzetti, Murphy, Walter, Koda, Bayet, Beirao,
  Bolatto, Bradford, Brinks, Hunt, Kennicutt, Knapen, Leroy, Rosolowsky,
  Vigroux, \& Hopwood}]{Pellegrini2013}
Pellegrini, E.~W., {Smith (PI)}, J.~D., Wolfire, M.~G., {et~al.} 2013, The
  Astrophysical Journal, 779, L19

\bibitem[{Pereira-Santaella {et~al.}(2013)Pereira-Santaella, Spinoglio,
  Busquet, Wilson, Glenn, Isaak, Kamenetzky, Rangwala, Schirm, Baes, Barlow,
  Boselli, Cooray, \& Cormier}]{Pereira-Santaella2013}
Pereira-Santaella, M., Spinoglio, L., Busquet, G., {et~al.} 2013, The
  Astrophysical Journal, 768, 55

\bibitem[{Petitpas \& Wilson(1998)}]{Petitpas1998}
Petitpas, G.~R. \& Wilson, C.~D. 1998, The Astrophysical Journal, 503, 219

\bibitem[{Pety {et~al.}(2013)Pety, Schinnerer, Leroy, Hughes, Meidt, Colombo,
  Dumas, Garc\'{\i}a-Burillo, Schuster, Kramer, Dobbs, \& Thompson}]{Pety2013}
Pety, J., Schinnerer, E., Leroy, A.~K., {et~al.} 2013, The Astrophysical
  Journal, 779, 43

\bibitem[{Pilbratt {et~al.}(2010)Pilbratt, Riedinger, Passvogel, Crone, Doyle,
  Gageur, Heras, Jewell, Metcalfe, Ott, \& Schmidt}]{Pilbratt2010}
Pilbratt, G.~L., Riedinger, J.~R., Passvogel, T., {et~al.} 2010, Astronomy and
  Astrophysics, 518, L1

\bibitem[{Pineda {et~al.}(2012)Pineda, Mizuno, R\"{o}llig, Stutzki, Kramer,
  Klein, Rubio, Kawamura, Minamidani, Benz, Burton, Fukui, Koo, \&
  Onishi}]{Pineda2012}
Pineda, J.~L., Mizuno, N., R\"{o}llig, M., {et~al.} 2012, Astronomy \&
  Astrophysics, 544, A84

\bibitem[{Rangwala {et~al.}(2011)Rangwala, Maloney, Glenn, Wilson, Rykala,
  Isaak, Baes, Bendo, Boselli, Bradford, Clements, Cooray, Fulton, Imhof,
  Kamenetzky, Madden, Mentuch, Sacchi, Sauvage, Schirm, Smith, Spinoglio, \&
  Wolfire}]{Rangwala2011}
Rangwala, N., Maloney, P.~R., Glenn, J., {et~al.} 2011, The Astrophysical
  Journal, 743, 94

\bibitem[{Rieke {et~al.}(2004)Rieke, Young, Engelbracht, Kelly, Low, Haller,
  Beeman, Gordon, Stansberry, Misselt, Cadien, Morrison, Rivlis, Latter,
  Noriega‐Crespo, Padgett, Stapelfeldt, Hines, Egami, Muzerolle,
  Alonso‐Herrero, Blaylock, Dole, Hinz, {Le Floc’h}, Papovich,
  Perez‐Gonzalez, Smith, Su, Bennett, Frayer, Henderson, Lu, Masci, Pesenson,
  Rebull, Rho, Keene, Stolovy, Wachter, Wheaton, Werner, \&
  Richards}]{Rieke2004}
Rieke, G.~H., Young, E.~T., Engelbracht, C.~W., {et~al.} 2004, The
  Astrophysical Journal Supplement Series, 154, 25

\bibitem[{Rigopoulou {et~al.}(2013)Rigopoulou, Hurley, Swinyard, Virdee,
  Croxall, Hopwood, Lim, Magdis, Pearson, Pellegrini, Polehampton, \&
  Smith}]{Rigopoulou2013}
Rigopoulou, D., Hurley, P.~D., Swinyard, B.~M., {et~al.} 2013, Monthly Notices
  of \ldots, 9

\bibitem[{Robitaille {et~al.}(2013)Robitaille, Tollerud, Greenfield,
  Droettboom, Bray, Aldcroft, Davis, Ginsburg, Price-Whelan, Kerzendorf,
  Conley, Crighton, Barbary, Muna, Ferguson, Grollier, Parikh, Nair,
  G\"{u}nther, Deil, Woillez, Conseil, Kramer, Turner, Singer, Fox, Weaver,
  Zabalza, Edwards, {Azalee Bostroem}, Burke, Casey, Crawford, Dencheva, Ely,
  Jenness, Labrie, Lim, Pierfederici, Pontzen, Ptak, Refsdal, Servillat, \&
  Streicher}]{Robitaille2013}
Robitaille, T.~P., Tollerud, E.~J., Greenfield, P., {et~al.} 2013, Astronomy \&
  Astrophysics, 558, A33

\bibitem[{Rubin {et~al.}(2007)Rubin, Simpson, Colgan, Dufour, Ray, Erickson,
  Haas, Pauldrach, \& Citron}]{Rubin:2007ig}
Rubin, R.~H., Simpson, J.~P., Colgan, S. W.~J., {et~al.} 2007, Monthly Notices
  of the Royal Astronomical Society, 377, 1407

\bibitem[{Sakamoto {et~al.}(2004)Sakamoto, Matsushita, Peck, Wiedner, \&
  Iono}]{Sakamoto2004}
Sakamoto, K., Matsushita, S., Peck, A.~B., Wiedner, M.~C., \& Iono, D. 2004,
  The Astrophysical Journal, 616, L59

\bibitem[{Sandstrom {et~al.}(2013)Sandstrom, Leroy, Walter, Bolatto, Croxall,
  Draine, Wilson, Wolfire, Calzetti, Kennicutt, Aniano, {Donovan Meyer}, Usero,
  Bigiel, Brinks, de~Blok, Crocker, Dale, Engelbracht, Galametz, Groves, Hunt,
  Koda, Kreckel, Linz, Meidt, Pellegrini, Rix, Roussel, Schinnerer, Schruba,
  Schuster, Skibba, van~der Laan, Appleton, Armus, Brandl, Gordon, Hinz,
  Krause, Montiel, Sauvage, Schmiedeke, Smith, \& Vigroux}]{Sandstrom2013}
Sandstrom, K.~M., Leroy, A.~K., Walter, F., {et~al.} 2013, The Astrophysical
  Journal, 777, 5

\bibitem[{Schirm {et~al.}(2013)Schirm, Wilson, Parkin, Kamenetzky, Glenn,
  Rangwala, Spinoglio, Pereira-Santaella, Baes, Barlow, Clements, Cooray, {De
  Looze}, Karczewski, Madden, R\'{e}my-Ruyer, \& Wu}]{Schirm2013}
Schirm, M. R.~P., Wilson, C.~D., Parkin, T.~J., {et~al.} 2013, Astrophysical
  Journal, 50

\bibitem[{Schruba {et~al.}(2011)Schruba, Leroy, Walter, Bigiel, Brinks,
  de~Blok, Dumas, Kramer, Rosolowsky, Sandstrom, Schuster, Usero, Weiss, \&
  Wiesemeyer}]{Schruba2011}
Schruba, A., Leroy, A.~K., Walter, F., {et~al.} 2011, The Astronomical Journal,
  142, 37

\bibitem[{Solomon \& Rivolo(1987)}]{Solomon1987}
Solomon, P. \& Rivolo, A. 1987, The Astrophysical \ldots, 319, 730

\bibitem[{Soria \& Wu(2002)}]{Soria:2002dn}
Soria, R. \& Wu, K. 2002, Astronomy and Astrophysics, 384, 15

\bibitem[{Soria \& Wu(2003)}]{Soria:2003cx}
Soria, R. \& Wu, K. 2003, Astronomy and Astrophysics, 410, 53

\bibitem[{Spinoglio {et~al.}(2012)Spinoglio, Pereira-Santaella, Busquet,
  Schirm, Wilson, Glenn, Kamenetzky, Rangwala, Maloney, Parkin, Bendo, Madden,
  Wolfire, Boselli, Cooray, \& Page}]{Spinoglio2012}
Spinoglio, L., Pereira-Santaella, M., Busquet, G., {et~al.} 2012, The
  Astrophysical Journal, 758, 108

\bibitem[{{SPIRE Instrument Control Center (ICC)}(2014)}]{observersmanual}
{SPIRE Instrument Control Center (ICC)}. 2014, {The Spectral and Photometric
  Imaging Receiver (SPIRE) Handbook, HERSCHEL-HSC-DOC-0798 version 2.5,
  accessed from http://herschel.esac.esa.int/Documentation.shtml}

\bibitem[{Strong \& Mattox(1996)}]{Strong1996}
Strong, A. \& Mattox, J. 1996, Astronomy and Astrophysics, 308, L21

\bibitem[{Swinyard {et~al.}(2014)Swinyard, Polehampton, Hopwood, Valtchanov,
  Lu, Fulton, Benielli, Imhof, Marchili, Baluteau, Bendo, Ferlet, Griffin, Lim,
  Makiwa, Naylor, Orton, Papageorgiou, Pearson, Schulz, Sidher, Spencer,
  van~der Wiel, \& Wu}]{Swinyard2014}
Swinyard, B.~M., Polehampton, E.~T., Hopwood, R., {et~al.} 2014, Monthly
  Notices of \ldots, 18, 20

\bibitem[{Thatte {et~al.}(2000)Thatte, Tecza, \& Genzel}]{Thatte:2000uo}
Thatte, N., Tecza, M., \& Genzel, R. 2000, Astronomy and Astrophysics, 364, 8

\bibitem[{Thim {et~al.}(2003)Thim, Tammann, Saha, Dolphin, Sandage, Tolstoy, \&
  Labhardt}]{Thim2003}
Thim, F., Tammann, G.~A., Saha, A., {et~al.} 2003, The Astrophysical Journal,
  590, 256

\bibitem[{van~der Tak {et~al.}(2007)van~der Tak, Black, Sch\"{o}ier, Jansen, \&
  van Dishoeck}]{vanderTak2007}
van~der Tak, F. F.~S., Black, J.~H., Sch\"{o}ier, F.~L., Jansen, D.~J., \& van
  Dishoeck, E.~F. 2007, Astronomy and Astrophysics, 468, 627

\bibitem[{van Dishoeck \& Black(1988)}]{Dishoeck1988}
van Dishoeck, E.~F. \& Black, J.~H. 1988, The Astrophysical Journal, 334, 771

\bibitem[{Vogler {et~al.}(2005)Vogler, Madden, Beck, Lundgren, Sauvage,
  Vigroux, \& Ehle}]{Vogler2005}
Vogler, a., Madden, S.~C., Beck, R., {et~al.} 2005, Astronomy and Astrophysics,
  441, 491

\bibitem[{Walter {et~al.}(2008)Walter, Brinks, de~Blok, Bigiel, Kennicutt,
  Thornley, \& Leroy}]{Walter2008}
Walter, F., Brinks, E., de~Blok, W. J.~G., {et~al.} 2008, The Astronomical
  Journal, 136, 2563

\bibitem[{Wolfire {et~al.}(2010)Wolfire, Hollenbach, \& McKee}]{Wolfire2010}
Wolfire, M.~G., Hollenbach, D., \& McKee, C.~F. 2010, The Astrophysical
  Journal, 716, 1191

\bibitem[{Wright {et~al.}(1991)Wright, Mather, Bennett, Cheng, Shafer, Fixsen,
  {Eplee, R. E.}, Isaacman, Read, Boggess, Gulkis, Hauser, Janssen, Kelsall,
  Lubin, Meyer, {Moseley, S. H.}, Murdock, Silverberg, Smoot, Weiss, \&
  Wilkinson}]{Wright1991}
Wright, E.~L., Mather, J.~C., Bennett, C.~L., {et~al.} 1991, The Astrophysical
  Journal, 381, 200

\bibitem[{Wu {et~al.}(2011)Wu, Hogg, \& Moustakas}]{Wu2011}
Wu, R., Hogg, D.~W., \& Moustakas, J. 2011, The Astrophysical Journal, 730, 111

\bibitem[{Wu {et~al.}(2013)Wu, Polehampton, Etxaluze, Makiwa, Naylor, Salji,
  Swinyard, Ferlet, van~der Wiel, Smith, Fulton, Griffin, Baluteau, Benielli,
  Glenn, Hopwood, Imhof, Lim, Lu, Panuzzo, Pearson, Sidher, \&
  Valtchanov}]{Wu2013}
Wu, R., Polehampton, E.~T., Etxaluze, M., {et~al.} 2013, Astronomy \&
  Astrophysics, 556, A116

\bibitem[{Yang {et~al.}(2010)Yang, Stancil, Balakrishnan, \& Forrey}]{Yang2010}
Yang, B., Stancil, P.~C., Balakrishnan, N., \& Forrey, R.~C. 2010, The
  Astrophysical Journal, 718, 1062

\bibitem[{Young \& Scoville(1991)}]{Young1991}
Young, J.~S. \& Scoville, N.~Z. 1991, Annual Review of Astronomy and
  Astrophysics, 29, 581

\bibitem[{Zhao {et~al.}(2013)Zhao, Lu, Xu, Gao, Lord, Howell, Isaak,
  Charmandaris, Diaz-Santos, Appleton, Evans, Iwasawa, Leech, Mazzarella,
  Petric, Sanders, Schulz, Surace, \& van~der Werf}]{Zhao2013}
Zhao, Y., Lu, N., Xu, C.~K., {et~al.} 2013, The Astrophysical Journal, 765, L13

\bibitem[{Zubko {et~al.}(2004)Zubko, Dwek, \& Arendt}]{Zubko2004}
Zubko, V., Dwek, E., \& Arendt, R.~G. 2004, The Astrophysical Journal
  Supplement Series, 152, 211

\end{thebibliography}
\clearpage
\appendix
\section{Line maps observed by the \textsl{Herschel} SPIRE FTS in M83}
\label{appmaps}
\par{We show the observed maps of the lines listed in Table~\ref{linetable} by the SPIRE FTS. All the maps are displayed in the observed spatial resolution and in the units of $\mathrm{W\,m^{-2}\,sr^{-1} GHz^{-1}}$~(line profile maps) and $\mathrm{W\,m^{-2}\,sr^{-1}}$~(integrated intensity maps).}
		\begin{figure*}[htbp!]
			\centering
			\subfloat[][]{
				\includegraphics[width=0.45\textwidth]{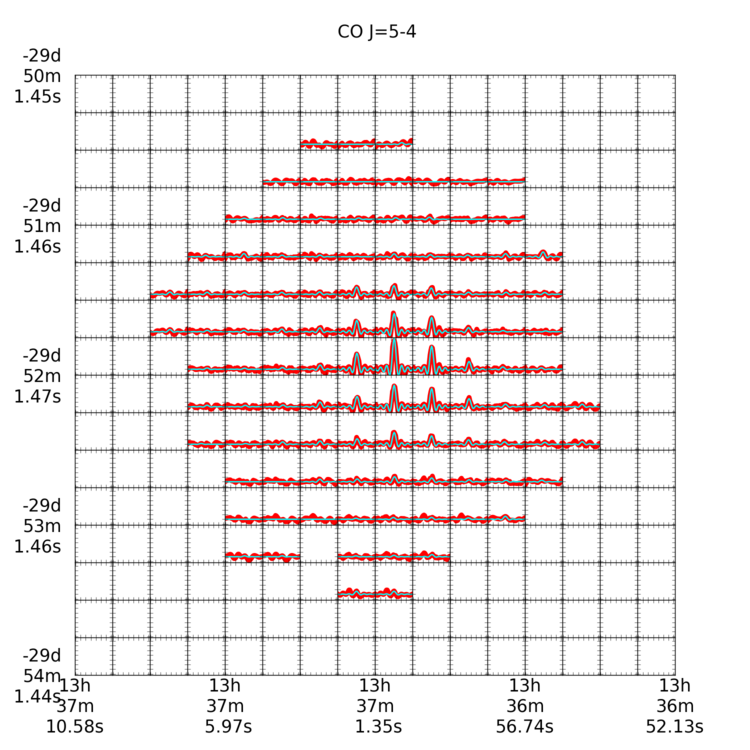}
				\label{co54_linestack}
			}
			\quad
			\subfloat[][]{
				\includegraphics[width=0.45\textwidth]{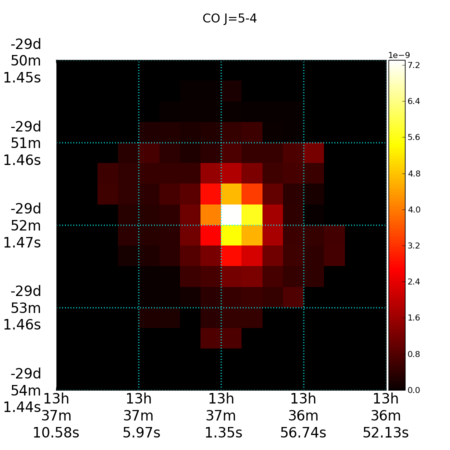}
				\label{co54_intmap}
			}
			\caption{Illustration of the spatial distribution of the observed CO $\mathrm{J}=5-4$ line. The left panel shows the continuum-removed coadded spectrum on every pixel within a range of $568<\nu<582\ \mathrm{GHz}$. The vertical axis in each pixel ranges between $-1.5\,\times\,10^{-18}$ and $8.2\,\times\,10^{-18}\,\mathrm{W\ m^{-2}\ sr^{-1}\ Hz^{-1}}$. The color map on the right is in the units of $\mathrm{W\,m^{-2}\,sr^{-1}}$.}			
		\end{figure*}
		\begin{figure*}[htbp!]
			\centering
			\subfloat[][]{
				\includegraphics[width=0.45\textwidth]{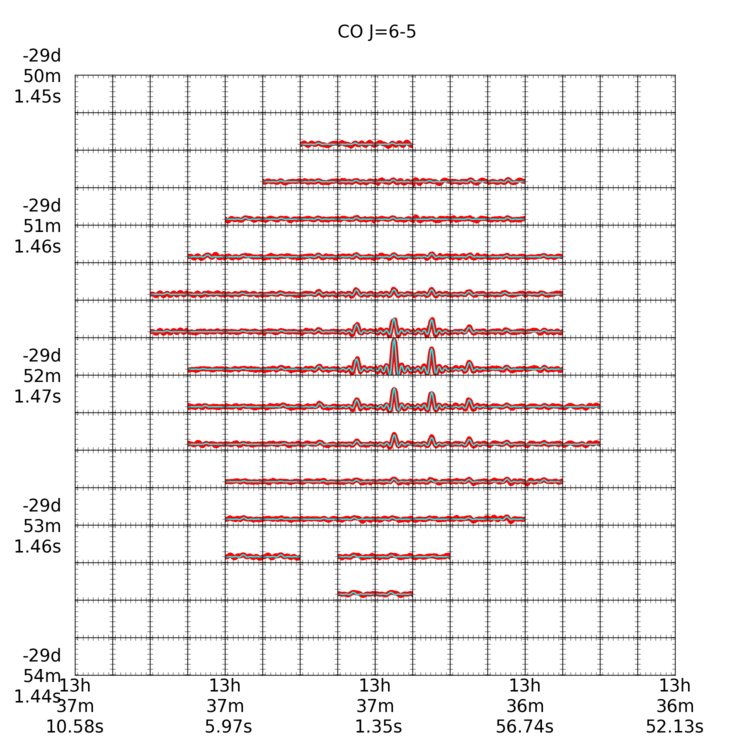}
				\label{co65_linestack}
			}
			\quad
			\subfloat[][]{
				\includegraphics[width=0.45\textwidth]{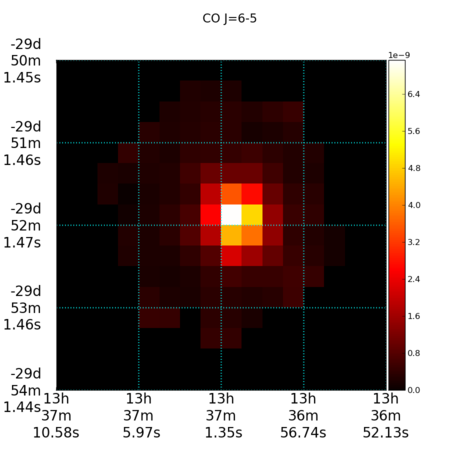}
				\label{co65_intmap}
			}
			\caption{Illustration of the spatial distribution of the observed CO $\mathrm{J}=6-5$ line. The left panel shows the continuum-removed coadded spectrum on every pixel within a range of $683<\nu<698\ \mathrm{GHz}$. The vertical axis in each pixel ranges between $-1.5\,\times\,10^{-18}$ and $8.2\,\times\,10^{-18}\,\mathrm{W\ m^{-2}\ sr^{-1}\ Hz^{-1}}$. The color map on the right is in the units of $\mathrm{W\,m^{-2}\,sr^{-1}}$.}			
		\end{figure*}
		\begin{figure*}[htbp!]
			\centering
			\subfloat[][]{
				\includegraphics[width=0.45\textwidth]{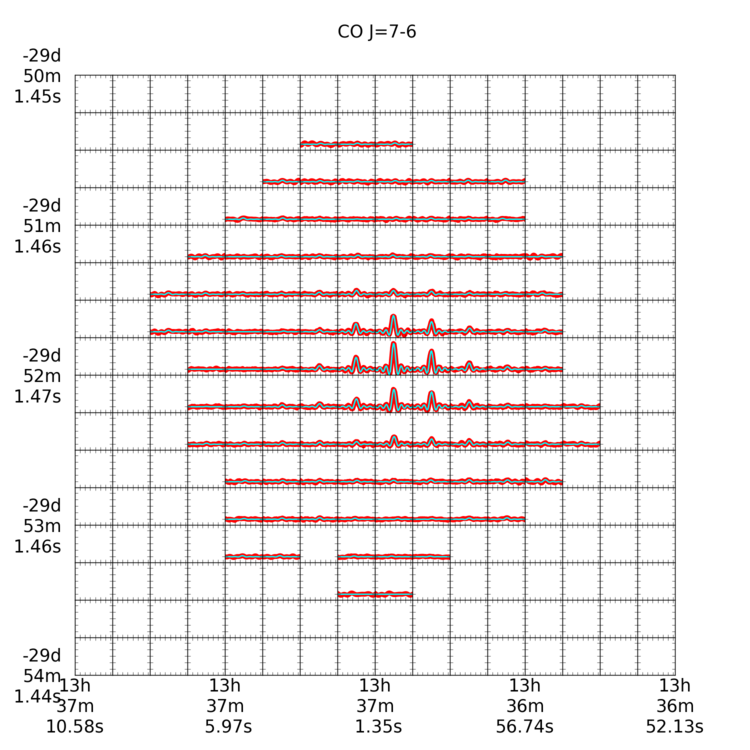}
				\label{co76_linestack}
			}
			\quad
			\subfloat[][]{
				\includegraphics[width=0.45\textwidth]{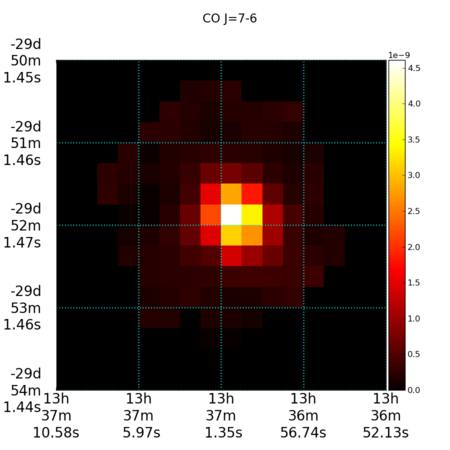}
				\label{co76_intmap}
			}
			\caption{Illustration of the spatial distribution of the observed CO $\mathrm{J}=7-6$ line. The left panel shows the continuum-removed coadded spectrum on every pixel within a range of $798<\nu<813\ \mathrm{GHz}$. The vertical axis in each pixel ranges between $-9.0\,\times\,10^{-19}$ and $5.0\,\times\,10^{-18}\,\mathrm{W\ m^{-2}\ sr^{-1}\ Hz^{-1}}$. The color map on the right is in the units of $\mathrm{W\,m^{-2}\,sr^{-1}}$.}			
		\end{figure*}
		\begin{figure*}[htbp!]
			\centering
			\subfloat[][]{
				\includegraphics[width=0.45\textwidth]{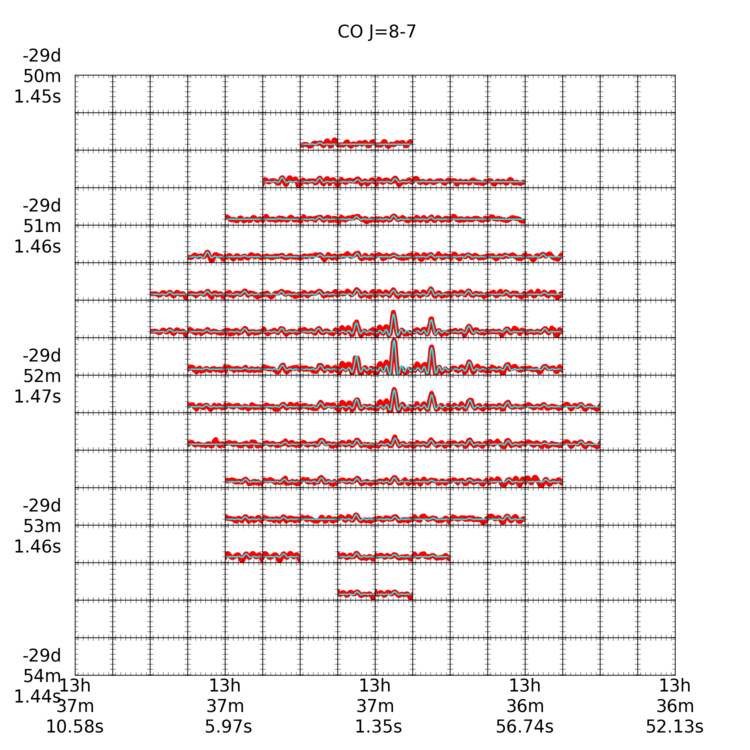}
				\label{co87_linestack}
			}
			\quad
			\subfloat[][]{
				\includegraphics[width=0.45\textwidth]{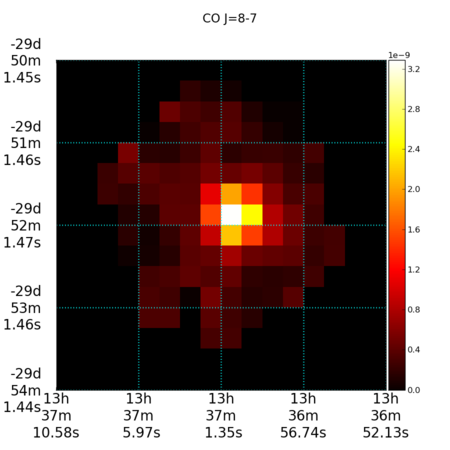}
				\label{co87_intmap}
			}
			\caption{Illustration of the spatial distribution of the observed CO $\mathrm{J}=8-7$ line. The left panel shows the continuum-removed coadded spectrum on every pixel within a range of $913<\nu<928\ \mathrm{GHz}$. The vertical axis in each pixel ranges between $-6.3\,\times\,10^{-19}$ and $3.5\,\times\,10^{-18}\,\mathrm{W\ m^{-2}\ sr^{-1}\ Hz^{-1}}$. The color map on the right is in the units of $\mathrm{W\,m^{-2}\,sr^{-1}}$.}			
		\end{figure*}
		\begin{figure*}[htbp!]
			\centering
			\subfloat[][]{
				\includegraphics[width=0.45\textwidth]{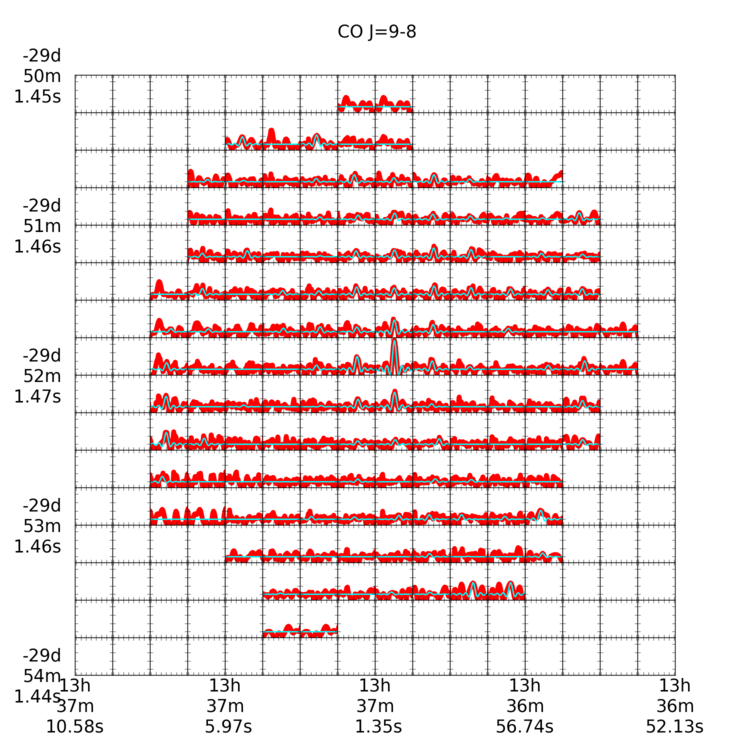}
				\label{co98_linestack}
			}
			\quad
			\subfloat[][]{
				\includegraphics[width=0.45\textwidth]{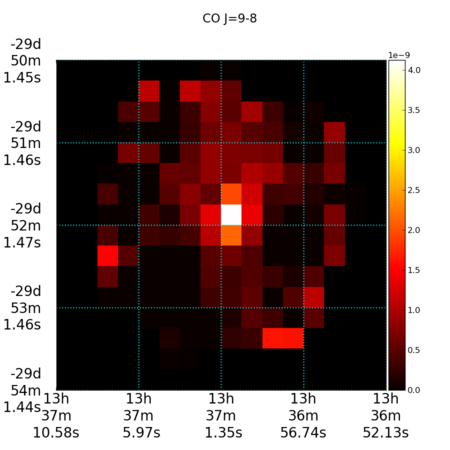}
				\label{co98_intmap}
			}
			\caption{Illustration of the spatial distribution of the observed CO $\mathrm{J}=9-8$ line. The left panel shows the continuum-removed coadded spectrum on every pixel within a range of $1028<\nu<1042\ \mathrm{GHz}$. The vertical axis in each pixel ranges between $-1.3\,\times\,10^{-18}$ and $7.1\,\times\,10^{-18}\,\mathrm{W\ m^{-2}\ sr^{-1}\ Hz^{-1}}$. The color map on the right is in the units of $\mathrm{W\,m^{-2}\,sr^{-1}}$.}			
		\end{figure*}
		\begin{figure*}[htbp!]
			\centering
			\subfloat[][]{
				\includegraphics[width=0.45\textwidth]{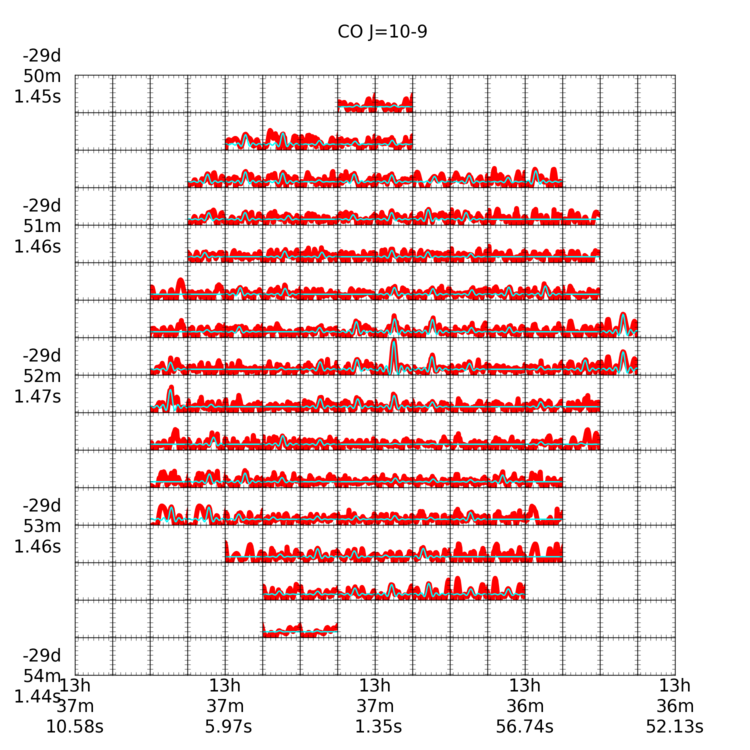}
				\label{co109_linestack}
			}
			\quad
			\subfloat[][]{
				\includegraphics[width=0.45\textwidth]{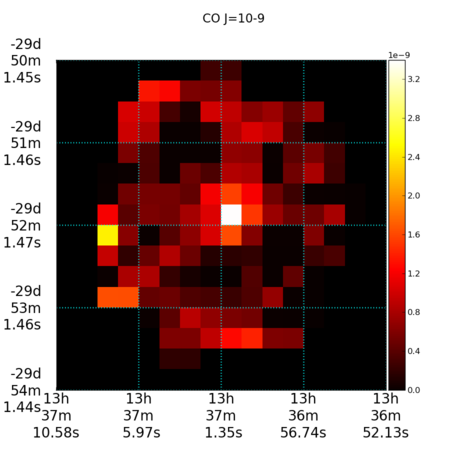}
				\label{co109_intmap}
			}
			\caption{Illustration of the spatial distribution of the observed CO $\mathrm{J}=10-9$ line. The left panel shows the continuum-removed coadded spectrum on every pixel within a range of $1143<\nu<1157\ \mathrm{GHz}$. The vertical axis in each pixel ranges between $-1.1\,\times\,10^{-18}$ and $6.1\,\times\,10^{-18}\,\mathrm{W\ m^{-2}\ sr^{-1}\ Hz^{-1}}$. The color map on the right is in the units of $\mathrm{W\,m^{-2}\,sr^{-1}}$.}			
		\end{figure*}
		\begin{figure*}[htbp!]
			\centering
			\subfloat[][]{
				\includegraphics[width=0.45\textwidth]{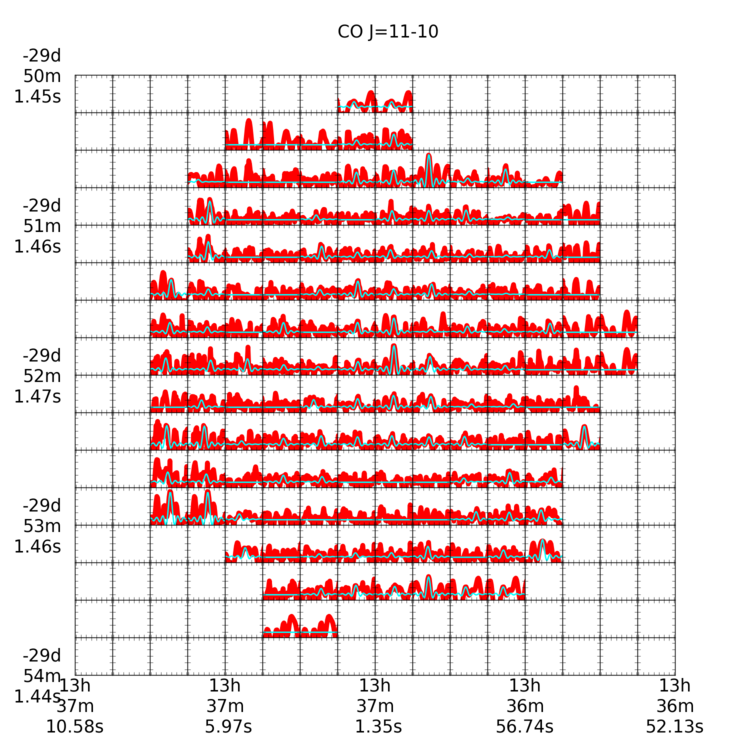}
				\label{co1110_linestack}
			}
			\quad
			\subfloat[][]{
				\includegraphics[width=0.45\textwidth]{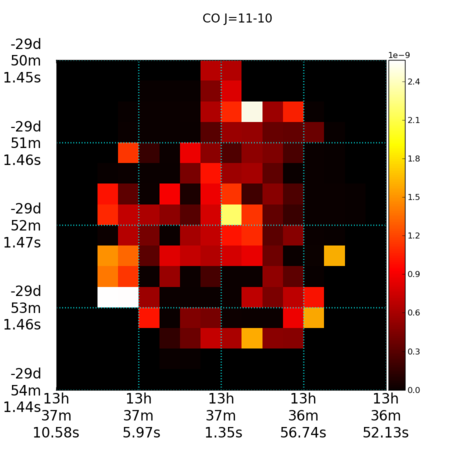}
				\label{co1110_intmap}
			}
			\caption{Illustration of the spatial distribution of the observed CO $\mathrm{J}=11-10$ line. The left panel shows the continuum-removed coadded spectrum on every pixel within a range of $1258<\nu<1272\ \mathrm{GHz}$. The vertical axis in each pixel ranges between $-9.2\,\times\,10^{-19}$ and $5.1\,\times\,10^{-18}\,\mathrm{W\ m^{-2}\ sr^{-1}\ Hz^{-1}}$. The color map on the right is in the units of $\mathrm{W\,m^{-2}\,sr^{-1}}$.}			
		\end{figure*}
		\begin{figure*}[htbp!]
			\centering
			\subfloat[][]{
				\includegraphics[width=0.45\textwidth]{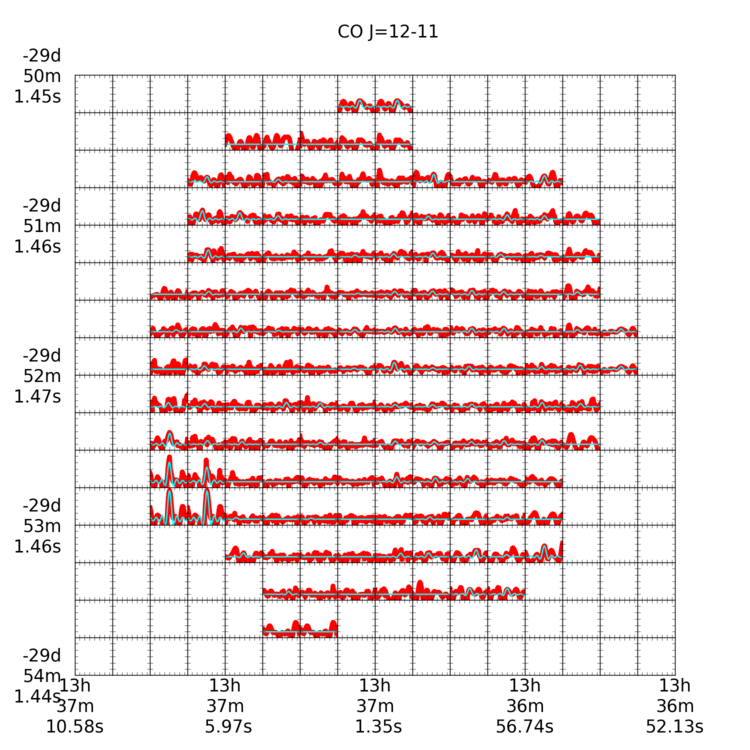}
				\label{co1211_linestack}
			}
			\quad
			\subfloat[][]{
				\includegraphics[width=0.45\textwidth]{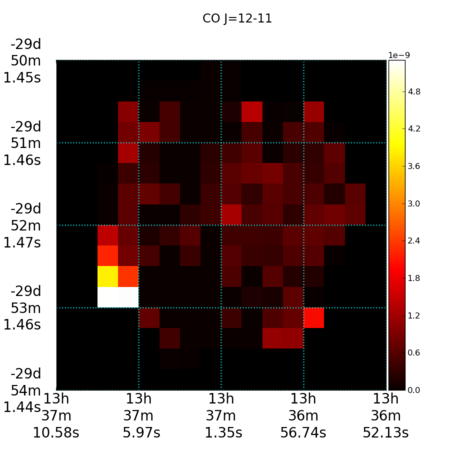}
				\label{co1211_intmap}
			}
			\caption{Illustration of the spatial distribution of the observed CO $\mathrm{J}=12-11$ line. The left panel shows the continuum-removed coadded spectrum on every pixel within a range of $1372<\nu<1387\ \mathrm{GHz}$. The vertical axis in each pixel ranges between $-8.9\,\times\,10^{-19}$ and $4.9\,\times\,10^{-18}\,\mathrm{W\ m^{-2}\ sr^{-1}\ Hz^{-1}}$. The color map on the right is in the units of $\mathrm{W\,m^{-2}\,sr^{-1}}$.}			
		\end{figure*}
		\begin{figure*}[htbp!]
			\centering
			\subfloat[][]{
				\includegraphics[width=0.45\textwidth]{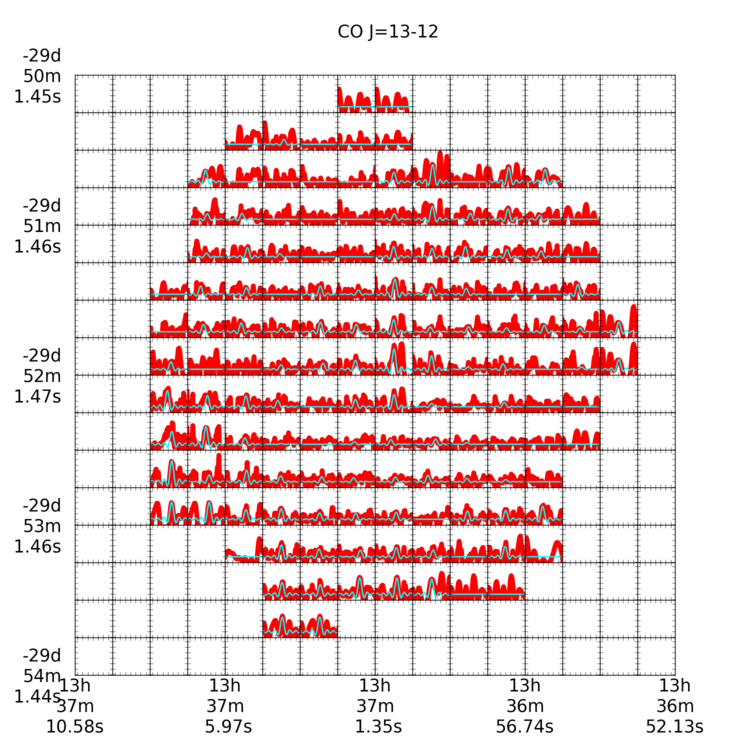}
				\label{co1312_linestack}
			}
			\quad
			\subfloat[][]{
				\includegraphics[width=0.45\textwidth]{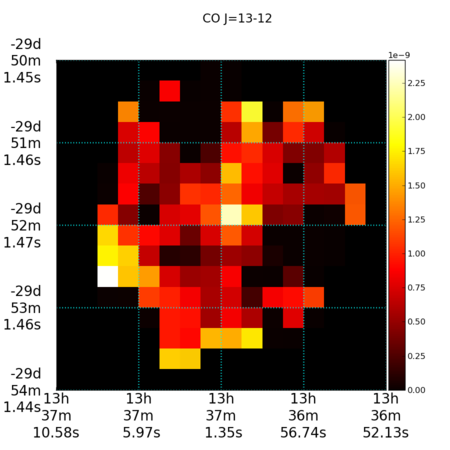}
				\label{co1312_intmap}
			}
			\caption{Illustration of the spatial distribution of the observed CO $\mathrm{J}=13-12$ line. The left panel shows the continuum-removed coadded spectrum on every pixel within a range of $1487<\nu<1502\ \mathrm{GHz}$. The vertical axis in each pixel ranges between $-9.7\,\times\,10^{-19}$ and $5.4\,\times\,10^{-18}\,\mathrm{W\ m^{-2}\ sr^{-1}\ Hz^{-1}}$. The color map on the right is in the units of $\mathrm{W\,m^{-2}\,sr^{-1}}$.}			
		\end{figure*}
		\begin{figure*}[htbp!]
			\centering
			\subfloat[][]{
				\includegraphics[width=0.45\textwidth]{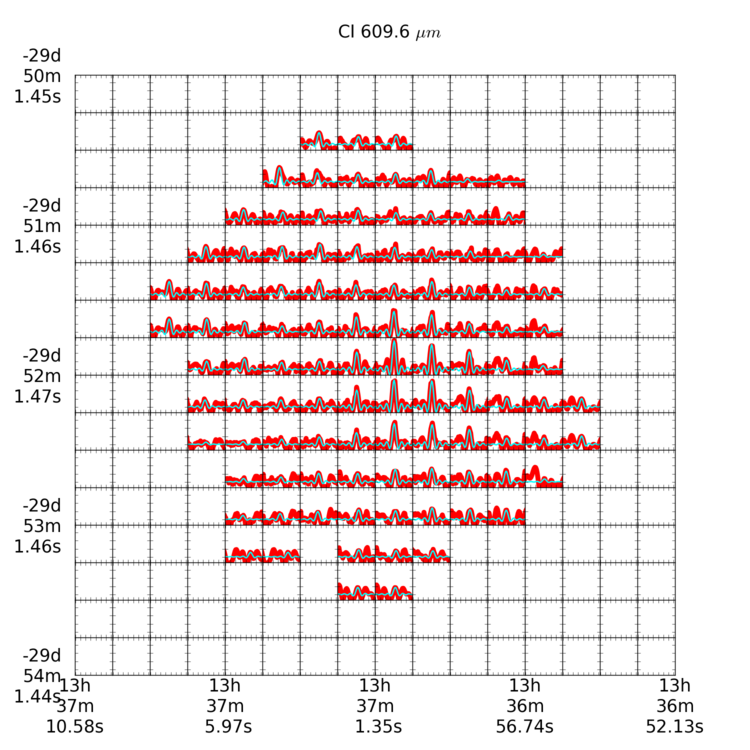}
				\label{ci609_linestack}
			}
			\quad
			\subfloat[][]{
				\includegraphics[width=0.45\textwidth]{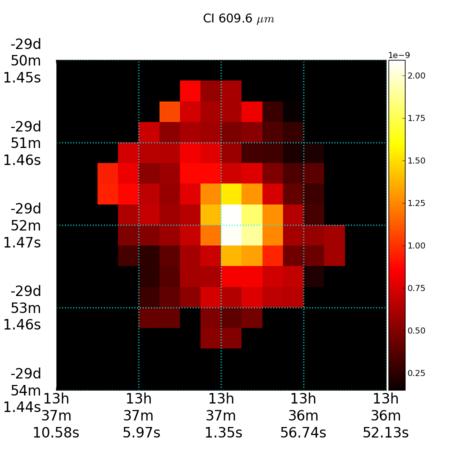}
				\label{ci609_intmap}
			}
			\caption{Illustration of the spatial distribution of the observed $\cison$ line. The left panel shows the continuum-removed coadded spectrum on every pixel within a range of $484<\nu<499\ \mathrm{GHz}$. The vertical axis in each pixel ranges between $-3.9\,\times\,10^{-19}$ and $2.2\,\times\,10^{-18}\,\mathrm{W\ m^{-2}\ sr^{-1}\ Hz^{-1}}$. The color map on the right is in the units of $\mathrm{W\,m^{-2}\,sr^{-1}}$.}			
		\end{figure*}
		\begin{figure*}[htbp!]
			\centering
			\subfloat[][]{
				\includegraphics[width=0.45\textwidth]{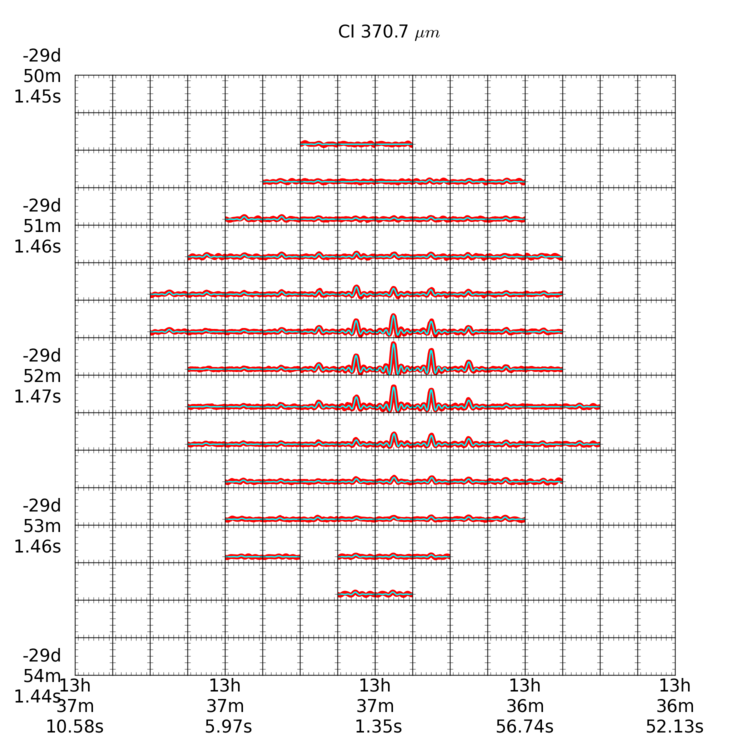}
				\label{ci370_linestack}
			}
			\quad
			\subfloat[][]{
				\includegraphics[width=0.45\textwidth]{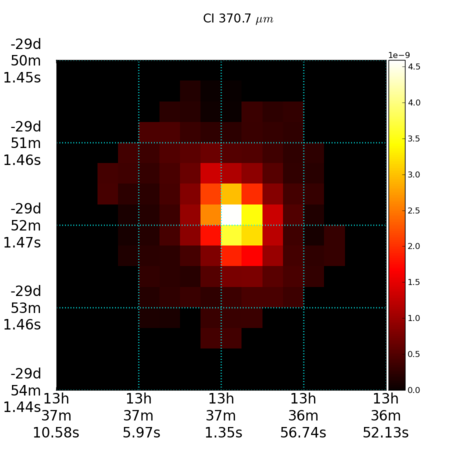}
				\label{ci370_intmap}
			}
			\caption{Illustration of the spatial distribution of the observed $\cits$ line. The left panel shows the continuum-removed coadded spectrum on every pixel within a range of $801<\nu<815\ \mathrm{GHz}$. The vertical axis in each pixel ranges between $-8.8\,\times\,10^{-19}$ and $4.9\,\times\,10^{-18}\,\mathrm{W\ m^{-2}\ sr^{-1}\ Hz^{-1}}$. The color map on the right is in the units of $\mathrm{W\,m^{-2}\,sr^{-1}}$.}			
		\end{figure*}

\end{document}